%
%
%
\def\unredoffs{} \def\redoffs{\voffset=-.31truein\hoffset=-.48truein}
\def\speclscape{}

%
%
%
%
%
\newbox\leftpage \newdimen\fullhsize \newdimen\hstitle \newdimen\hsbody
\tolerance=1000\hfuzz=2pt
\catcode`\@=11 
\ifx\hyperdef\UNd@FiNeD\def\hyperdef#1#2#3#4{#4}\def\hyperref#1#2#3#4{#4}\fi
\def\bigans{b }
\def\answ{b }
%
\ifx\answ\bigans\message{(This will come out unreduced.}
\magnification=1200\unredoffs\baselineskip=16pt plus 2pt minus 1pt
\hsbody=\hsize \hstitle=\hsize 
\else\message{(This will be reduced.} \let\l@r=L
\magnification=1000\baselineskip=16pt plus 2pt minus 1pt \vsize=7truein
\redoffs \hstitle=8truein\hsbody=4.75truein\fullhsize=10truein\hsize=\hsbody
\output={\ifnum\pageno=0 
  \shipout\vbox{\speclscape{\hsize\fullhsize\makeheadline}
    \hbox to \fullhsize{\hfill\pagebody\hfill}}\advancepageno
  \else
  \almostshipout{\leftline{\vbox{\pagebody\makefootline}}}\advancepageno
  \fi}
\def\almostshipout#1{\if L\l@r \count1=1 \message{[\the\count0.\the\count1]}
      \global\setbox\leftpage=#1 \global\let\l@r=R
 \else \count1=2
  \shipout\vbox{\speclscape{\hsize\fullhsize\makeheadline}
      \hbox to\fullhsize{\box\leftpage\hfil#1}}  \global\let\l@r=L\fi}
\fi
%
\newcount\yearltd\yearltd=\year\advance\yearltd by -2000

\def\Title#1#2{\nopagenumbers\abstractfont\hsize=\hstitle\rightline{#1}%
\vskip 1in\centerline{\titlefont #2}\abstractfont\vskip .5in\pageno=0}
\def\Date#1{\vfill\leftline{#1}\tenpoint\supereject\global\hsize=\hsbody%
\footline={\hss\tenrm\hyperdef\hypernoname{page}\folio\folio\hss}}%
%

\def\draftmode{\message{ DRAFTMODE }\def\draftdate{{\rm preliminary draft:
\number\month/\number\day/\number\yearltd\ \ \hourmin}}%
\headline={\hfil\draftdate}\writelabels\baselineskip=20pt plus 2pt minus 2pt
 {\count255=\time\divide\count255 by 60 \xdef\hourmin{\number\count255}
  \multiply\count255 by-60\advance\count255 by\time
  \xdef\hourmin{\hourmin:\ifnum\count255<10 0\fi\the\count255}}}
\def\nolabels{\def\wrlabeL##1{}\def\eqlabeL##1{}\def\reflabeL##1{}}
\def\writelabels{\def\wrlabeL##1{\leavevmode\vadjust{\rlap{\smash%
{\line{{\escapechar=` \hfill\rlap{\sevenrm\hskip.03in\string##1}}}}}}}%
\def\eqlabeL##1{{\escapechar-1\rlap{\sevenrm\hskip.05in\string##1}}}%
\def\reflabeL##1{\noexpand\llap{\noexpand\sevenrm\string\string\string##1}}}
\nolabels
%
\global\newcount\secno \global\secno=0
\global\newcount\meqno \global\meqno=1
\def\s@csym{}
\def\newsec#1{\global\advance\secno by1%
{\toks0{#1}\message{(\the\secno. \the\toks0)}}%
\global\subsecno=0\eqnres@t\let\s@csym\secsym\xdef\secn@m{\the\secno}\noindent
{\bf\hyperdef\hypernoname{section}{\the\secno}{\the\secno.} #1}%
\writetoca{{\string\hyperref{}{section}{\the\secno}{\the\secno.}} {#1}}%
\par\nobreak\medskip\nobreak}
\def\eqnres@t{\xdef\secsym{\the\secno.}\global\meqno=1\bigbreak\bigskip}
\def\sequentialequations{\def\eqnres@t{\bigbreak}}\xdef\secsym{}
\global\newcount\subsecno \global\subsecno=0
\def\subsec#1{\global\advance\subsecno by1%
{\toks0{#1}\message{(\s@csym\the\subsecno. \the\toks0)}}%
\ifnum\lastpenalty>9000\else\bigbreak\fi
\noindent{\it\hyperdef\hypernoname{subsection}{\secn@m.\the\subsecno}%
{\secn@m.\the\subsecno.} #1}\writetoca{\string\quad
{\string\hyperref{}{subsection}{\secn@m.\the\subsecno}{\secn@m.\the\subsecno.}}
{#1}}\par\nobreak\medskip\nobreak}
\def\appendix#1#2{\global\meqno=1\global\subsecno=0\xdef\secsym{\hbox{#1.}}%
\bigbreak\bigskip\noindent{\bf Appendix \hyperdef\hypernoname{appendix}{#1}%
{#1.} #2}{\toks0{(#1. #2)}\message{\the\toks0}}%
\xdef\s@csym{#1.}\xdef\secn@m{#1}%
\writetoca{\string\hyperref{}{appendix}{#1}{Appendix {#1.}} {#2}}%
\par\nobreak\medskip\nobreak}
%
%
\def\checkm@de#1#2{\ifmmode{\def\f@rst##1{##1}\hyperdef\hypernoname{equation}%
{#1}{#2}}\else\hyperref{}{equation}{#1}{#2}\fi}
\def\eqnn#1{\DefWarn#1\xdef #1{(\noexpand\relax\noexpand\checkm@de%
{\s@csym\the\meqno}{\secsym\the\meqno})}%
\wrlabeL#1\writedef{#1\leftbracket#1}\global\advance\meqno by1}
\def\f@rst#1{\c@t#1a\em@ark}\def\c@t#1#2\em@ark{#1}
\def\eqna#1{\DefWarn#1\wrlabeL{#1$\{\}$}%
\xdef #1##1{(\noexpand\relax\noexpand\checkm@de%
{\s@csym\the\meqno\noexpand\f@rst{##1}}{\hbox{$\secsym\the\meqno##1$}})}
\writedef{#1\numbersign1\leftbracket#1{\numbersign1}}\global\advance\meqno by1}
\def\eqn#1#2{\DefWarn#1%
\xdef #1{(\noexpand\hyperref{}{equation}{\s@csym\the\meqno}%
{\secsym\the\meqno})}$$#2\eqno(\hyperdef\hypernoname{equation}%
{\s@csym\the\meqno}{\secsym\the\meqno})\eqlabeL#1$$%
\writedef{#1\leftbracket#1}\global\advance\meqno by1}
\def\xeqn{\expandafter\xe@n}\def\xe@n(#1){#1}
\def\xeqna#1{\expandafter\xe@n#1}
\def\eqns#1{(\e@ns #1{\hbox{}})}
\def\e@ns#1{\ifx\UNd@FiNeD#1\message{eqnlabel \string#1 is undefined.}%
\xdef#1{(?.?)}\fi{\let\hyperref=\relax\xdef\next{#1}}%
\ifx\next\em@rk\def\next{}\else%
\ifx\next#1\xeqn#1\else\def\n@xt{#1}\ifx\n@xt\next#1\else\xeqna#1\fi
\fi\let\next=\e@ns\fi\next}

\def\DefWarn#1{\ifx\UNd@FiNeD#1\else
\immediate\write16{*** WARNING: the label \string#1 is already defined ***}\fi}
%
\newskip\footskip\footskip14pt plus 1pt minus 1pt 
\def\footnotefont{\ninepoint}\def\f@t#1{\footnotefont #1\@foot}
\def\f@@t{\baselineskip\footskip\bgroup\footnotefont\aftergroup\@foot\let\next}
\setbox\strutbox=\hbox{\vrule height9.5pt depth4.5pt width0pt}
\global\newcount\ftno \global\ftno=0
\def\foot{\global\advance\ftno by1\def\foot@rg{\hyperref{}{footnote}%
{\the\ftno}{\the\ftno}\xdef\foot@rg{\noexpand\hyperdef\noexpand\hypernoname%
{footnote}{\the\ftno}{\the\ftno}}}\footnote{$^{\foot@rg}$}}
%
\newwrite\ftfile
\def\footend{\def\foot{\global\advance\ftno by1\chardef\wfile=\ftfile
\hyperref{}{footnote}{\the\ftno}{$^{\the\ftno}$}%
\ifnum\ftno=1\immediate\openout\ftfile=\jobname.fts\fi%
\immediate\write\ftfile{\noexpand\smallskip%
\noexpand\item{\noexpand\hyperdef\noexpand\hypernoname{footnote}
{\the\ftno}{f\the\ftno}:\ }\pctsign}\findarg}%
\def\footatend{\vfill\eject\immediate\closeout\ftfile{\parindent=20pt
\centerline{\bf Footnotes}\nobreak\bigskip\input \jobname.fts }}}
\def\footatend{}
%
%
\global\newcount\refno \global\refno=1
\newwrite\rfile
\def\ref{[\hyperref{}{reference}{\the\refno}{\the\refno}]\nref}
\def\nref#1{\DefWarn#1%
\xdef#1{[\noexpand\hyperref{}{reference}{\the\refno}{\the\refno}]}%
\writedef{#1\leftbracket#1}%
\ifnum\refno=1\immediate\openout\rfile=\jobname.refs\fi
\chardef\wfile=\rfile\immediate\write\rfile{\noexpand\item{[\noexpand\hyperdef%
\noexpand\hypernoname{reference}{\the\refno}{\the\refno}]\ }%
\reflabeL{#1\hskip.31in}\pctsign}\global\advance\refno by1\findarg}
\def\findarg#1#{\begingroup\obeylines\newlinechar=`\^^M\pass@rg}
{\obeylines\gdef\pass@rg#1{\writ@line\relax #1^^M\hbox{}^^M}%
\gdef\writ@line#1^^M{\expandafter\toks0\expandafter{\striprel@x #1}%
\edef\next{\the\toks0}\ifx\next\em@rk\let\next=\endgroup\else\ifx\next\empty%
\else\immediate\write\wfile{\the\toks0}\fi\let\next=\writ@line\fi\next\relax}}
\def\striprel@x#1{} \def\em@rk{\hbox{}}
\def\lref{\begingroup\obeylines\lr@f}
\def\lr@f#1#2{\DefWarn#1\gdef#1{\let#1=\UNd@FiNeD\ref#1{#2}}\endgroup\unskip}

\def\addref#1{\immediate\write\rfile{\noexpand\item{}#1}} 
\def\listrefs{\footatend\vfill\supereject\immediate\closeout\rfile\writestoppt
\baselineskip=\footskip\centerline{{\bf References}}\bigskip{\parindent=20pt%
\frenchspacing\escapechar=` \input \jobname.refs\vfill\eject}\nonfrenchspacing}
\def\startrefs#1{\immediate\openout\rfile=\jobname.refs\refno=#1}
\def\xref{\expandafter\xr@f}\def\xr@f[#1]{#1}
\def\refs#1{\count255=1[\r@fs #1{\hbox{}}]}
\def\r@fs#1{\ifx\UNd@FiNeD#1\message{reflabel \string#1 is undefined.}%
\nref#1{need to supply reference \string#1.}\fi%
\vphantom{\hphantom{#1}}{\let\hyperref=\relax\xdef\next{#1}}%
\ifx\next\em@rk\def\next{}%
\else\ifx\next#1\ifodd\count255\relax\xref#1\count255=0\fi%
\else#1\count255=1\fi\let\next=\r@fs\fi\next}
%

%
\newwrite\ffile\global\newcount\figno \global\figno=1
\def\fig{fig.~\hyperref{}{figure}{\the\figno}{\the\figno}\nfig}
\def\nfig#1{\DefWarn#1%
\xdef#1{fig.~\noexpand\hyperref{}{figure}{\the\figno}{\the\figno}}%
\writedef{#1\leftbracket fig.\noexpand~\xfig#1}%
\ifnum\figno=1\immediate\openout\ffile=\jobname.figs\fi\chardef\wfile=\ffile%
{\let\hyperref=\relax
\immediate\write\ffile{\noexpand\medskip\noexpand\item{Fig.\ %
\noexpand\hyperdef\noexpand\hypernoname{figure}{\the\figno}{\the\figno}. }
\reflabeL{#1\hskip.55in}\pctsign}}\global\advance\figno by1\findarg}
\def\listfigs{\vfill\eject\immediate\closeout\ffile{\parindent40pt
\baselineskip14pt\centerline{{\bf Figure Captions}}\nobreak\medskip
\escapechar=` \input \jobname.figs\vfill\eject}}
\def\xfig{\expandafter\xf@g}\def\xf@g fig.\penalty\@M\ {}
\def\figs#1{figs.~\f@gs #1{\hbox{}}}
\def\f@gs#1{{\let\hyperref=\relax\xdef\next{#1}}\ifx\next\em@rk\def\next{}\else
\ifx\next#1\xfig #1\else#1\fi\let\next=\f@gs\fi\next}
\def\figin{\epsfcheck\figin}\def\figins{\epsfcheck\figins}
\def\epsfcheck{\ifx\epsfbox\UNd@FiNeD
\message{(NO epsf.tex, FIGURES WILL BE IGNORED)}
\gdef\figin##1{\vskip2in}\gdef\figins##1{\hskip.5in}
\else\message{(FIGURES WILL BE INCLUDED)}%
\gdef\figin##1{##1}\gdef\figins##1{##1}\fi}
\def\DefWarn#1{}
\def\figinsert{\goodbreak\midinsert}
\def\ifig#1#2#3{\DefWarn#1\xdef#1{fig.~\noexpand\hyperref{}{figure}%
{\the\figno}{\the\figno}}\writedef{#1\leftbracket fig.\noexpand~\xfig#1}%
\figinsert\figin{\centerline{#3}}\medskip\centerline{\vbox{\baselineskip12pt
\advance\hsize by -1truein\noindent\wrlabeL{#1=#1}\footnotefont%
{\bf Fig.~\hyperdef\hypernoname{figure}{\the\figno}{\the\figno}:} #2}}
\bigskip\endinsert\global\advance\figno by1}
\newwrite\lfile
{\escapechar-1\xdef\pctsign{\string\%}\xdef\leftbracket{\string\{}
\xdef\rightbracket{\string\}}\xdef\numbersign{\string\#}}
\def\writedefs{\immediate\openout\lfile=\jobname.defs \def\writedef##1{%
{\let\hyperref=\relax\let\hyperdef=\relax\let\hypernoname=\relax
 \immediate\write\lfile{\string\def\string##1\rightbracket}}}}%
\def\writestop{\def\writestoppt{\immediate\write\lfile{\string\pageno
 \the\pageno\string\startrefs\leftbracket\the\refno\rightbracket
 \string\def\string\secsym\leftbracket\secsym\rightbracket
 \string\secno\the\secno\string\meqno\the\meqno}\immediate\closeout\lfile}}
\def\writestoppt{}\def\writedef#1{}
\def\seclab#1{\DefWarn#1%
\xdef #1{\noexpand\hyperref{}{section}{\the\secno}{\the\secno}}%
\writedef{#1\leftbracket#1}\wrlabeL{#1=#1}}
\def\subseclab#1{\DefWarn#1%
\xdef #1{\noexpand\hyperref{}{subsection}{\secn@m.\the\subsecno}%
{\secn@m.\the\subsecno}}\writedef{#1\leftbracket#1}\wrlabeL{#1=#1}}
\def\applab#1{\DefWarn#1%
\xdef #1{\noexpand\hyperref{}{appendix}{\secn@m}{\secn@m}}%
\writedef{#1\leftbracket#1}\wrlabeL{#1=#1}}
\newwrite\tfile \def\writetoca#1{}
\def\leaderfill{\leaders\hbox to 1em{\hss.\hss}\hfill}
\def\writetoc{\immediate\openout\tfile=\jobname.toc
   \def\writetoca##1{{\edef\next{\write\tfile{\noindent ##1
   \string\leaderfill {\string\hyperref{}{page}{\noexpand\number\pageno}%
                       {\noexpand\number\pageno}} \par}}\next}}}
\newread\ch@ckfile
\def\listtoc{\immediate\closeout\tfile\immediate\openin\ch@ckfile=\jobname.toc
\ifeof\ch@ckfile\message{no file \jobname.toc, no table of contents this pass}%
\else\closein\ch@ckfile\centerline{\bf Contents}\nobreak\medskip%
{\baselineskip=12pt\footnotefont\parskip=0pt\catcode`\@=11\input\jobname.toc
\catcode`\@=12\bigbreak\bigskip}\fi}
\catcode`\@=12 
%
\edef\tfontsize{\ifx\answ\bigans scaled\magstep3\else scaled\magstep4\fi}
\font\titlerm=cmr10 \tfontsize \font\titlerms=cmr7 \tfontsize
\font\titlermss=cmr5 \tfontsize \font\titlei=cmmi10 \tfontsize
\font\titleis=cmmi7 \tfontsize \font\titleiss=cmmi5 \tfontsize
\font\titlesy=cmsy10 \tfontsize \font\titlesys=cmsy7 \tfontsize
\font\titlesyss=cmsy5 \tfontsize \font\titleit=cmti10 \tfontsize
\skewchar\titlei='177 \skewchar\titleis='177 \skewchar\titleiss='177
\skewchar\titlesy='60 \skewchar\titlesys='60 \skewchar\titlesyss='60
\def\titlefont{\def\rm{\fam0\titlerm}
\textfont0=\titlerm \scriptfont0=\titlerms \scriptscriptfont0=\titlermss
\textfont1=\titlei \scriptfont1=\titleis \scriptscriptfont1=\titleiss
\textfont2=\titlesy \scriptfont2=\titlesys \scriptscriptfont2=\titlesyss
\textfont\itfam=\titleit \def\it{\fam\itfam\titleit}\rm}
 \ifx\answ\bigans\else scaled\magstep1\fi
\ifx\answ\bigans\def\abstractfont{\tenpoint}\else
\font\absit=cmti10 scaled \magstep1
\font\abssl=cmsl10 scaled \magstep1
\font\absrm=cmr10 scaled\magstep1 \font\absrms=cmr7 scaled\magstep1
\font\absrmss=cmr5 scaled\magstep1 \font\absi=cmmi10 scaled\magstep1
\font\absis=cmmi7 scaled\magstep1 \font\absiss=cmmi5 scaled\magstep1
\font\abssy=cmsy10 scaled\magstep1 \font\abssys=cmsy7 scaled\magstep1
\font\abssyss=cmsy5 scaled\magstep1 \font\absbf=cmbx10 scaled\magstep1
\skewchar\absi='177 \skewchar\absis='177 \skewchar\absiss='177
\skewchar\abssy='60 \skewchar\abssys='60 \skewchar\abssyss='60
\def\abstractfont{\def\rm{\fam0\absrm}
\textfont0=\absrm \scriptfont0=\absrms \scriptscriptfont0=\absrmss
\textfont1=\absi \scriptfont1=\absis \scriptscriptfont1=\absiss
\textfont2=\abssy \scriptfont2=\abssys \scriptscriptfont2=\abssyss
\textfont\itfam=\absit \def\it{\fam\itfam\absit}\def\footnotefont{\tenpoint}%
\textfont\slfam=\abssl \def\sl{\fam\slfam\abssl}%
\textfont\bffam=\absbf \def\bf
{\fam\bffam\absbf}\rm}\fi
\def\tenpoint{\def\rm{\fam0\tenrm}
\textfont0=\tenrm \scriptfont0=\sevenrm \scriptscriptfont0=\fiverm
\textfont1=\teni  \scriptfont1=\seveni  \scriptscriptfont1=\fivei
\textfont2=\tensy \scriptfont2=\sevensy \scriptscriptfont2=\fivesy
\textfont\itfam=\tenit \def\it{\fam\itfam\tenit}\def\footnotefont{\ninepoint}%
\textfont\bffam=\tenbf \def\bf{\fam\bffam\tenbf}\def\sl{\fam\slfam\tensl}\rm}
\font\ninerm=cmr9 \font\sixrm=cmr6 \font\ninei=cmmi9 \font\sixi=cmmi6
\font\ninesy=cmsy9 \font\sixsy=cmsy6 \font\ninebf=cmbx9
\font\nineit=cmti9 \font\ninesl=cmsl9 \skewchar\ninei='177
\skewchar\sixi='177 \skewchar\ninesy='60 \skewchar\sixsy='60
\def\ninepoint{\def\rm{\fam0\ninerm}
\textfont0=\ninerm \scriptfont0=\sixrm \scriptscriptfont0=\fiverm
\textfont1=\ninei \scriptfont1=\sixi \scriptscriptfont1=\fivei
\textfont2=\ninesy \scriptfont2=\sixsy \scriptscriptfont2=\fivesy
\textfont\itfam=\ninei \def\it{\fam\itfam\nineit}\def\sl{\fam\slfam\ninesl}%
\textfont\bffam=\ninebf \def\bf{\fam\bffam\ninebf}\rm}
%
%
\def\noblackbox{\overfullrule=0pt}
\hyphenation{anom-aly anom-alies coun-ter-term coun-ter-terms}
\def\inv{^{\raise.15ex\hbox{${\scriptscriptstyle -}$}\kern-.05em 1}}

\def\Dsl{\,\raise.15ex\hbox{/}\mkern-13.5mu D} 
\def\dsl{\raise.15ex\hbox{/}\kern-.57em\partial}
\def\del{\partial}

\def\lspace{\ifx\answ\bigans{}\else\qquad\fi}
\def\lbspace{\ifx\answ\bigans{}\else\hskip-.2in\fi} 

\def\boxeqn#1{\vcenter{\vbox{\hrule\hbox{\vrule\kern3pt\vbox{\kern3pt
	\hbox{${\displaystyle #1}$}\kern3pt}\kern3pt\vrule}\hrule}}}
\def\mbox#1#2{\vcenter{\hrule \hbox{\vrule height#2in
		\kern#1in \vrule} \hrule}}  

\def\darr#1{\raise1.5ex\hbox{$\leftrightarrow$}\mkern-16.5mu #1}

\def\roughly#1{\raise.3ex\hbox{$#1$\kern-.75em\lower1ex\hbox{$\sim$}}}

\input amssym
\input epsf

\def\IZ{\relax\ifmmode\mathchoice
{\hbox{\cmss Z\kern-.4em Z}}{\hbox{\cmss Z\kern-.4em Z}} {\lower.9pt\hbox{\cmsss Z\kern-.4em Z}}
{\lower1.2pt\hbox{\cmsss Z\kern-.4em Z}}\else{\cmss Z\kern-.4em Z}\fi}

\newif\ifdraft\draftfalse
\newif\ifinter\interfalse
\ifdraft\draftmode\else\interfalse\fi
\def\journal#1&#2(#3){\unskip, \sl #1\ \bf #2 \rm(19#3) }
\def\andjournal#1&#2(#3){\sl #1~\bf #2 \rm (19#3) }

\def\frac#1#2{{#1\over#2}}

\def\ds{\displaystyle}

\def\inbar{\,\vrule height1.5ex width.4pt depth0pt}
\def\IC{\relax\hbox{$\inbar\kern-.3em{\rm C}$}}
\def\IR{\relax{\rm I\kern-.18em R}}
\def\IP{\relax{\rm I\kern-.18em P}}

%
%


%
\catcode`\@=11
\def\slash#1{\mathord{\mathpalette\c@ncel{#1}}}
\overfullrule=0pt

\def\Z{\hbox{$\bb Z$}}
\def\R{\hbox{$\bb R$}}

\def\underrel#1\over#2{\mathrel{\mathop{\kern\z@#1}\limits_{#2}}}

\catcode`\@=12


%

\def\mod{{\rm mod}}

\def\sgn{{\rm sgn}}


\def\[{[}
\def\]{]}

\def\comment#1{ }

\global\newcount\subsubsecno \global\subsubsecno=0
\let\oldsubsec=\subsec
\def\subsec{\global\subsubsecno=0\relax\oldsubsec}
\def\thesubsubsec{\secsym\the\subsecno.\the\subsubsecno.}
\def\subsubsec#1{\global\advance\subsubsecno by1
  \message{(\thesubsubsec\ #1)}%
  \ifnum\lastpenalty>9000\else\bigbreak\fi
\noindent{\bf\thesubsubsec\ #1}%
  \writetoca{\string\qquad {\thesubsubsec} {#1}}\par\nobreak\smallskip\nobreak}

%
\def\draftnote#1{\ifdraft{\baselineskip2ex
                 \vbox{\kern1em\hrule\hbox{\vrule\kern1em\vbox{\kern1ex
                 \noindent \underbar{NOTE}: #1
             \vskip1ex}\kern1em\vrule}\hrule}}\fi}
\def\internote#1{\ifinter{\baselineskip2ex
                 \vbox{\kern1em\hrule\hbox{\vrule\kern1em\vbox{\kern1ex
                 \noindent \underbar{Internal Note}: #1
             \vskip1ex}\kern1em\vrule}\hrule}}\fi}

%

%
%

%

\def\inv{^{-1}}



\def\b{\beta}


\def\bb{
\font\tenmsb=msbm10
\font\sevenmsb=msbm7
\font\fivemsb=msbm5
\textfont1=\tenmsb
\scriptfont1=\sevenmsb
\scriptscriptfont1=\fivemsb
}





\def\bar{\overline}
\def\b{\bar}
\def\bsq#1{{{\b{#1}}^{\lower 2.5pt\hbox{$\scriptstyle 2$}}}}
\def\bexp#1#2{{{\b{#1}}^{\lower 2.5pt\hbox{$\scriptstyle #2$}}}}
\def\dotexp#1#2{{{#1}^{\lower 2.5pt\hbox{$\scriptstyle #2$}}}}


\def\Re{\mathop{\rm Re}}
\def\Im{\mathop{\rm Im}}
\def\det{\mathop{\rm det}}

\def\Tr{\mathop{\rm Tr}}

\def\rt2{\sqrt{2}}

\def\mod{{\rm mod}}



\def\CB{{\cal B}}

\def\CL{{\cal L}}

\def\CN{{\cal N}}

\def\CS{{\cal S}}


\def\1{{\ds 1}}
\def\R{\hbox{$\bb R$}}

\def\Z{\hbox{$\bb Z$}}


\noblackbox

\def\unit{\relax{\rm 1\kern-.26em I}}
\def\nada{\relax{\rm 0\kern-.30em l}}

\def\mod{{\rm \ mod \ }}

\noblackbox
\def\IL{\relax{\rm I\kern-.18em L}}
\def\IH{\relax{\rm I\kern-.18em H}}
\def\IR{\relax{\rm I\kern-.18em R}}
\def\IC{\relax\hbox{$\inbar\kern-.3em{\rm C}$}}
\def\IZ{\relax\ifmmode\mathchoice
{\hbox{\cmss Z\kern-.4em Z}}{\hbox{\cmss Z\kern-.4em Z}} {\lower.9pt\hbox{\cmsss Z\kern-.4em Z}}
{\lower1.2pt\hbox{\cmsss Z\kern-.4em Z}}\else{\cmss Z\kern-.4em Z}\fi}

\def\partialslash{\not{\hbox{\kern-2pt $\partial$}}}

\font\manual=manfnt \def\dbend{\lower3.5pt\hbox{\manual\char127}}

\def\IZ{\relax\ifmmode\mathchoice
{\hbox{\cmss Z\kern-.4em Z}}{\hbox{\cmss Z\kern-.4em Z}} {\lower.9pt\hbox{\cmsss Z\kern-.4em Z}}
{\lower1.2pt\hbox{\cmsss Z\kern-.4em Z}}\else{\cmss Z\kern-.4em Z}\fi}
\def\half{{1\over 2}}

\def\bar{\overline}

\def\rt2{\sqrt{2}}
\def\irt2{{1\over\sqrt{2}}}

\def\slashchar#1{\setbox0=\hbox{$#1$}           
   \dimen0=\wd0                                 
   \setbox1=\hbox{/} \dimen1=\wd1               
   \ifdim\dimen0>\dimen1                        
      \rlap{\hbox to \dimen0{\hfil/\hfil}}      
      #1                                        
   \else                                        
      \rlap{\hbox to \dimen1{\hfil$#1$\hfil}}   
      /                                         
   \fi}


\def\figcaption#1#2{\DefWarn#1\xdef#1{Figure~\noexpand\hyperref{}{figure}%
{\the\figno}{\the\figno}}\writedef{#1\leftbracket Figure\noexpand~\xfig#1}%
\medskip\centerline{{\footnotefont\bf Figure~\hyperdef\hypernoname{figure}{\the\figno}{\the\figno}:}  #2 \wrlabeL{#1=#1}}%
\global\advance\figno by1}


\def\savefig{\expandafter\savefigaux\expandafter{\the\figno}}
\def\savefigaux#1#2#3#4{\DefWarn#2%
 \gdef#2{fig.~\hyperref{}{figure}{#1}{#1}}%
 \writedef{#2\leftbracket fig.\noexpand~\xfig#2}%
 \expandafter\gdef\csname savedfig-\string#2\endcsname{%
   \figinsert\figin{\centerline{#4}}%
   \medskip\centerline{\vbox{\baselineskip12pt
       \advance\hsize by -1truein\noindent\wrlabeL{#2=#2}
       \footnotefont%
       {\bf Fig.~\hyperdef\hypernoname{figure}{#1}{#1}:} #3}}%
   \bigskip\endinsert}%
 \global\advance\figno by1}
\def\putfig#1{\csname savedfig-\string#1\endcsname}

\lref\MooreSS{
  G.~W.~Moore and N.~Seiberg,
  ``Naturality in Conformal Field Theory,''
Nucl.\ Phys.\ B {\bf 313}, 16 (1989).
}

\lref\MooreYH{
  G.~W.~Moore and N.~Seiberg,
  ``Taming the Conformal Zoo,''
Phys.\ Lett.\ B {\bf 220}, 422 (1989).
}

\lref\VenezianoYB{
  G.~Veneziano,
  ``Construction of a crossing - symmetric, Regge behaved amplitude for linearly rising trajectories,''
Nuovo Cim.\ A {\bf 57}, 190 (1968).
}

\lref\GrossBR{
  D.~J.~Gross, R.~D.~Pisarski and L.~G.~Yaffe,
  ``QCD and Instantons at Finite Temperature,''
Rev.\ Mod.\ Phys.\  {\bf 53}, 43 (1981).
}

\lref\SvetitskyGS{
  B.~Svetitsky and L.~G.~Yaffe,
  ``Critical Behavior at Finite Temperature Confinement Transitions,''
Nucl.\ Phys.\ B {\bf 210}, 423 (1982).
}

\lref\SvetitskyYE{
  B.~Svetitsky,
  ``Symmetry Aspects of Finite Temperature Confinement Transitions,''
Phys.\ Rept.\  {\bf 132}, 1 (1986).
}

\lref\JafferisNS{
  D.~Jafferis and X.~Yin,
  ``A Duality Appetizer,''
[arXiv:1103.5700 [hep-th]].
}

\lref\SiversIG{
  D.~Sivers and J.~Yellin,
  ``Review of recent work on narrow resonance models,''
Rev.\ Mod.\ Phys.\  {\bf 43}, 125 (1971).
}

\lref\WittenEY{
  E.~Witten,
  ``Dyons of Charge e theta/2 pi,''
Phys.\ Lett.\ B {\bf 86}, 283 (1979).
}

\lref\GreenSG{
  M.~B.~Green and J.~H.~Schwarz,
  ``Anomaly Cancellation in Supersymmetric D=10 Gauge Theory and Superstring Theory,''
Phys.\ Lett.\  {\bf 149B}, 117 (1984).
}

\lref\AffleckCH{
  I.~Affleck and F.~D.~M.~Haldane,
  ``Critical Theory of Quantum Spin Chains,''
Phys.\ Rev.\ B {\bf 36}, 5291 (1987).
}

\lref\BilloJDA{
  M.~Bill�, M.~Caselle, D.~Gaiotto, F.~Gliozzi, M.~Meineri and R.~Pellegrini,
  ``Line defects in the 3d Ising model,''
JHEP {\bf 1307}, 055 (2013).
[arXiv:1304.4110 [hep-th]].
}

\lref\WittenABA{
  E.~Witten,
  ``Fermion Path Integrals And Topological Phases,''
Rev.\ Mod.\ Phys.\  {\bf 88}, no. 3, 035001 (2016).
[arXiv:1508.04715 [cond-mat.mes-hall]].
}

\lref\GaiottoYUP{
  D.~Gaiotto, A.~Kapustin, Z.~Komargodski and N.~Seiberg,
  ``Theta, Time Reversal, and Temperature,''
JHEP {\bf 1705}, 091 (2017).
[arXiv:1703.00501 [hep-th]].
}

\lref\GaiottoNVA{
  D.~Gaiotto, D.~Mazac and M.~F.~Paulos,
  min
  ``Bootstrapping the 3d Ising twist defect,''
JHEP {\bf 1403}, 100 (2014).
[arXiv:1310.5078 [hep-th]].
}

\lref\BrowerEA{
  R.~C.~Brower, J.~Polchinski, M.~J.~Strassler and C.~I.~Tan,
  ``The Pomeron and gauge/string duality,''
JHEP {\bf 0712}, 005 (2007).
[hep-th/0603115].
}

\lref\inprogress{
  In progress.
}

\lref\inprogressi{
  In progress.
}

\lref\mandelstam{
S.~Mandelstam, ``Dual-resonance models." Physics Reports 13.6 (1974): 259-353.
}

\lref\FreundHW{
  P.~G.~O.~Freund,
  ``Finite energy sum rules and bootstraps,''
Phys.\ Rev.\ Lett.\  {\bf 20}, 235 (1968).
}

\lref\MeyerJC{
  H.~B.~Meyer and M.~J.~Teper,
  ``Glueball Regge trajectories and the pomeron: A Lattice study,''
Phys.\ Lett.\ B {\bf 605}, 344 (2005).
[hep-ph/0409183].
}

\lref\CoonYW{
  D.~D.~Coon,
Phys.\ Lett.\ B {\bf 29}, 669 (1969).
}

\lref\FairlieAD{
  D.~B.~Fairlie and J.~Nuyts,
Nucl.\ Phys.\ B {\bf 433}, 26 (1995).
[hep-th/9406043].
}

\lref\RedlichDV{
  A.~N.~Redlich,
  ``Parity Violation and Gauge Noninvariance of the Effective Gauge Field Action in Three-Dimensions,''
Phys.\ Rev.\ D {\bf 29}, 2366 (1984).
}

\lref\RedlichKN{
  A.~N.~Redlich,
  ``Gauge Noninvariance and Parity Violation of Three-Dimensional Fermions,''
Phys.\ Rev.\ Lett.\  {\bf 52}, 18 (1984).
}

\lref\PonomarevJQK{
  D.~Ponomarev and A.~A.~Tseytlin,
[arXiv:1603.06273 [hep-th]].
}

\lref\StromingerTalk{
  A.~Strominger, Talk at Strings 2014, Princeton.
}

\lref\CostaMG{
  M.~S.~Costa, J.~Penedones, D.~Poland and S.~Rychkov,
JHEP {\bf 1111}, 071 (2011).
[arXiv:1107.3554 [hep-th]].
}

\lref\AbanovQZ{
  A.~G.~Abanov and P.~B.~Wiegmann,
  ``Theta terms in nonlinear sigma models,''
Nucl.\ Phys.\ B {\bf 570}, 685 (2000).
[hep-th/9911025].
}

\lref\AffleckAS{
  I.~Affleck, J.~A.~Harvey and E.~Witten,
  ``Instantons and (Super)Symmetry Breaking in (2+1)-Dimensions,''
Nucl.\ Phys.\ B {\bf 206}, 413 (1982).
}

\lref\AharonyBX{
  O.~Aharony, A.~Hanany, K.~A.~Intriligator, N.~Seiberg and M.~J.~Strassler,
  ``Aspects of $N=2$ supersymmetric gauge theories in three-dimensions,''
Nucl.\ Phys.\ B {\bf 499}, 67 (1997).
[hep-th/9703110].
}

\lref\AharonyGP{
  O.~Aharony,
  ``IR duality in $d = 3$ $N=2$ supersymmetric $USp(2N_c)$ and $U(N_c)$ gauge theories,''
Phys.\ Lett.\ B {\bf 404}, 71 (1997).
[hep-th/9703215].
}

\lref\BarkeshliRZD{
  M.~Barkeshli and M.~Cheng,
  ``Time-reversal and spatial reflection symmetry localization anomalies in (2+1)D topological phases of matter,''
[arXiv:1706.09464 [cond-mat.str-el]].
}

\lref\TachikawaXVS{
  Y.~Tachikawa and K.~Yonekura,
  ``Gauge interactions and topological phases of matter,''
PTEP {\bf 2016}, no. 9, 093B07 (2016).
[arXiv:1604.06184 [hep-th]].
}

\lref\TachikawaCHA{
  Y.~Tachikawa and K.~Yonekura,
  ``On time-reversal anomaly of 2+1d topological phases,''
PTEP {\bf 2017}, no. 3, 033B04 (2017).
[arXiv:1610.07010 [hep-th]].
}

\lref\SeibergRSG{
  N.~Seiberg and E.~Witten,
  ``Gapped Boundary Phases of Topological Insulators via Weak Coupling,''
PTEP {\bf 2016}, 12C101 (2016). 
[arXiv:1602.04251 [cond-mat.str-el]].
}

\lref\SeibergGMD{
  N.~Seiberg, T.~Senthil, C.~Wang and E.~Witten,
  ``A Duality Web in 2+1 Dimensions and Condensed Matter Physics,''
Annals Phys.\  {\bf 374}, 395 (2016).
[arXiv:1606.01989 [hep-th]].
}

\lref\TachikawaNMO{
  Y.~Tachikawa and K.~Yonekura,
  ``More on time-reversal anomaly of 2+1d topological phases,''
[arXiv:1611.01601 [hep-th]].
}

\lref\AharonyCI{
  O.~Aharony and I.~Shamir,
  ``On $O(N_c)$ $d=3$ ${\cal N}{=}2$ supersymmetric QCD Theories,''
JHEP {\bf 1112}, 043 (2011).
[arXiv:1109.5081 [hep-th]].
}

\lref\AharonyJZ{
  O.~Aharony, G.~Gur-Ari and R.~Yacoby,
  ``$d=3$ Bosonic Vector Models Coupled to Chern-Simons Gauge Theories,''
JHEP {\bf 1203}, 037 (2012).
[arXiv:1110.4382 [hep-th]].
}

\lref\AharonyNH{
  O.~Aharony, G.~Gur-Ari and R.~Yacoby,
  ``Correlation Functions of Large $N$ Chern-Simons-Matter Theories and Bosonization in Three Dimensions,''
JHEP {\bf 1212}, 028 (2012).
[arXiv:1207.4593 [hep-th]].
}

\lref\AharonyNS{
  O.~Aharony, S.~Giombi, G.~Gur-Ari, J.~Maldacena and R.~Yacoby,
  ``The Thermal Free Energy in Large $N$ Chern-Simons-Matter Theories,''
JHEP {\bf 1303}, 121 (2013).
[arXiv:1211.4843 [hep-th]].
}

\lref\AharonyHDA{
  O.~Aharony, N.~Seiberg and Y.~Tachikawa,
  ``Reading between the lines of four-dimensional gauge theories,''
JHEP {\bf 1308}, 115 (2013).
[arXiv:1305.0318 [hep-th]].
}

\lref\AharonyDHA{
  O.~Aharony, S.~S.~Razamat, N.~Seiberg and B.~Willett,
  ``3d dualities from 4d dualities,''
JHEP {\bf 1307}, 149 (2013).
[arXiv:1305.3924 [hep-th]].
}

\lref\AharonyKMA{
  O.~Aharony, S.~S.~Razamat, N.~Seiberg and B.~Willett,
  ``3d dualities from 4d dualities for orthogonal groups,''
JHEP {\bf 1308}, 099 (2013)
[arXiv:1307.0511 [hep-th]].
}

\lref\AharonyMJS{
  O.~Aharony,
  ``Baryons, monopoles and dualities in Chern-Simons-matter theories,''
JHEP {\bf 1602}, 093 (2016).
[arXiv:1512.00161 [hep-th]].
}

\lref\AharonyJVV{
  O.~Aharony, F.~Benini, P.~S.~Hsin and N.~Seiberg,
  ``Chern-Simons-matter dualities with $SO$ and $USp$ gauge groups,''
JHEP {\bf 1702}, 072 (2017).
[arXiv:1611.07874 [cond-mat.str-el]].
}

\lref\MetlitskiDHT{
  M.~A.~Metlitski, A.~Vishwanath and C.~Xu,
   ``Duality and bosonization of (2+1) -dimensional Majorana fermions,''
Phys.\ Rev.\ B {\bf 95}, no. 20, 205137 (2017).
[arXiv:1611.05049 [cond-mat.str-el]].
}

\lref\WangTXT{
  C.~Wang, A.~Nahum, M.~A.~Metlitski, C.~Xu and T.~Senthil,
  ``Deconfined quantum critical points: symmetries and dualities,''
[arXiv:1703.02426 [cond-mat.str-el]].
}

\lref\AlvarezGaumeNF{
  L.~Alvarez-Gaume, S.~Della Pietra and G.~W.~Moore,
  ``Anomalies and Odd Dimensions,''
Annals Phys.\  {\bf 163}, 288 (1985).
}

\lref\AnninosUI{
  D.~Anninos, T.~Hartman and A.~Strominger,
  ``Higher Spin Realization of the dS/CFT Correspondence,''
[arXiv:1108.5735 [hep-th]].
}

\lref\AnninosHIA{
  D.~Anninos, R.~Mahajan, D.~Radicevic and E.~Shaghoulian,
  ``Chern-Simons-Ghost Theories and de Sitter Space,''
JHEP {\bf 1501}, 074 (2015).
[arXiv:1405.1424 [hep-th]].
}

\lref\RadicevicWQN{
  D.~Radicevic, D.~Tong and C.~Turner,
   ``Non-Abelian 3d Bosonization and Quantum Hall States,''
JHEP {\bf 1612}, 067 (2016).
[arXiv:1608.04732 [hep-th]].
}

\lref\AtiyahJF{
  M.~F.~Atiyah, V.~K.~Patodi and I.~M.~Singer,
  ``Spectral Asymmetry in Riemannian Geometry, I,''
  Math.\ Proc.\ Camb.\ Phil.\ Soc.\ {\bf 77} (1975) 43--69.
}

\lref\BanksZN{
  T.~Banks and N.~Seiberg,
  ``Symmetries and Strings in Field Theory and Gravity,''
Phys.\ Rev.\ D {\bf 83}, 084019 (2011).
[arXiv:1011.5120 [hep-th]].
}

\lref\BarkeshliIDA{
  M.~Barkeshli and J.~McGreevy,
  ``Continuous transition between fractional quantum Hall and superfluid states,''
Phys.\ Rev.\ B {\bf 89}, 235116 (2014).
}

\lref\BeemMB{
  C.~Beem, T.~Dimofte and S.~Pasquetti,
  ``Holomorphic Blocks in Three Dimensions,''
[arXiv:1211.1986 [hep-th]].
}
\lref\VafaXG{
  C.~Vafa and E.~Witten,
  ``Parity Conservation in QCD,''
Phys.\ Rev.\ Lett.\  {\bf 53}, 535 (1984).
}

\lref\BeniniMF{
  F.~Benini, C.~Closset and S.~Cremonesi,
  ``Comments on 3d Seiberg-like dualities,''
JHEP {\bf 1110}, 075 (2011).
[arXiv:1108.5373 [hep-th]].
}

\lref\BernardXY{
  D.~Bernard,
  ``String Characters From {Kac-Moody} Automorphisms,''
  Nucl.\ Phys.\ B {\bf 288}, 628 (1987).
}

\lref\BhattacharyaZY{
  J.~Bhattacharya, S.~Bhattacharyya, S.~Minwalla and S.~Raju,
  ``Indices for Superconformal Field Theories in 3,5 and 6 Dimensions,''
JHEP {\bf 0802}, 064 (2008).
[arXiv:0801.1435 [hep-th]].
}

\lref\deBoerMP{
  J.~de Boer, K.~Hori, H.~Ooguri and Y.~Oz,
  ``Mirror symmetry in three-dimensional gauge theories, quivers and D-branes,''
Nucl.\ Phys.\ B {\bf 493}, 101 (1997).
[hep-th/9611063].
}

\lref\deBoerKA{
  J.~de Boer, K.~Hori, Y.~Oz and Z.~Yin,
  ``Branes and mirror symmetry in N=2 supersymmetric gauge theories in three-dimensions,''
Nucl.\ Phys.\ B {\bf 502}, 107 (1997).
[hep-th/9702154].
}

\lref\WilczekCY{
  F.~Wilczek and A.~Zee,
  ``Linking Numbers, Spin, and Statistics of Solitons,''
Phys.\ Rev.\ Lett.\  {\bf 51}, 2250 (1983).
}

\lref\BondersonPLA{
  P.~Bonderson, C.~Nayak and X.~L.~Qi,
  ``A time-reversal invariant topological phase at the surface of a 3D topological insulator,''
J.\ Stat.\ Mech.\  {\bf 2013}, P09016 (2013).
}

\lref\BorokhovIB{
  V.~Borokhov, A.~Kapustin and X.~k.~Wu,
  ``Topological disorder operators in three-dimensional conformal field theory,''
JHEP {\bf 0211}, 049 (2002).
[hep-th/0206054].
}

\lref\WittenTW{
  E.~Witten,
  ``Global Aspects of Current Algebra,''
Nucl.\ Phys.\ B {\bf 223}, 422 (1983).
}
\lref\WittenTX{
  E.~Witten,
  ``Current Algebra, Baryons, and Quark Confinement,''
Nucl.\ Phys.\ B {\bf 223}, 433 (1983).
}

\lref\GaiottoYUP{
  D.~Gaiotto, A.~Kapustin, Z.~Komargodski and N.~Seiberg,
  ``Theta, Time Reversal, and Temperature,''
[arXiv:1703.00501 [hep-th]].
}

\lref\BorokhovCG{
  V.~Borokhov, A.~Kapustin and X.~k.~Wu,
  ``Monopole operators and mirror symmetry in three-dimensions,''
JHEP {\bf 0212}, 044 (2002).
[hep-th/0207074].
}

\lref\KomargodskiKEH{
  Z.~Komargodski and N.~Seiberg,
  ``A Symmetry Breaking Scenario for QCD$_3$,''
[arXiv:1706.08755 [hep-th]].
}

\lref\KomargodskiDMC{
  Z.~Komargodski, A.~Sharon, R.~Thorngren and X.~Zhou,
  ``Comments on Abelian Higgs Models and Persistent Order,''
[arXiv:1705.04786 [hep-th]].
}

\lref\KomargodskiSMK{
  Z.~Komargodski, T.~Sulejmanpasic and M.~Unsal,
  ``Walls, Anomalies, and (De) Confinement in Quantum Anti-Ferromagnets,''
[arXiv:1706.05731 [cond-mat.str-el]].
}

\lref\Browder{
  W.~Browder and E.~Thomas,
  ``Axioms for the generalized Pontryagin cohomology operations,''
  Quart.\ J.\ Math.\ Oxford {\bf 13}, 55--60 (1962).
}

\lref\debult{
  F.~van~de~Bult,
  ``Hyperbolic Hypergeometric Functions,''
University of Amsterdam Ph.D. thesis
}

\lref\Camperi{
  M.~Camperi, F.~Levstein and G.~Zemba,
  ``The Large N Limit Of Chern-simons Gauge Theory,''
  Phys.\ Lett.\ B {\bf 247} (1990) 549.
}

\lref\ChenCD{
  W.~Chen, M.~P.~A.~Fisher and Y.~S.~Wu,
  ``Mott transition in an anyon gas,''
Phys.\ Rev.\ B {\bf 48}, 13749 (1993).
[cond-mat/9301037].
}

\lref\WittenDS{
  E.~Witten,
  ``Supersymmetric index of three-dimensional gauge theory,''
In *Shifman, M.A. (ed.): The many faces of the superworld* 156-184.
[hep-th/9903005].
}

\lref\TachikawaStrings{
  Y.~Tachikawa,
 Talk At Strings 2017, 
``Time-reversal Anomalies of 2+1d Topological Phases.''
}

\lref\ChenJHA{
  X.~Chen, L.~Fidkowski and A.~Vishwanath,
  ``Symmetry Enforced Non-Abelian Topological Order at the Surface of a Topological Insulator,''
Phys.\ Rev.\ B {\bf 89}, no. 16, 165132 (2014).
[arXiv:1306.3250 [cond-mat.str-el]].
}

\lref\AharonyJVV{
  O.~Aharony, F.~Benini, P.~S.~Hsin and N.~Seiberg,
  ``Chern-Simons-matter dualities with $SO$ and $USp$ gauge groups,''
JHEP {\bf 1702}, 072 (2017).
[arXiv:1611.07874 [cond-mat.str-el]].
}

\lref\IntriligatorLCA{
  K.~Intriligator and N.~Seiberg,
  ``Aspects of 3d N=2 Chern-Simons-Matter Theories,''
JHEP {\bf 1307}, 079 (2013).
[arXiv:1305.1633 [hep-th]].
}

\lref\ChengPDN{
  M.~Cheng and C.~Xu,
  ``Series of (2+1)-dimensional stable self-dual interacting conformal field theories,''
Phys.\ Rev.\ B {\bf 94}, 214415 (2016). 
[arXiv:1609.02560 [cond-mat.str-el]].
}

\lref\AvdeevZA{
  L.~V.~Avdeev, G.~V.~Grigorev and D.~I.~Kazakov,
  ``Renormalizations in Abelian Chern-Simons field theories with matter,''
Nucl.\ Phys.\ B {\bf 382}, 561 (1992).
}

\lref\AvdeevJT{
  L.~V.~Avdeev, D.~I.~Kazakov and I.~N.~Kondrashuk,
  ``Renormalizations in supersymmetric and nonsupersymmetric nonAbelian Chern-Simons field theories with matter,''
Nucl.\ Phys.\ B {\bf 391}, 333 (1993).
}

\lref\ClossetVG{
  C.~Closset, T.~T.~Dumitrescu, G.~Festuccia, Z.~Komargodski and N.~Seiberg,
  ``Contact Terms, Unitarity, and F-Maximization in Three-Dimensional Superconformal Theories,''
JHEP {\bf 1210}, 053 (2012).
[arXiv:1205.4142 [hep-th]].
}

\lref\ClossetVP{
  C.~Closset, T.~T.~Dumitrescu, G.~Festuccia, Z.~Komargodski and N.~Seiberg,
  ``Comments on Chern-Simons Contact Terms in Three Dimensions,''
JHEP {\bf 1209}, 091 (2012).
[arXiv:1206.5218 [hep-th]].
}

\lref\ClossetRU{
  C.~Closset, T.~T.~Dumitrescu, G.~Festuccia and Z.~Komargodski,
  ``Supersymmetric Field Theories on Three-Manifolds,''
JHEP {\bf 1305}, 017 (2013).
[arXiv:1212.3388 [hep-th]].
}

\lref\CveticXN{
  M.~Cvetic, T.~W.~Grimm and D.~Klevers,
  ``Anomaly Cancellation And Abelian Gauge Symmetries In F-theory,''
JHEP {\bf 1302}, 101 (2013).
[arXiv:1210.6034 [hep-th]].
}

\lref\DaiKQ{
  X.~z.~Dai and D.~S.~Freed,
  ``eta invariants and determinant lines,''
J.\ Math.\ Phys.\  {\bf 35}, 5155 (1994), Erratum: [J.\ Math.\ Phys.\  {\bf 42}, 2343 (2001)].
[hep-th/9405012].
}

\lref\DasguptaZZ{
  C.~Dasgupta and B.~I.~Halperin,
  ``Phase Transition in a Lattice Model of Superconductivity,''
Phys.\ Rev.\ Lett.\  {\bf 47}, 1556 (1981).
}

\lref\DaviesUW{
  N.~M.~Davies, T.~J.~Hollowood, V.~V.~Khoze and M.~P.~Mattis,
  ``Gluino condensate and magnetic monopoles in supersymmetric gluodynamics,''
Nucl.\ Phys.\ B {\bf 559}, 123 (1999).
[hep-th/9905015].
}

\lref\DaviesNW{
  N.~M.~Davies, T.~J.~Hollowood and V.~V.~Khoze,
  ``Monopoles, affine algebras and the gluino condensate,''
J.\ Math.\ Phys.\  {\bf 44}, 3640 (2003).
[hep-th/0006011].
}

\lref\DimoftePY{
  T.~Dimofte, D.~Gaiotto and S.~Gukov,
  ``3-Manifolds and 3d Indices,''
[arXiv:1112.5179 [hep-th]].
}

\lref\DolanQI{
  F.~A.~Dolan and H.~Osborn,
  ``Applications of the Superconformal Index for Protected Operators and q-Hypergeometric Identities to N=1 Dual Theories,''
Nucl.\ Phys.\ B {\bf 818}, 137 (2009).
[arXiv:0801.4947 [hep-th]].
}

\lref\DolanRP{
  F.~A.~H.~Dolan, V.~P.~Spiridonov and G.~S.~Vartanov,
  ``From 4d superconformal indices to 3d partition functions,''
Phys.\ Lett.\ B {\bf 704}, 234 (2011).
[arXiv:1104.1787 [hep-th]].
}

\lref\DouglasEX{
  M.~R.~Douglas,
  ``Chern-Simons-Witten theory as a topological Fermi liquid,''
[hep-th/9403119].
}

\lref\EagerHX{
  R.~Eager, J.~Schmude and Y.~Tachikawa,
  ``Superconformal Indices, Sasaki-Einstein Manifolds, and Cyclic Homologies,''
[arXiv:1207.0573 [hep-th]].
}

\lref\VenezianoYB{
  G.~Veneziano,
  ``Construction of a crossing - symmetric, Regge behaved amplitude for linearly rising trajectories,''
Nuovo Cim.\ A {\bf 57}, 190 (1968).
}

\lref\GrossBR{
  D.~J.~Gross, R.~D.~Pisarski and L.~G.~Yaffe,
  ``QCD and Instantons at Finite Temperature,''
Rev.\ Mod.\ Phys.\  {\bf 53}, 43 (1981).
}

\lref\SvetitskyGS{
  B.~Svetitsky and L.~G.~Yaffe,
  ``Critical Behavior at Finite Temperature Confinement Transitions,''
Nucl.\ Phys.\ B {\bf 210}, 423 (1982).
}

\lref\SvetitskyYE{
  B.~Svetitsky,
  ``Symmetry Aspects of Finite Temperature Confinement Transitions,''
Phys.\ Rept.\  {\bf 132}, 1 (1986).
}

\lref\WittenUKA{
  E.~Witten,
  ``Theta dependence in the large N limit of four-dimensional gauge theories,''
Phys.\ Rev.\ Lett.\  {\bf 81}, 2862 (1998).
[hep-th/9807109].
}
\lref\KachruRUI{
  S.~Kachru, M.~Mulligan, G.~Torroba and H.~Wang,
   ``Bosonization and Mirror Symmetry,''
Phys.\ Rev.\ D {\bf 94}, no. 8, 085009 (2016).
[arXiv:1608.05077 [hep-th]].
}
\lref\KachruAON{
  S.~Kachru, M.~Mulligan, G.~Torroba and H.~Wang,
   ``Nonsupersymmetric dualities from mirror symmetry,''
Phys.\ Rev.\ Lett.\  {\bf 118}, no. 1, 011602 (2017).
[arXiv:1609.02149 [hep-th]].
}

\lref\SiversIG{
  D.~Sivers and J.~Yellin,
  ``Review of recent work on narrow resonance models,''
Rev.\ Mod.\ Phys.\  {\bf 43}, 125 (1971).
}
\lref\WittenNV{
  E.~Witten,
  ``Supersymmetric index in four-dimensional gauge theories,''
Adv.\ Theor.\ Math.\ Phys.\  {\bf 5}, 841 (2002).
[hep-th/0006010].
 }
  
\lref\WittenDF{
  E.~Witten,
   ``Constraints on Supersymmetry Breaking,''
Nucl.\ Phys.\ B {\bf 202}, 253 (1982).
}

\lref\WittenEY{
  E.~Witten,
  ``Dyons of Charge e theta/2 pi,''
Phys.\ Lett.\ B {\bf 86}, 283 (1979).
}

\lref\GreenSG{
  M.~B.~Green and J.~H.~Schwarz,
  ``Anomaly Cancellation in Supersymmetric D=10 Gauge Theory and Superstring Theory,''
Phys.\ Lett.\  {\bf 149B}, 117 (1984).
}

\lref\ColemanUZ{
  S.~R.~Coleman,
  ``More About the Massive Schwinger Model,''
Annals Phys.\  {\bf 101}, 239 (1976).
}

\lref\BaluniRF{
  V.~Baluni,
  ``CP Violating Effects in QCD,''
Phys.\ Rev.\ D {\bf 19}, 2227 (1979).
}

\lref\DashenET{
  R.~F.~Dashen,
  ``Some features of chiral symmetry breaking,''
Phys.\ Rev.\ D {\bf 3}, 1879 (1971).
}

\lref\WittenBC{
  E.~Witten,
  ``Instantons, the Quark Model, and the 1/n Expansion,''
Nucl.\ Phys.\ B {\bf 149}, 285 (1979).
}

\lref\WittenVV{
  E.~Witten,
  ``Current Algebra Theorems for the U(1) Goldstone Boson,''
Nucl.\ Phys.\ B {\bf 156}, 269 (1979).
}

\lref\WittenSP{
  E.~Witten,
  ``Large N Chiral Dynamics,''
Annals Phys.\  {\bf 128}, 363 (1980).
}

\lref\DAddaVBW{
  A.~D'Adda, M.~Luscher and P.~Di Vecchia,
  ``A 1/n Expandable Series of Nonlinear Sigma Models with Instantons,''
Nucl.\ Phys.\ B {\bf 146}, 63 (1978).
}

\lref\AcharyaDZ{
  B.~S.~Acharya and C.~Vafa,
  ``On domain walls of N=1 supersymmetric Yang-Mills in four-dimensions,''
[hep-th/0103011].
}

\lref\AffleckCH{
  I.~Affleck and F.~D.~M.~Haldane,
  ``Critical Theory of Quantum Spin Chains,''
Phys.\ Rev.\ B {\bf 36}, 5291 (1987).
}

\lref\BilloJDA{
  M.~Bill�, M.~Caselle, D.~Gaiotto, F.~Gliozzi, M.~Meineri and R.~Pellegrini,
  ``Line defects in the 3d Ising model,''
JHEP {\bf 1307}, 055 (2013).
[arXiv:1304.4110 [hep-th]].
}

\lref\WittenABA{
  E.~Witten,
  ``Fermion Path Integrals And Topological Phases,''
Rev.\ Mod.\ Phys.\  {\bf 88}, no. 3, 035001 (2016).
[arXiv:1508.04715 [cond-mat.mes-hall]].
}

\lref\GaiottoNVA{
  D.~Gaiotto, D.~Mazac and M.~F.~Paulos,
  ``Bootstrapping the 3d Ising twist defect,''
JHEP {\bf 1403}, 100 (2014).
[arXiv:1310.5078 [hep-th]].
}

\lref\BrowerEA{
  R.~C.~Brower, J.~Polchinski, M.~J.~Strassler and C.~I.~Tan,
  ``The Pomeron and gauge/string duality,''
JHEP {\bf 0712}, 005 (2007).
[hep-th/0603115].
}

\lref\inprogress{
  In progress.
}

\lref\inprogressi{
  In progress.
}

\lref\mandelstam{
S.~Mandelstam, ``Dual-resonance models." Physics Reports 13.6 (1974): 259-353.
}

\lref\FreundHW{
  P.~G.~O.~Freund,
  ``Finite energy sum rules and bootstraps,''
Phys.\ Rev.\ Lett.\  {\bf 20}, 235 (1968).
}

\lref\MeyerJC{
  H.~B.~Meyer and M.~J.~Teper,
  ``Glueball Regge trajectories and the pomeron: A Lattice study,''
Phys.\ Lett.\ B {\bf 605}, 344 (2005).
[hep-ph/0409183].
}

\lref\CoonYW{
  D.~D.~Coon,
  ``Uniqueness of the veneziano representation,''
Phys.\ Lett.\ B {\bf 29}, 669 (1969).
}

\lref\FairlieAD{
  D.~B.~Fairlie and J.~Nuyts,
  ``A fresh look at generalized Veneziano amplitudes,''
Nucl.\ Phys.\ B {\bf 433}, 26 (1995).
[hep-th/9406043].
}

\lref\CWpot{
  S.~R.~Coleman and E.~J.~Weinberg,
  ``Radiative Corrections as the Origin of Spontaneous Symmetry Breaking,''
Phys.\ Rev.\ D {\bf 7}, 1888 (1973).
}

\lref\RedlichDV{
  A.~N.~Redlich,
  ``Parity Violation and Gauge Noninvariance of the Effective Gauge Field Action in Three-Dimensions,''
Phys.\ Rev.\ D {\bf 29}, 2366 (1984).
}

\lref\RedlichKN{
  A.~N.~Redlich,
  ``Gauge Noninvariance and Parity Violation of Three-Dimensional Fermions,''
Phys.\ Rev.\ Lett.\  {\bf 52}, 18 (1984).
}

\lref\PonomarevJQK{
  D.~Ponomarev and A.~A.~Tseytlin,
[arXiv:1603.06273 [hep-th]].
}

\lref\StromingerTalk{
  A.~Strominger, Talk at Strings 2014, Princeton.
}

\lref\PoppitzNZ{
  E.~Poppitz, T.~Sch�fer and M.~�nsal,
  ``Universal mechanism of (semi-classical) deconfinement and theta-dependence for all simple groups,''
JHEP {\bf 1303}, 087 (2013).
[arXiv:1212.1238 [hep-th]].
}

\lref\UnsalZJ{
  M.~Unsal,
  ``Theta dependence, sign problems and topological interference,''
Phys.\ Rev.\ D {\bf 86}, 105012 (2012).
[arXiv:1201.6426 [hep-th]].
}

\lref\CostaMG{
  M.~S.~Costa, J.~Penedones, D.~Poland and S.~Rychkov,
JHEP {\bf 1111}, 071 (2011).
[arXiv:1107.3554 [hep-th]].
}

\lref\TachikawaXVS{
  Y.~Tachikawa and K.~Yonekura,
  ``Gauge interactions and topological phases of matter,''
PTEP {\bf 2016}, no. 9, 093B07 (2016).
[arXiv:1604.06184 [hep-th]].
}

\lref\CamanhoAPA{
  X.~O.~Camanho, J.~D.~Edelstein, J.~Maldacena and A.~Zhiboedov,
JHEP {\bf 1602}, 020 (2016).
[arXiv:1407.5597 [hep-th]].
}

\lref\LandsteinerCP{
  K.~Landsteiner, E.~Megias and F.~Pena-Benitez,
  ``Gravitational Anomaly and Transport,''
Phys.\ Rev.\ Lett.\  {\bf 107}, 021601 (2011).
[arXiv:1103.5006 [hep-ph]].
}

\lref\BanerjeeIZ{
  N.~Banerjee, J.~Bhattacharya, S.~Bhattacharyya, S.~Jain, S.~Minwalla and T.~Sharma,
  ``Constraints on Fluid Dynamics from Equilibrium Partition Functions,''
JHEP {\bf 1209}, 046 (2012).
[arXiv:1203.3544 [hep-th]].
}

\lref\JensenKJ{
  K.~Jensen, R.~Loganayagam and A.~Yarom,
  ``Thermodynamics, gravitational anomalies and cones,''
JHEP {\bf 1302}, 088 (2013).
[arXiv:1207.5824 [hep-th]].
}

\lref\BonettiELA{
  F.~Bonetti, T.~W.~Grimm and S.~Hohenegger,
  ``One-loop Chern-Simons terms in five dimensions,''
JHEP {\bf 1307}, 043 (2013).
[arXiv:1302.2918 [hep-th]].
}

\lref\DiPietroBCA{
  L.~Di Pietro and Z.~Komargodski,
  ``Cardy formulae for SUSY theories in $d =$ 4 and $d =$ 6,''
JHEP {\bf 1412}, 031 (2014).
[arXiv:1407.6061 [hep-th]].
}

\lref\ArdehaliHYA{
  A.~Arabi Ardehali, J.~T.~Liu and P.~Szepietowski,
  ``High-Temperature Expansion of Supersymmetric Partition Functions,''
JHEP {\bf 1507}, 113 (2015).
[arXiv:1502.07737 [hep-th]].
}

\lref\FeiOHA{
  L.~Fei, S.~Giombi, I.~R.~Klebanov and G.~Tarnopolsky,
  ``Generalized $F$-Theorem and the $\epsilon$ Expansion,''
JHEP {\bf 1512}, 155 (2015).
[arXiv:1507.01960 [hep-th]].
}

\lref\AcharyaDZ{
  B.~S.~Acharya and C.~Vafa,
  ``On domain walls of N=1 supersymmetric Yang-Mills in four-dimensions,''
[hep-th/0103011].
}

\lref\GiombiXXA{
  S.~Giombi and I.~R.~Klebanov,
  ``Interpolating between $a$ and $F$,''
JHEP {\bf 1503}, 117 (2015).
[arXiv:1409.1937 [hep-th]].
}

\lref\FrohlichWE{
  J.~Frohlich and E.~Thiran,
  ``Integral quadratic forms, Kac-Moody algebras, and fractional quantum Hall effect: An ADE-O classification,''
}

\lref\KapustinGUA{
  A.~Kapustin and N.~Seiberg,
  ``Coupling a QFT to a TQFT and Duality,''
JHEP {\bf 1404}, 001 (2014).
[arXiv:1401.0740 [hep-th]].
}

\lref\GaiottoKFA{
  D.~Gaiotto, A.~Kapustin, N.~Seiberg and B.~Willett,
  ``Generalized Global Symmetries,''
JHEP {\bf 1502}, 172 (2015).
[arXiv:1412.5148 [hep-th]].
}

\lref\DijkgraafPZ{
  R.~Dijkgraaf and E.~Witten,
  ``Topological Gauge Theories and Group Cohomology,''
Commun.\ Math.\ Phys.\  {\bf 129}, 393 (1990).
}

\lref\GaiottoTNE{
  D.~Gaiotto, Z.~Komargodski and N.~Seiberg,
   ``Time-Reversal Breaking in QCD$_4$, Walls, and Dualities in 2+1 Dimensions,''
[arXiv:1708.06806 [hep-th]].
}

\lref\KadanoffKZ{
  L.~P.~Kadanoff and H.~Ceva,
  ``Determination of an opeator algebra for the two-dimensional Ising model,''
Phys.\ Rev.\ B {\bf 3}, 3918 (1971).
}

\lref\GrossKZA{
  D.~J.~Gross and P.~F.~Mende,
Phys.\ Lett.\ B {\bf 197}, 129 (1987).
}

\lref\KarlinerHD{
  M.~Karliner, I.~R.~Klebanov and L.~Susskind,
Int.\ J.\ Mod.\ Phys.\ A {\bf 3}, 1981 (1988).
}

\lref\Shankar{
  R.~Shankar and N.~Read,
  ``The $\theta = \pi$ Nonlinear $\sigma$ Model Is Massless,''
Nucl.\ Phys.\ B {\bf 336}, 457 (1990).
}

\lref\CachazoZK{
  F.~Cachazo, N.~Seiberg and E.~Witten,
  ``Phases of N=1 supersymmetric gauge theories and matrices,''
JHEP {\bf 0302}, 042 (2003).
[hep-th/0301006].
}

\lref\WeissRJ{
  N.~Weiss,
  ``The Effective Potential for the Order Parameter of Gauge Theories at Finite Temperature,''
Phys.\ Rev.\ D {\bf 24}, 475 (1981).
}

\lref\ColemanJX{
  S.~R.~Coleman and E.~J.~Weinberg,
  ``Radiative Corrections as the Origin of Spontaneous Symmetry Breaking,''
Phys.\ Rev.\ D {\bf 7}, 1888 (1973).
}

\lref\KaoGF{
  H.~C.~Kao, K.~M.~Lee and T.~Lee,
  ``The Chern-Simons coefficient in supersymmetric Yang-Mills Chern-Simons theories,''
Phys.\ Lett.\ B {\bf 373}, 94 (1996).
[hep-th/9506170].
}

\lref\DashenET{
  R.~F.~Dashen,
  ``Some features of chiral symmetry breaking,''
Phys.\ Rev.\ D {\bf 3}, 1879 (1971).
}

\lref\Hitoshi{
http://hitoshi.berkeley.edu/221b/scattering3.pdf
}

\lref\GreensiteZZ{
  J.~Greensite,
  ``An introduction to the confinement problem,''
Lect.\ Notes Phys.\  {\bf 821}, 1 (2011).
}

\lref\ArmoniJSA{
  A.~Armoni and V.~Niarchos,
  ``Defects in Chern-Simons theory, gauged WZW models on the brane, and level-rank duality,''
JHEP {\bf 1507}, 062 (2015).
[arXiv:1505.02916 [hep-th]].
}
\lref\ArmoniZZ{
  A.~Armoni,
Int.\ J.\ Mod.\ Phys.\ A {\bf 25}, 470 (2010)..
}
\lref\ArmoniVV{
  A.~Armoni, A.~Giveon, D.~Israel and V.~Niarchos,
  ``Brane Dynamics and 3D Seiberg Duality on the Domain Walls of 4D N=1 SYM,''
JHEP {\bf 0907}, 061 (2009).
[arXiv:0905.3195 [hep-th]].
}

\lref\DraperMPJ{
  P.~Draper,
  ``Domain Walls and the $CP$ Anomaly in Softly Broken Supersymmetric QCD,''
[arXiv:1801.05477 [hep-th]].
}

\lref\SeibergRS{
  N.~Seiberg and E.~Witten,
  ``Electric - magnetic duality, monopole condensation, and confinement in N=2 supersymmetric Yang-Mills theory,''
Nucl.\ Phys.\ B {\bf 426}, 19 (1994), Erratum: [Nucl.\ Phys.\ B {\bf 430}, 485 (1994)].
[hep-th/9407087].
}

\lref\ColemanUZ{
  S.~R.~Coleman,
  ``More About the Massive Schwinger Model,''
Annals Phys.\  {\bf 101}, 239 (1976).
}
\lref\IntriligatorAU{
  K.~A.~Intriligator and N.~Seiberg,
  ``Lectures on supersymmetric gauge theories and electric-magnetic duality,''
Nucl.\ Phys.\ Proc.\ Suppl.\  {\bf 45BC}, 1 (1996), [Subnucl.\ Ser.\  {\bf 34}, 237 (1997)].
[hep-th/9509066].
}

\lref\toappear{
    To Appear.

}

\lref\SeibergAJ{
  N.~Seiberg and E.~Witten,
  ``Monopoles, duality and chiral symmetry breaking in N=2 supersymmetric QCD,''
Nucl.\ Phys.\ B {\bf 431}, 484 (1994).
[hep-th/9408099].
}

\lref\Deepak{
D.~Naidu, ``Categorical Morita equivalence for group-theoretical categories, " Communications in Algebra 35.11 (2007): 3544-3565.
APA	
}
\lref\AharonyDHA{
  O.~Aharony, S.~S.~Razamat, N.~Seiberg and B.~Willett,
  ``3d dualities from 4d dualities,''
JHEP {\bf 1307}, 149 (2013).
[arXiv:1305.3924 [hep-th]].
}

\lref\ArgyresJJ{
  P.~C.~Argyres and M.~R.~Douglas,
  ``New phenomena in SU(3) supersymmetric gauge theory,''
Nucl.\ Phys.\ B {\bf 448}, 93 (1995).
[hep-th/9505062].
}

\lref\SusskindAA{
  L.~Susskind,
Phys.\ Rev.\ D {\bf 49}, 6606 (1994).
[hep-th/9308139].
}

\lref\DubovskySH{
  S.~Dubovsky, R.~Flauger and V.~Gorbenko,
  ``Effective String Theory Revisited,''
JHEP {\bf 1209}, 044 (2012).
[arXiv:1203.1054 [hep-th]].
}

\lref\AharonyIPA{
  O.~Aharony and Z.~Komargodski,
  ``The Effective Theory of Long Strings,''
JHEP {\bf 1305}, 118 (2013).
[arXiv:1302.6257 [hep-th]].
}

\lref\HellermanCBA{
  S.~Hellerman, S.~Maeda, J.~Maltz and I.~Swanson,
  ``Effective String Theory Simplified,''
JHEP {\bf 1409}, 183 (2014).
[arXiv:1405.6197 [hep-th]].
}

\lref\AthenodorouCS{
  A.~Athenodorou, B.~Bringoltz and M.~Teper,
  ``Closed flux tubes and their string description in D=3+1 SU(N) gauge theories,''
JHEP {\bf 1102}, 030 (2011).
[arXiv:1007.4720 [hep-lat]].
}

\lref\CordovaKUE{
  C.~Cordova, P.~S.~Hsin and N.~Seiberg,
   ``Time-Reversal Symmetry, Anomalies, and Dualities in (2+1)$d$,''
[arXiv:1712.08639 [cond-mat.str-el]].
}

\lref\WittenDF{
  E.~Witten,
  ``Constraints on Supersymmetry Breaking,''
Nucl.\ Phys.\ B {\bf 202}, 253 (1982)..
}

\lref\AharonyHDA{
  O.~Aharony, N.~Seiberg and Y.~Tachikawa,
  ``Reading between the lines of four-dimensional gauge theories,''
JHEP {\bf 1308}, 115 (2013).
[arXiv:1305.0318 [hep-th]].
}

\lref\OhtaIV{
  K.~Ohta,
   ``Supersymmetric index and s rule for type IIB branes,''
JHEP {\bf 9910}, 006 (1999).
[hep-th/9908120].
}

\lref\Oferfuture{
O.~Aharony, S.~Jain, and S.~Minwalla,
  ``Flows, Fixed Points and Duality in
Chern-Simons-matter theories,''
  to appear.
}

\lref\CallanSA{
  C.~G.~Callan, Jr. and J.~A.~Harvey,
  ``Anomalies and Fermion Zero Modes on Strings and Domain Walls,''
Nucl.\ Phys.\ B {\bf 250}, 427 (1985).
}

\lref\tHooftXSS{
  G.~'t Hooft et al.,
    ``Recent Developments in Gauge Theories. Proceedings, Nato Advanced Study Institute, Cargese, France, August 26 - September 8, 1979,''
NATO Sci.\ Ser.\ B {\bf 59}, pp.1 (1980).
}

\lref\AppelquistVG{
  T.~Appelquist and R.~D.~Pisarski,
  ``High-Temperature Yang-Mills Theories and Three-Dimensional Quantum Chromodynamics,''
Phys.\ Rev.\ D {\bf 23}, 2305 (1981).
}
\lref\AharonyMJS{
  O.~Aharony,
  ``Baryons, monopoles and dualities in Chern-Simons-matter theories,''
JHEP {\bf 1602}, 093 (2016).
[arXiv:1512.00161 [hep-th]].
}
\lref\HarveyIT{
  J.~A.~Harvey,
  ``TASI 2003 lectures on anomalies,''
[hep-th/0509097].
}
\lref\JainGZA{
  S.~Jain, S.~Minwalla and S.~Yokoyama,
   ``Chern Simons duality with a fundamental boson and fermion,''
JHEP {\bf 1311}, 037 (2013).
[arXiv:1305.7235 [hep-th]].
}

\lref\KonishiIZ{
  K.~Konishi,
  ``Confinement, supersymmetry breaking and theta parameter dependence in the Seiberg-Witten model,''
Phys.\ Lett.\ B {\bf 392}, 101 (1997).
[hep-th/9609021].
}

\lref\KarchAUX{
  A.~Karch, B.~Robinson and D.~Tong,
   ``More Abelian Dualities in 2+1 Dimensions,''
JHEP {\bf 1701}, 017 (2017).
[arXiv:1609.04012 [hep-th]].
}

\lref\ColemanZI{
  S.~R.~Coleman and B.~R.~Hill,
  ``No More Corrections to the Topological Mass Term in QED in Three-Dimensions,''
Phys.\ Lett.\  {\bf 159B}, 184 (1985)..
}

\lref\DineSGQ{
  M.~Dine, P.~Draper, L.~Stephenson-Haskins and D.~Xu,
  ``$\theta$ and the $\eta^\prime$ in Large $N$ Supersymmetric QCD,''
[arXiv:1612.05770 [hep-th]].
}
\lref\MuruganZAL{
  J.~Murugan and H.~Nastase,
   ``Particle-vortex duality in topological insulators and superconductors,''
JHEP {\bf 1705}, 159 (2017).
[arXiv:1606.01912 [hep-th]].
}

\lref\MaldacenaPB{
  J.~M.~Maldacena and H.~S.~Nastase,
   ``The Supergravity dual of a theory with dynamical supersymmetry breaking,''
JHEP {\bf 0109}, 024 (2001).
[hep-th/0105049].
}

\lref\GomisIXY{
  J.~Gomis, Z.~Komargodski and N.~Seiberg,
  ``Phases Of Adjoint QCD$_3$ And Dualities,''
[arXiv:1710.03258 [hep-th]].
}

\lref\GomisXW{
  J.~Gomis,
   ``On SUSY breaking and chi**SB from string duals,''
Nucl.\ Phys.\ B {\bf 624}, 181 (2002).
[hep-th/0111060].
}

\lref\BeniniDUS{
  F.~Benini, P.~S.~Hsin and N.~Seiberg,
  ``Comments on Global Symmetries, Anomalies, and Duality in (2+1)d,''
[arXiv:1702.07035 [cond-mat.str-el]].
}

\lref\AmatiWQ{
  D.~Amati, M.~Ciafaloni and G.~Veneziano,
Phys.\ Lett.\ B {\bf 197}, 81 (1987).
}

\lref\HsinBLU{
  P.~S.~Hsin and N.~Seiberg,
  ``Level/rank Duality and Chern-Simons-Matter Theories,''
JHEP {\bf 1609}, 095 (2016).
[arXiv:1607.07457 [hep-th]].
}

\lref\WittenCIO{
  E.~Witten,
   ``The "Parity" Anomaly On An Unorientable Manifold,''
Phys.\ Rev.\ B {\bf 94}, no. 19, 195150 (2016).
[arXiv:1605.02391 [hep-th]].
}

\lref\KapustinLWA{
  A.~Kapustin and R.~Thorngren,
  ``Anomalies of discrete symmetries in three dimensions and group cohomology,''
Phys.\ Rev.\ Lett.\  {\bf 112}, no. 23, 231602 (2014).
[arXiv:1403.0617 [hep-th]].
}
\lref\HsiehXAA{
  C.~T.~Hsieh, G.~Y.~Cho and S.~Ryu,
   ``Global anomalies on the surface of fermionic symmetry-protected topological phases in (3+1) dimensions,''
Phys.\ Rev.\ B {\bf 93}, no. 7, 075135 (2016).
[arXiv:1503.01411 [cond-mat.str-el]].
}

\lref\ElitzurXJ{
  S.~Elitzur, Y.~Frishman, E.~Rabinovici and A.~Schwimmer,
  ``Origins of Global Anomalies in Quantum Mechanics,''
Nucl.\ Phys.\ B {\bf 273}, 93 (1986).
}

\lref\KapustinVZ{
	A.~Kapustin, H.~Kim and J.~Park,
	``Dualities for 3d Theories with Tensor Matter,''
	JHEP {\bf 1112}, 087 (2011).
	[arXiv:1110.2547 [hep-th]].
}

\lref\KapustinKZ{
	A.~Kapustin, B.~Willett and I.~Yaakov,
	``Exact Results for Wilson Loops in Superconformal Chern-Simons Theories with Matter,''
	JHEP {\bf 1003}, 089 (2010).
	[arXiv:0909.4559 [hep-th]].
}

\lref\KapustinZVA{
  A.~Kapustin and R.~Thorngren,
  ``Anomalies of discrete symmetries in various dimensions and group cohomology,''
[arXiv:1404.3230 [hep-th]].
}

\lref\DunneQY{
  G.~V.~Dunne,
  ``Aspects of Chern-Simons theory,''
[hep-th/9902115].
}

\lref\KarchSXI{
  A.~Karch and D.~Tong,
   ``Particle-Vortex Duality from 3d Bosonization,''
Phys.\ Rev.\ X {\bf 6}, no. 3, 031043 (2016).
[arXiv:1606.01893 [hep-th]].
}

\lref\LevinYB{
  M.~Levin and Z.~C.~Gu,
  ``Braiding statistics approach to symmetry-protected topological phases,''
Phys.\ Rev.\ B {\bf 86}, 115109 (2012).
[arXiv:1202.3120 [cond-mat.str-el]].
}

\lref\BhardwajCLT{
  L.~Bhardwaj, D.~Gaiotto and A.~Kapustin,
  ``State sum constructions of spin-TFTs and string net constructions of fermionic phases of matter,''
JHEP {\bf 1704}, 096 (2017).
[arXiv:1605.01640 [cond-mat.str-el]].
}

\lref\tHooftUJ{
  G.~'t Hooft,
  ``A Property of Electric and Magnetic Flux in Nonabelian Gauge Theories,''
Nucl.\ Phys.\ B {\bf 153}, 141 (1979).
}

\lref\Brower{
R.~C.~Brower and J.~Harte.
Physical Review 164.5 (1967): 1841.
}

\lref\ArmoniEE{
  A.~Armoni and T.~J.~Hollowood,
  ``Interactions of domain walls of SUSY Yang-Mills as D-branes,''
JHEP {\bf 0602}, 072 (2006).
[hep-th/0601150].
}
\lref\ArmoniSP{
  A.~Armoni and T.~J.~Hollowood,
  ``Sitting on the domain walls of N=1 super Yang-Mills,''
JHEP {\bf 0507}, 043 (2005).
[hep-th/0505213].
}
\lref\BarkeshliIDA{
  M.~Barkeshli and J.~McGreevy,
   ``Continuous transition between fractional quantum Hall and superfluid states,''
Phys.\ Rev.\ B {\bf 89}, no. 23, 235116 (2014).
[arXiv:1201.4393 [cond-mat.str-el]].
}

\lref\WittenHF{
  E.~Witten,
  ``Quantum Field Theory and the Jones Polynomial,''
Commun.\ Math.\ Phys.\  {\bf 121}, 351 (1989).
}

\lref\ArgyresXN{
  P.~C.~Argyres, M.~R.~Plesser, N.~Seiberg and E.~Witten,
  ``New N=2 superconformal field theories in four-dimensions,''
Nucl.\ Phys.\ B {\bf 461}, 71 (1996).
[hep-th/9511154].
}

\lref\GatesNR{
  S.~J.~Gates, M.~T.~Grisaru, M.~Rocek and W.~Siegel,
Front.\ Phys.\  {\bf 58}, 1 (1983).
[hep-th/0108200].
}

\lref\JensenXBS{
  K.~Jensen and A.~Karch,
  ``Embedding three-dimensional bosonization dualities into string theory,''
[arXiv:1709.07872 [hep-th]].
}

\lref\JensenDSO{
  K.~Jensen and A.~Karch,
  ``Bosonizing three-dimensional quiver gauge theories,''
[arXiv:1709.01083 [hep-th]].
}

\lref\JafferisUN{
  D.~L.~Jafferis,
   ``The Exact Superconformal R-Symmetry Extremizes Z,''
JHEP {\bf 1205}, 159 (2012).
[arXiv:1012.3210 [hep-th]].
}

\lref\HamaEA{
  N.~Hama, K.~Hosomichi and S.~Lee,
  ``SUSY Gauge Theories on Squashed Three-Spheres,''
JHEP {\bf 1105}, 014 (2011).
[arXiv:1102.4716 [hep-th]].
}
\lref\BeniniAED{
  F.~Benini,
   ``Three-dimensional dualities with bosons and fermions,''
[arXiv:1712.00020 [hep-th]].
}

\lref\DieriglXTA{
  M.~Dierigl and A.~Pritzel,
  ``Topological Model for Domain Walls in (Super-)Yang-Mills Theories,''
Phys.\ Rev.\ D {\bf 90}, no. 10, 105008 (2014).
[arXiv:1405.4291 [hep-th]].
}

\lref\HamaAV{
  N.~Hama, K.~Hosomichi and S.~Lee,
  ``Notes on SUSY Gauge Theories on Three-Sphere,''
JHEP {\bf 1103}, 127 (2011).
[arXiv:1012.3512 [hep-th]].
}
\lref\ImamuraSU{
  Y.~Imamura and S.~Yokoyama,
   ``Index for three dimensional superconformal field theories with general R-charge assignments,''
JHEP {\bf 1104}, 007 (2011).
[arXiv:1101.0557 [hep-th]].
}
\lref\FreedRLK{
  D.~S.~Freed, Z.~Komargodski and N.~Seiberg,
  ``The Sum Over Topological Sectors and $\theta$ in the 2+1-Dimensional $CP^1$ $\sigma$-Model,''
[arXiv:1707.05448 [cond-mat.str-el]].
}

\lref\NiemiRQ{
  A.~J.~Niemi and G.~W.~Semenoff,
   ``Axial Anomaly Induced Fermion Fractionization and Effective Gauge Theory Actions in Odd Dimensional Space-Times,''
Phys.\ Rev.\ Lett.\  {\bf 51}, 2077 (1983).
}

\lref\KimWB{
  S.~Kim,
   ``The Complete superconformal index for N=6 Chern-Simons theory,''
Nucl.\ Phys.\ B {\bf 821}, 241 (2009), Erratum: [Nucl.\ Phys.\ B {\bf 864}, 884 (2012)].
[arXiv:0903.4172 [hep-th]].
}
\lref\GaiottoTNE{
  D.~Gaiotto, Z.~Komargodski and N.~Seiberg,
   ``Time-Reversal Breaking in QCD$_4$, Walls, and Dualities in 2+1 Dimensions,''
[arXiv:1708.06806 [hep-th]].
}
\lref\JensenBJO{
  K.~Jensen,
   ``A master bosonization duality,''
JHEP {\bf 1801}, 031 (2018).
[arXiv:1712.04933 [hep-th]].
}
\lref\CordovaVAB{
  C.~Cordova, P.~S.~Hsin and N.~Seiberg,
   ``Global Symmetries, Counterterms, and Duality in Chern-Simons Matter Theories with Orthogonal Gauge Groups,''
[arXiv:1711.10008 [hep-th]].
}

\lref\GiveonZN{
  A.~Giveon and D.~Kutasov,
Nucl.\ Phys.\ B {\bf 812}, 1 (2009).
[arXiv:0808.0360 [hep-th]].
}

\lref\deBoerKR{
  J.~de Boer, K.~Hori and Y.~Oz,
Nucl.\ Phys.\ B {\bf 500}, 163 (1997).
[hep-th/9703100].
}

\lref\HamaAV{
  N.~Hama, K.~Hosomichi and S.~Lee,
JHEP {\bf 1103}, 127 (2011).
[arXiv:1012.3512 [hep-th]].
}

\lref\JafferisUN{
  D.~L.~Jafferis,
JHEP {\bf 1205}, 159 (2012).
[arXiv:1012.3210 [hep-th]].
}

\lref\RadicevicYLA{
  D.~Radicevic,
  ``Disorder Operators in Chern-Simons-Fermion Theories,''
JHEP {\bf 1603}, 131 (2016).
[arXiv:1511.01902 [hep-th]].
}

\lref\ChenCD{
  W.~Chen, M.~P.~A.~Fisher and Y.~S.~Wu,
   ``Mott transition in an anyon gas,''
Phys.\ Rev.\ B {\bf 48}, 13749 (1993).
[cond-mat/9301037].
}

\lref\TachikawaCHA{
  Y.~Tachikawa and K.~Yonekura,
  ``On time-reversal anomaly of 2+1d topological phases,''
PTEP {\bf 2017}, no. 3, 033B04 (2017).
[arXiv:1610.07010 [hep-th]].
}

\lref\GiombiTXG{
  S.~Giombi,
  ``Testing the Boson/Fermion Duality on the Three-Sphere,''
[arXiv:1707.06604 [hep-th]].
}

\lref\DiPietroKCD{
  L.~Di Pietro and E.~Stamou,
  ``Scaling dimensions in QED$_3$ from the $\epsilon$-expansion,''
JHEP {\bf 1712}, 054 (2017).
[arXiv:1708.03740 [hep-th]].
}

\lref\ChesterVDH{
  S.~M.~Chester, L.~V.~Iliesiu, M.~Mezei and S.~S.~Pufu,
  ``Monopole Operators in $U(1)$ Chern-Simons-Matter Theories,''
[arXiv:1710.00654 [hep-th]].
}

\lref\IntriligatorDD{
  K.~A.~Intriligator, N.~Seiberg and D.~Shih,
JHEP {\bf 0604}, 021 (2006).
[hep-th/0602239].
}
\lref\JainQI{
  S.~Jain, S.~P.~Trivedi, S.~R.~Wadia and S.~Yokoyama,
  ``Supersymmetric Chern-Simons Theories with Vector Matter,''
JHEP {\bf 1210}, 194 (2012).
[arXiv:1207.4750 [hep-th]].
}

\lref\ArmoniJKL{
  A.~Armoni and V.~Niarchos,
  ``Phases of QCD$_3$ from Non-SUSY Seiberg Duality and Brane Dynamics,''
[arXiv:1711.04832 [hep-th]].
}

\lref\ChermanDWT{
  A.~Cherman and M.~Unsal,
  ``Critical behavior of gauge theories and Coulomb gases in three and four dimensions,''
[arXiv:1711.10567 [hep-th]].
}

\lref\KomargodskiDMC{
  Z.~Komargodski, A.~Sharon, R.~Thorngren and X.~Zhou,
  ``Comments on Abelian Higgs Models and Persistent Order,''
[arXiv:1705.04786 [hep-th]].
}

\lref\SeibergPQ{
  N.~Seiberg,
Nucl.\ Phys.\ B {\bf 435}, 129 (1995).
[hep-th/9411149].
}

\lref\GiveonZN{
  A.~Giveon and D.~Kutasov,
  ``Seiberg Duality in Chern-Simons Theory,''
Nucl.\ Phys.\ B {\bf 812}, 1 (2009).
[arXiv:0808.0360 [hep-th]].
}


\def\savefig{\expandafter\savefigaux\expandafter{\the\figno}}
\def\savefigaux#1#2#3#4{\DefWarn#2%
 \gdef#2{fig.~\hyperref{}{figure}{#1}{#1}}%
 \writedef{#2\leftbracket fig.\noexpand~\xfig#2}%
 \expandafter\gdef\csname savedfig-\string#2\endcsname{%
   \figinsert\figin{\centerline{#4}}%
   \medskip\centerline{\vbox{\baselineskip12pt
       \advance\hsize by -1truein\noindent\wrlabeL{#2=#2}
       \footnotefont%
       {\bf Fig.~\hyperdef\hypernoname{figure}{#1}{#1}:} #3}}%
   \bigskip\endinsert}%
 \global\advance\figno by1}
\def\putfig#1{\csname savedfig-\string#1\endcsname}

\savefig\Wall{New vacua can emerge from infinity in field space at a wall in parameter space where the asymptotics of the superpotential changes (one such vacuum appears on the other side of the wall in the figure). The new vacua can be reliably exhibited by a perturbative two-loop computation. This  combined with our proposed infrared dynamics of the vacua results in a description of the phase diagram of $\CN=1$ theories as a function of superpotential couplings. 
}%
{\epsfxsize2in\epsfbox{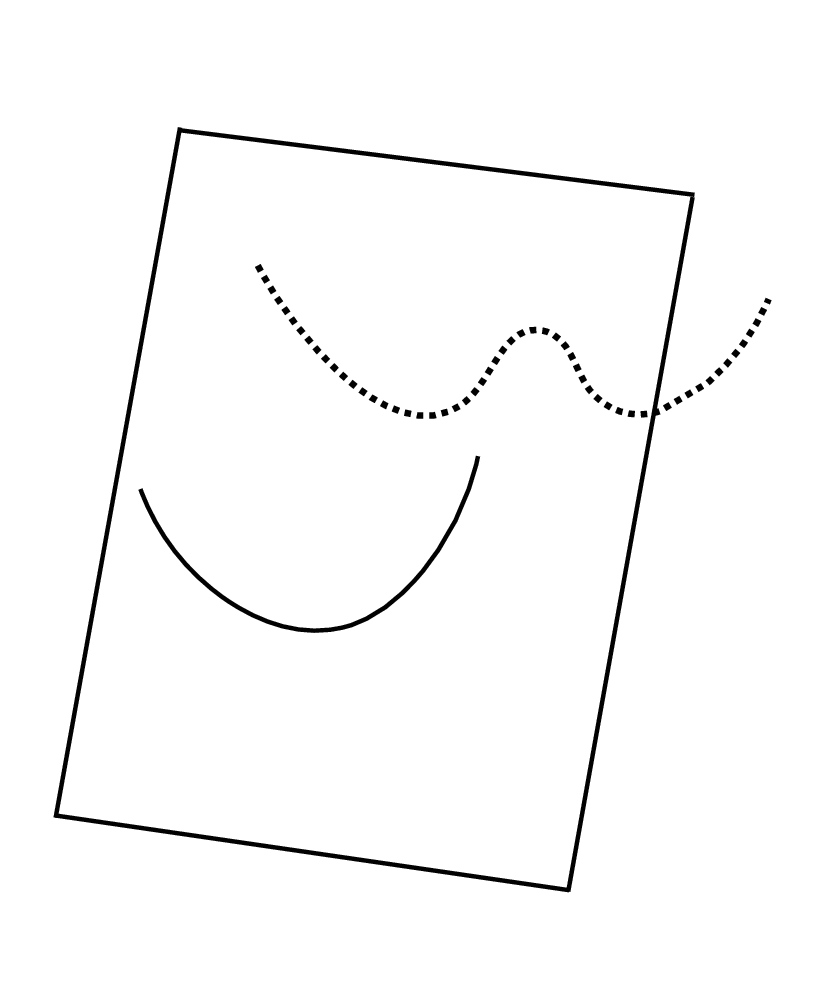}}

\savefig\Largek{Proposed phase diagram for $k\geq N$.  On the right hand side of the wall at $m=0$ the theory 
develops new supersymmetric vacua that come in from infinity in field space. These vacua flow to specific TQFTs. As the mass is further increased vacua merge in a sequence of second order phase transitions. At the end of the sequence the physics is described by the large and positive mass asymptotic phase. On the left-hand side of the wall at $m=0$ there is a unique supersymmetric vacuum with a TQFT that coincides with that describing the large and negative mass asymptotic phase. The new   vacua at small positive mass account for the jump in the Witten between the asymptotic large mass phases. 
}%
{\epsfxsize3.5in\epsfbox{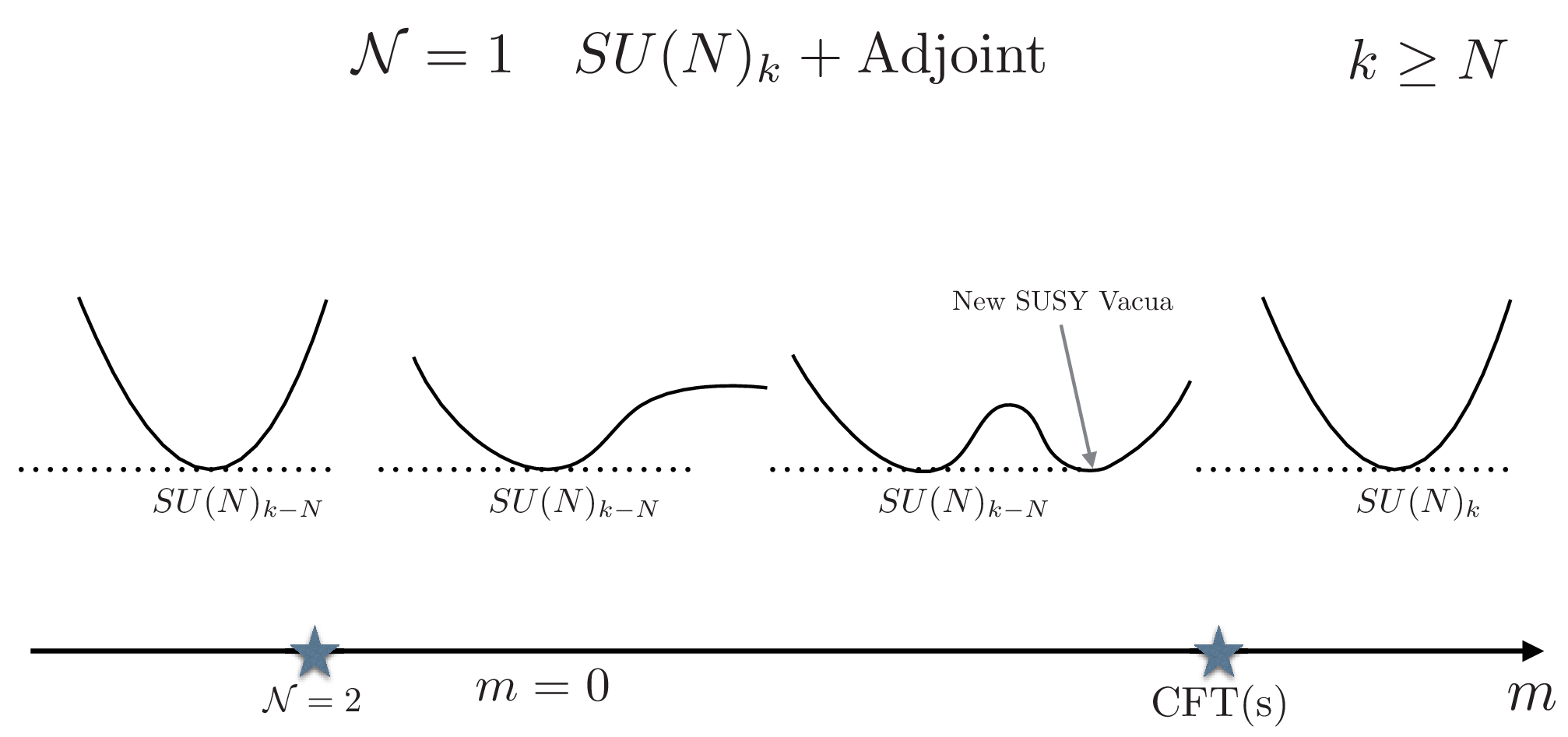}}

\savefig\Smallk{Proposed phase diagram for $0<k<N$.  The asymptotic large and negative mass
phase breaks supersymmetry and is described by  a Majorana Goldstino with a TQFT. This TQFT persists all the way down to the wall at $m=0$, with an additional massless Majorana Goldstino appearing at the $\CN=2$ point. 
On the right-hand side of the wall at $m=0$ the theory 
develops new supersymmetric vacua with novel TQFTs that come in from infinity in field space. 
The  vacuum state at the origin in field space is a metastable supersymmetry breaking state. 
As the mass is further increased vacua merge in a sequence of second order phase transitions. At the end of the sequence the physics is described by the large and positive mass asymptotic phase, where supersymmetry is unbroken and described by a distinct TQFT.  The new   vacua at small positive mass account for the jump in the Witten between the asymptotic large mass phases. 
}%
{\epsfxsize3.5in\epsfbox{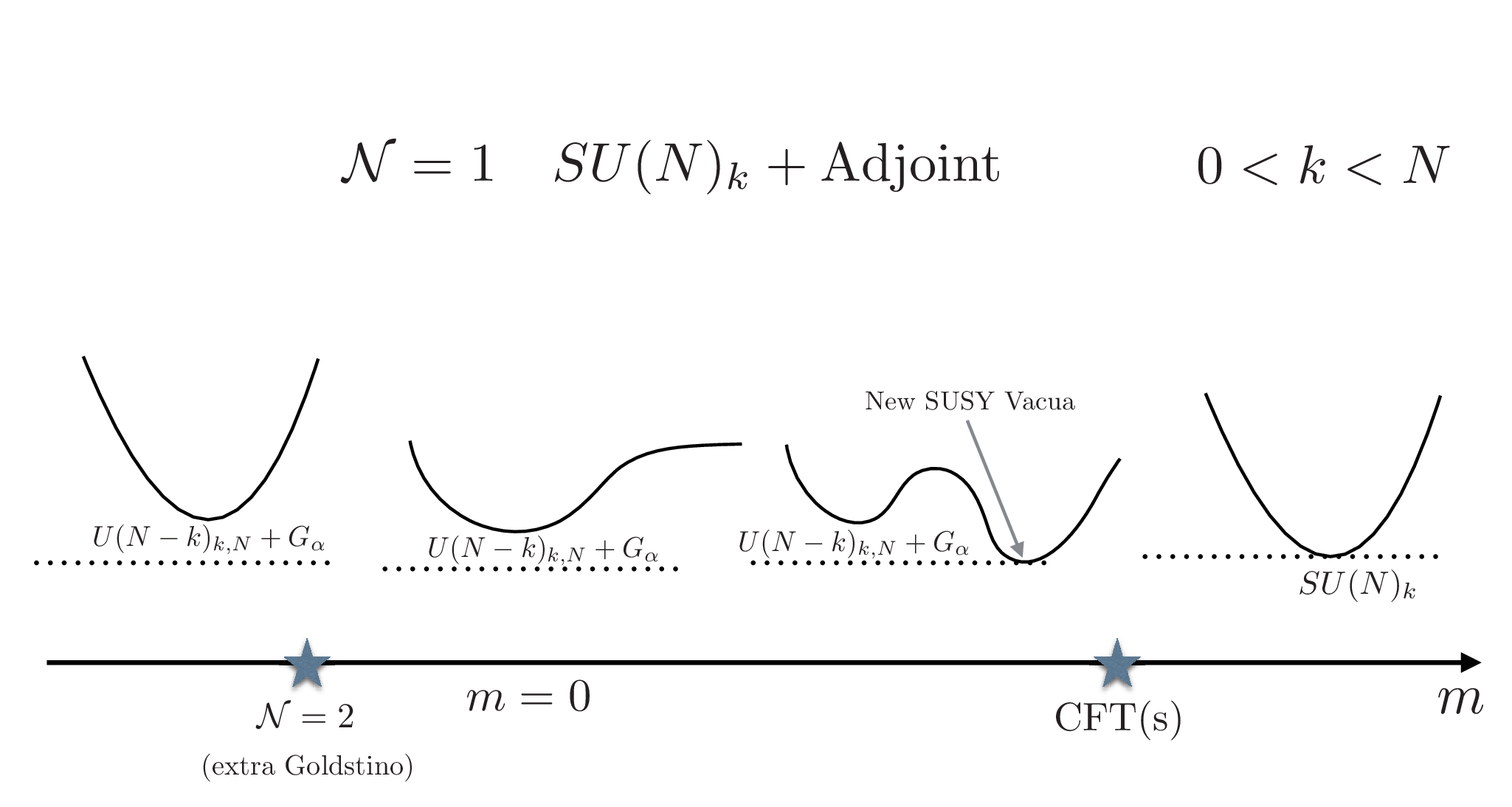}}

\savefig\Zerok{Phase diagram for $k=0$. The theory flows to a supersymmetric trivial gapped vacuum everywhere, except at  the $\CN=2$ supersymmetric point $m=0$,    where there is a runaway due to 
the non-perturbatively induced $\CN=2$ superpotential. The Witten index does not jump.
}%
{\epsfxsize3.5in\epsfbox{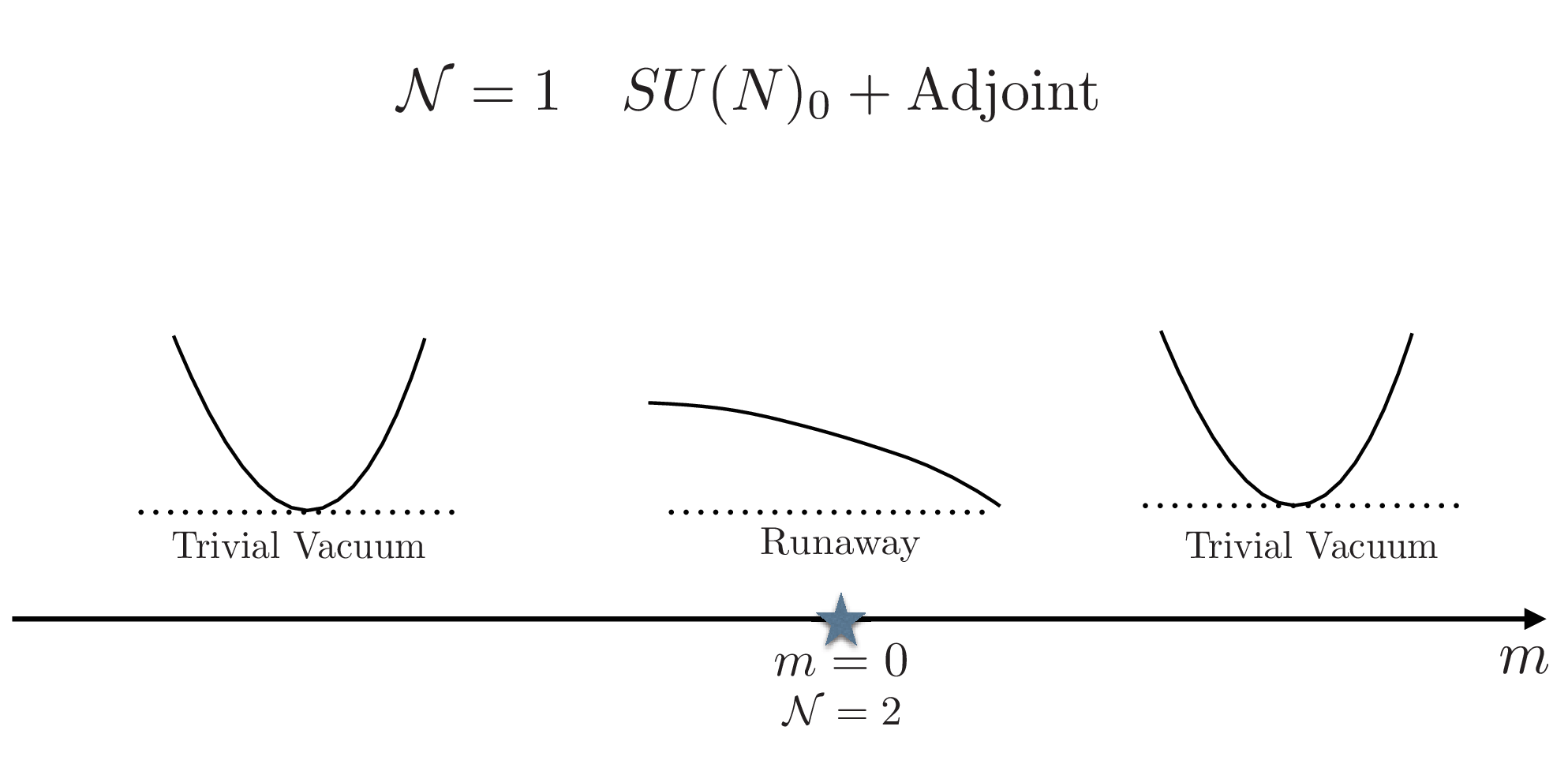}}

\savefig\Dualityfig{A duality between $SU(k+1)_{-N-k/2}$ gauge theory with a fundamental matter multiplet and $U(N)_{k+N/2+1/2,k+1/2}$ with a 
fundamental matter multiplet.  The duality exchanges the vacuum that exists asymptotically in the phase
diagram and the vacuum that appears from infinity near $m=m'=0$. Note that the sign of $m'$ is reversed. The phases match by virtue of level/rank duality.
}%
{\epsfxsize3.8in\epsfbox{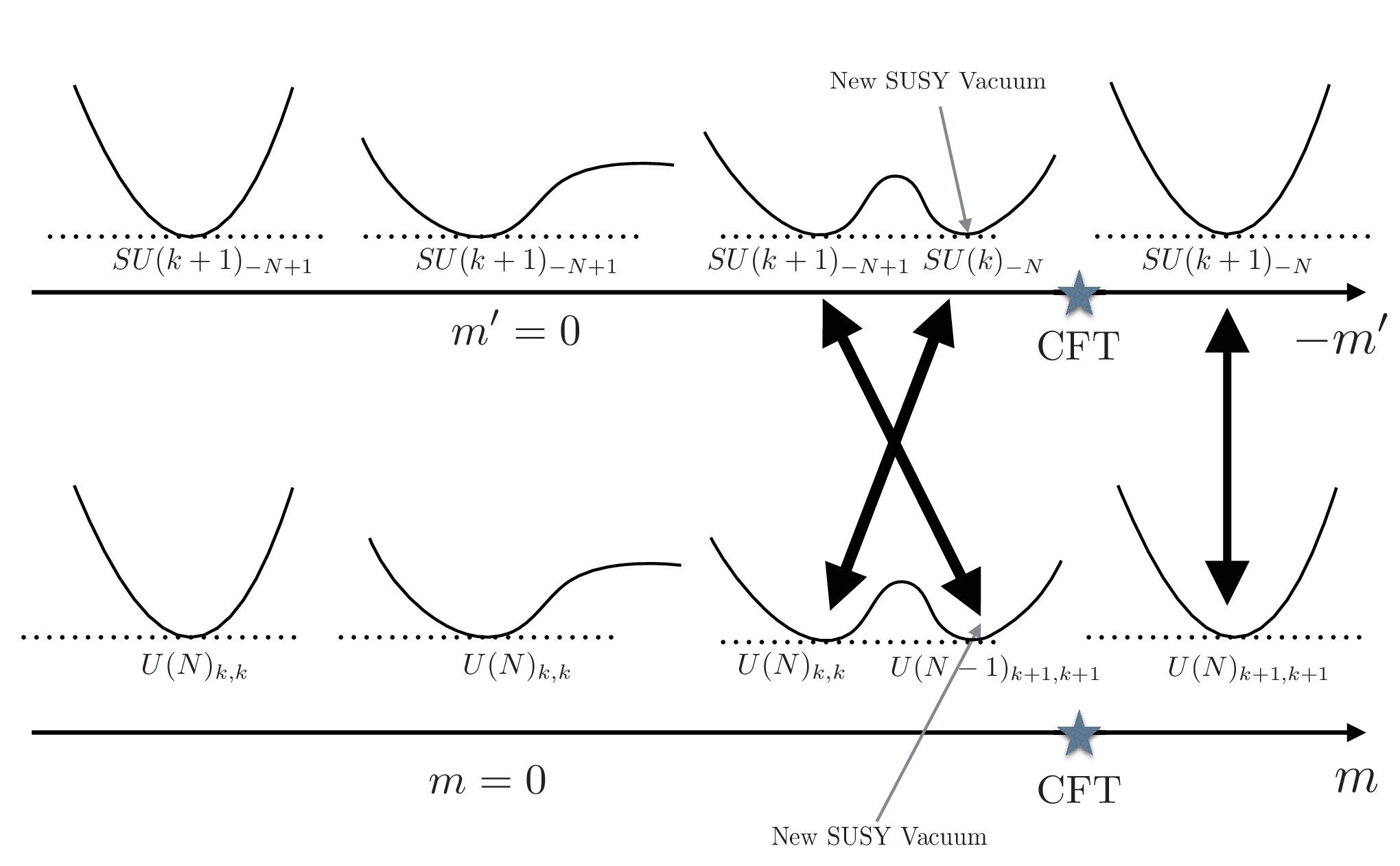}}

\Title{
} {\vbox{\centerline{Phases of ${\cal N}=1$ Theories in 2+1 Dimensions}}}
{\centerline{}}

\vskip+5pt
\centerline{Vladimir Bashmakov,${}^1$ Jaume Gomis,${}^2$ Zohar Komargodski,${}^{3,4}$ and Adar Sharon${}^3$ }
\vskip25pt
 \centerline{\it ${}^1$ SISSA and INFN - Via Bonomea 265; I 34136 Trieste, Italy }
  \centerline{\it ${}^2$ Perimeter Institute for Theoretical Physics, Waterloo, Ontario, N2L 2Y5, Canada}
\centerline{\it ${}^3$ Department of Particle Physics and Astrophysics, Weizmann Institute of Science, Israel }
\centerline{\it ${}^4$   Simons Center for Geometry and Physics, Stony Brook University, Stony Brook, NY}
\vskip25pt

\noindent

We study the dynamics of 2+1 dimensional theories with ${\cal N}=1$ supersymmetry.
In these theories the supersymmetric ground states behave discontinuously at co-dimension one walls in the
space of couplings, with new vacua coming in from infinity in field space. We show that the dynamics
near these walls is calculable: the two-loop effective potential yields exact results about the ground states near the walls. 
Far away from the walls the ground states can be inferred by decoupling arguments. In this way, we are able to follow the
ground states of ${\cal N}=1$  theories in 2+1 dimensions and construct the infrared phases of these theories.
 We study two examples in detail: Adjoint SQCD and SQCD with one fundamental quark. In Adjoint QCD we show that for sufficiently small Chern-Simons level the theory has a non-perturbative metastable supersymmetry-breaking
ground state. We also briefly discuss the critical points of this theory. For SQCD with one quark we establish an infrared duality between a $U(N)$ gauge theory and an $SU(N)$ gauge theory. The duality crucially involves the vacua
that appear from infinity near the walls.

\Date{}

\listtoc
\writetoc

\newsec{Introduction and Summary}

There has been substantial progress recently on the dynamics of 2+1 dimensional gauge theories. Many interesting phenomena such as  confining phases, symmetry breaking phases, phases with topological order and dualities were uncovered. These
results are supported by many nontrivial consistency checks, such as the matching of various discrete anomalies (including anomalies of space-time symmetries), consistency with renormalization group flows, various other non-perturbative constraints (such as the Vafa-Witten theorem, the study of counterterms, etc.) and the rigorous study of weakly coupled limits such as the large $N$ limit, etc.  For recent work   
on the phases of 2+1 dimensional gauge theories see~\refs{\AharonyNH\JainGZA\RadicevicYLA\AharonyMJS\KarchSXI\MuruganZAL\SeibergGMD\HsinBLU\RadicevicWQN\KachruRUI\KachruAON\KarchAUX\MetlitskiDHT\AharonyJVV\KomargodskiDMC\KomargodskiSMK\KomargodskiKEH\FreedRLK\GiombiTXG\DiPietroKCD\GaiottoTNE\JensenDSO\ChesterVDH\GomisIXY\ArmoniJKL\CordovaVAB\ChermanDWT\BeniniAED\JensenBJO-\CordovaKUE} and references therein.

Our main goal in this paper is to revisit some questions about the dynamics of supersymmetric gauge theories in 2+1 dimensions. We will investigate  theories with $\CN=1$ supersymmetry, that is  theories with two real supercharges. 
Theories with $\CN=1$ supersymmetry in 2+1 dimensions have received relatively little attention over the years. This is mostly because $\CN=1$ theories do not enjoy the powerful constraints on the  dynamics implied by holomorphy. In addition, localization techniques essentially do not apply and indeed relatively little has hitherto been known about these theories, cf.~\refs{\WittenDS,\GomisIXY}. In this paper we introduce some new tools to study  $\CN=1$ supersymmetric theories.

Since the superpotential parameters reside in real superfields, the superpotential is not protected against quantum corrections. Furthermore the number of supersymmetric ground states can jump across walls   in the space of couplings  where the behaviour of the potential at infinity in field space changes.\foot{This is in contrast to $\CN=2$ theories,  where the number of supersymmetric ground states cannot jump as a function of real masses, cf.~\IntriligatorLCA.} 
This can happen at co-dimension one surfaces in parameter space, which we refer to as walls. 
On one of the sides  of the wall there can be   supersymmetric ground states that ``appear from infinity'' in field space (see \Wall).  The main point is that since these vacua appear from infinity, and since the underlying model is typically super-renormalizable, these vacua can be reliably studied by computing radiative corrections to the effective potential on the wall in parameter space. We show that a two-loop computation is necessary to establish the existence of these vacua (and sufficient for our purposes). We will see that this quantum-mechanically generated superpotential, together with our proposed infrared dynamics of each new vacuum, precisely reproduces   the Witten index~\WittenDF\ of the theory farther away from the wall (where the index must remain constant, as the asymptotics of the potential does not change away from the wall). Together with an analysis of the physics far from these walls, one can therefore determine the phases of $\CN=1$ theories as a function of the superpotential parameters. In particular, one can identify the phase transitions that take place.

 \putfig\Wall

We will carry out this analysis explicitly in two cases which demonstrate rather different mechanisms and principles. The methods we introduce are general and can be applied to a wide variety of $\CN=1$ models.

We turn now to a brief summary of the theories we analyze and outline our results (the derivations and details are presented in the bulk of the paper). The gauge theories we study have a Yang-Mills term   and  a Chern-Simons term. We label  a theory by the gauge group,    the matter content  and by the  Chern-Simons level\foot{
The level $k$, upon which time-reversal acts   by $k\rightarrow -k$, is given in terms of the Chern-Simons level $k_{bare}$ appearing in the classical Lagrangian by 
\eqn\levelshift{
k=k_{bare}-{1\over 2}\sum_{f} T(R_f)\,,}
where the sum is over  the charged Majorana fermions in the theory and $T(R)$ is the index of the real representation $R$. The shift can be thought of as the contribution from the (massless) fermion determinant. Integrating out a  massive Majorana fermion of mass $m$ in a real representation $R$ shifts the level $k$ by~\refs{\NiemiRQ\RedlichKN-\RedlichDV}
\eqn\levelmassshi{
k\rightarrow k+{1\over 2}\hbox{sign}(m)\,T(R)\,.}
  This  makes manifest the action of time-reversal, which flips the sign of the mass of a fermion. For a   fermion in a complex or pseudoreal representation $R$ the shift in  \levelshift\levelmassshi\ should be multiplied by a factor of 2. We note that while the level $k\in \Z/2$, the ultraviolet and infrared levels $k_{bare}$ and $k+{1\over 2}\hbox{sign}(m)\,T(R)$ are always integrally quantized.} $k$,
 which henceforth we   take to be non-negative since the theory with $k<0$ can be  obtained by acting with time-reversal on the theory with $k> 0$. The theory with $k=0$ and  with massless fermions is time-reversal invariant and typically needs a separate treatment.

\bigskip
\centerline{$\underline{\hbox{$\CN=1$ \ $SU(N)_k$ Vector Multiplet}}$}
\bigskip

The gauge multiplet consists of an $SU(N)$ gauge field  $A$, and a Majorana fermion $\lambda$. All the terms in the Lagrangian are the standard minimal couplings with a Chern-Simons term. $\CN=1$ supersymmetry requires that the adjoint fermion mass is $-{kg^2\over 2\pi}$, i.e. proportional to the Chern-Simons level $k$.  The $\CN=1$ Lagrangian is\foot{In this paper we follow   standard practice and write $k$ in the Lagrangian, keeping in mind that it is integer for $N$ even and half-integer for $N$ odd.}
\eqn\IntroVM{{\CL}=-{1\over 4g^2} \Tr F^2+i\Tr \lambda\slash D \lambda+{k\over 4\pi} \Tr\left( AdA-{2i\over 3}A^3\right)-{kg^2\over 2\pi}\Tr \lambda\lambda~. } 
 
This model has been studied in detail in~\refs{\WittenDS,\GomisIXY} and  we  review here those results   as we build on them in this paper. This model has no adjustable continuous $\CN=1$ preserving  parameters and hence there is no wall for the Witten index to jump. 
 For $k\gg1$, the gauge fields and $\lambda$ are both classically much heavier than the scale of interactions $kg^2\gg g^2$ and hence we can simply integrate them out. What remains after we integrate them out is a pure Chern-Simons theory. In particular, supersymmetry is unbroken.  Upon integrating out the heavy particles and by virtue of~\levelmassshi\ the theory flows\foot{We recall that
$T(\hbox{adjoint})=N$ and $T(\hbox{fundamental})=1/2$ for $SU(N)$.}
to the $SU(N)_{k-N/2}$ Chern-Simons TQFT at long distances.
 This   analysis is reliable for $k\gg 1$. It turns out that $SU(N)_{k-N/2}$  is the correct description at long distances for $k\geq N/2$. In other words, for $k\geq N/2$ supersymmetry is unbroken and the low-energy theory is the  $SU(N)_{k-N/2}$ TQFT. The Witten index~\refs{\WittenDS} for $k\geq N/2$ agrees (possibly up to a sign) with the partition function of the   $SU(N)_{k-N/2}$ TQFT on the torus.

For $0\leq k<{N\over 2}$ the index vanishes, supersymmetry is spontaneously broken \WittenDS\ and the infrared of the theory consists of~\GomisIXY\ a Majorana Goldstino particle $G_\alpha$ accompanied by a Chern-Simons TQFT\foot{We  use the standard notation $$U(M)_{P,Q}={SU(M)_P\times U(1)_{MQ}\over \Z_M}~.$$ Consistency  requires that 
$Q=P \mod M$. }
 (the two sectors do not interact)
\eqn\infraredintro{G_\alpha+U\left({N\over 2}-k\right)_{{N\over 2}+k,N} ~.}
The Goldstino operator $G$ is the low energy limit of $\Tr(F\lambda)$, which is defined in the microscopic theory~\IntroVM. This is simply because $\Tr(F\lambda)$ is the conserved supercurrent.
It is much harder to understand the origin of the $U\left({N\over 2}-k\right)_{{N\over 2}+k,N}$ Chern-Simons theory  in terms of the microscopic degrees of freedom.\foot{In~\GomisIXY\ a duality was put forward that sheds some light on the origin of this TQFT.} Nevertheless, this TQFT has many desirable properties and it passes several nontrivial checks (such as having the correct one-form symmetry and anomaly, time-reversal anomaly for $k=0$ etc.).

\bigskip

\centerline{$\underline{\hbox{$\CN=1$ \ $SU(N)_k+$Adjoint Multiplet}}$}

\bigskip

The gauge multiplet consists of an $SU(N)$ gauge field  $A$, and a Majorana fermion, $\lambda$. We briefly reviewed its dynamics above. 
The additional adjoint multiplet includes a Majorana fermion and a real scalar $(X,\psi_X)$, both in the adjoint representation.

 This theory has an interesting family of $\CN=1$ supersymmetry-preserving deformations, namely, a general (real) superpotential  
\eqn\gensupint{W=\sum_l \alpha_l\, \hbox{Tr}(X^l)~.}
For one particular choice $\alpha_2=-{kg^2\over 2\pi}$  (and $\alpha_{l>2}=0$) the theory has enhanced $\CN=2$ supersymmetry, corresponding to a pure $\CN=2$ vector multiplet. We will study     the dynamics of the softly mass deformed $\CN=2$ theory, namely, we only activate $\alpha_2$ and denote $\alpha_2\equiv m$. Therefore, we only include the superpotential mass term
\eqn\supint{W=m\Tr(X^2)~.}
  We study the dynamics of the theory as a function of $k$, $N$, and $m$ (in this notation, for $m=-
{kg^2\over 2\pi}$ the theory enjoys $\CN=2$ supersymmetry).

We find surprisingly rich dynamics both for large $k$ and for small $k$. 
For sufficiently large $|m|$ we can always integrate out the adjoint multiplet and the theory flows to   a pure $\CN=1$ vector multiplet with gauge group  $SU(N)$ and level $k\pm N/2$ depending on whether $m$ is large and positive or large and negative, respectively (cf.~\levelmassshi). Then, according to what we found  above, the discussion splits depending  whether $k\geq N$,  $0<k<N$ or $k=0$.  

Let us now summarize the different phenomena we encounter in these various cases.  A 
key element in our analysis is that  the Witten index  of $\CN=1$ theories can jump as a function of $m$ unlike in theories with more supersymmetry, where holomorphy forbids the index from jumping, cf.~\IntriligatorLCA.
 We find a new mechanism  that allows the index to jump. This leads to new supersymmetric vacua (and in particular, one that supports an Abelian TQFT in the infrared) and also it leads to a mechanism for metastable supersymmetry breaking.
  Let us now summarize our findings in a little bit more detail:

\medskip

\item{1.} $k\geq N\, $: Supersymmetry is unbroken for large $|m|$ and the theory  flows to the TQFTs $SU(N)_k$ and $SU(N)_{k-N}$ for large positive $m$ and large negative $m$, respectively. The Witten index therefore jumps as   $m$ is varied.\foot{The Witten index coincides (up to a sign) with the torus partition functions of the TQFTs.} 
The $\CN=2$ supersymmetric point is at $m=-{g^2k\over 2\pi}$ and it  flows to a supersymmetric vacuum with the $SU(N)_{k-N}$ TQFT, a result that follows\foot{The two Majorana fermions in the $\CN=2$ massive vector multiplet have spin 1/2  and since  in 2+1d the sign of the spin of a fermion is determined by the sign of its mass, the fermions in the $\CN=2$ vector multiplet shift the Chern-Simons level additively, cf.~\levelmassshi.} for large $k$ from~\levelmassshi\ and   holds all the way down  to $k=N$. (This is   consistent with the index~\OhtaIV\ and the
 exact computation  of the $S^3$ and $S^2\times S^1$ partition function of  $\CN=2$ theories.)  The transition between the two asymptotic large mass phases happens in two steps. The wall where the index jumps is at $m=0$. At $m=0$ (where there is a classical non-compact space of supersymmetric vacua) a radiatively induced asymptotically flat  (non-supersymmetric)  direction with non-zero energy density  opens up.\foot{The point $m=0$ does not have new symmetries. But it is still a special point -- it is where the superpotential is asymptotically linear rather than quadratic.} For infinitesimal positive $m$, the effective superpotential deformed by the mass term \supint\ supports $2^{N-1}-1$ new supersymmetric vacua which come in from infinity in field space and are described
  by  new infrared Chern-Simons TQFTs.
  The jump in the Witten index
between the two asymptotic phases is now fully accounted for by these new vacua. For infinitesimal negative $m$ there is only one vacuum near the origin, supporting the $SU(N)_{k-N}$ TQFT.
As    $m$ is increased these new vacua approach the original supersymmetric  vacuum with the $SU(N)_{k-N}$ TQFT. Then these vacua merge in a series of second order phase transitions. When these transitions are completed, we get the $SU(N)_k$ TQFT in a supersymmetric vacuum describing the asymptotic positive mass phase.   See   \Largek.
  
   \putfig\Largek

\item{2.} $0<k<N\,$: 
At large positive $m$ supersymmetry is unbroken and the theory has a vacuum supporting the $SU(N)_k$ TQFT. On the other hand, at large negative $m$ supersymmetry is spontaneously broken and the supersymmetry-breaking vacuum has a Majorana Goldstino $G_\alpha$ accompanied by the $U(N-k)_{k,N}$ TQFT  (cf. \infraredintro). The $\CN=2$ theory at $m=-{g^2k\over 2\pi}$ is likewise in this phase, i.e. it breaks $\CN=2$ supersymmetry and the Majorana Goldstino fermion is joined by another Majorana fermion so that the low energy theory is a  Dirac Goldstino particle    with a  decoupled $U(N-k)_{k,N}$ TQFT. (This is consistent with  the vanishing of the Witten index~\OhtaIV\  and of the $S^3$ and $S^1\times S^2$ partition function of   the $\CN=2$ theory.\foot{Both the $S^3$ and $S^1\times S^2$ partition function of the $\CN=2$ $SU(N)_k$   theory vanish  for $k=1,2,\ldots, N-1$ (but not for $k\geq N$). This is consistent with supersymmetry breaking since an  $\CN=2$ Goldstino on the $S^3$ and $S^1\times S^2$ background has a fermionic zero mode implying that  the partition function vanishes. In spite of that we will  
be able to provide strong evidence for our proposed infrared description of the $\CN=2$ $SU(N)_k$ theory for $k<N$.})
The additional Majorana Goldstino becomes massive as the theory is deformed away from the $\CN=2$ point. The transition between the asymptotic phases at large negative and large positive $m$ again happens in two stages. Around $m=0$ an asymptotically flat direction opens up and new supersymmetry-preserving vacua with various Chern-Simons TQFTs come in  from infinity for $m>0$. Near the origin there is a metastable supersymmetry breaking state. As we increase $m$ we encounter phase transitions and we eventually get to the supersymmetric vacuum with $SU(N)_k$ TQFT describing the large positive mass asymptotic phase. See   \Smallk.

   \putfig\Smallk

\item{3.} $k=0\,$: There is a trivial supersymmetric vacuum (i.e no TQFT)  at both $m\to\pm\infty$ and a non-perturbative runaway direction opens up at 
$m=0$, where there is a  classical non-compact space of vacua. The runaway at 
$m=0$ is stabilized for both positive and negative $m$ into a trivial supersymmetric vacuum, and there are no further phase transitions as $m$ is varied. The $m=0$ point for $k=0$ is also the $\CN=2$ supersymmetric point and the runaway potential is due to the familiar Affleck-Harvey-Witten monopole-instanton mechanism~\AffleckAS. The index remains constant everywhere where it is well-defined. In this example therefore the Witten index does not jump. The phase diagram is summarized in \Zerok.
 \putfig\Zerok

 \medskip

The dynamics of these theories is thus governed by the wall at $m=0$, where the Witten index jumps. This happens in a pretty elaborate fashion, by the appearance of many new supersymmetric Chern-Simons vacua, which by virtue of being semiclassical match all anomalies. In particular, a vacuum with an Abelian TQFT appears.
In addition, this phenomenon comes hand in hand with the appearance of a metastable supersymmetry  breaking vacuum.\foot{Loosely speaking, one can think of a crude gravitational dual picture where the metastable supersymmetry breaking vacuum corresponds to $N-k$ anti-branes on top of each other. Then, there is a non-perturbative effect which allows this state to tunnel to a supersymmetric ground state.} Certain conformal field theories  govern the mergers of these  vacua that appear from infinity. We analyse these conformal field theories in the simplest cases.
We emphasize the appearance of an asymptotically flat direction with constant energy density on the wall (point) at $m=0$, the way by which the Witten index jumps, and the implications for metastable supersymmetry breaking.

\bigskip

\centerline{$\underline{\hbox{$\CN=1$ \ ${\rm Theories \ with } \ $Fundamental  Matter}}$}

\bigskip

The study of theories with fundamental matter will be carried out systematically elsewhere. Here we study just two examples with a single matter multiplet, demonstrating how the tools we have introduced allow to determine the phases of these theories and infer  $\CN=1$ dualities. We will see
that the dynamics near the walls is necessary to understand these $\CN=1$ dualities.

Our first example is an $\CN=1$ $SU(N)$ gauge theory coupled to a matter multiplet in the fundamental representation. The matter multiplet consists of a complex scalar and a Dirac fermion $(\Psi,\psi_\Psi)$ in the fundamental representation of $SU(N)$.  

The theory has an $\CN=1$ supersymmetry-preserving mass deformation by the superpotential
\eqn\superfund{
W=m |\Psi|^2\,.}
We  solve for the infrared dynamics of this model as a function of $k$ and $m$.

For large $|m|$ the matter multiplet can be integrated out and we are left with a pure $\CN=1$ vector multiplet with $SU(N)$ gauge group and level $k\pm1/2$,  depending on whether the mass is large and positive or large and negative.  At $m=0$ the theory has a classical moduli space of vacua that is lifted by a radiatively induced two-loop superpotential. For infinitesimal small positive $m$ the theory has a new supersymmetric vacuum state coming in from infinity in field space. This vacuum supports at low energies  a pure  $\CN=1$ $SU(N-1)_{k}$ vector multiplet. For infinitesimal negative $m$ there is only the vacuum near the origin, continuously connected to the vacuum that we see at large negative mass. Therefore, the phase diagram  depends on whether $k\geq {N+1\over 2}$, $k={N-1\over 2}$ or $k<{N-1\over 2}$.
 
\medskip

\item{1.}  $k\geq {N+1\over 2}$: There is a supersymmetric vacuum at large negative and large positive mass described by the TQFTs $SU(N)_{k-{N+1\over 2}}$ and  $SU(N)_{k-{N-1\over 2}}$ respectively. At $m=0$ the theory has a classical moduli space of vacua that is lifted by a radiatively induced two-loop superpotential.
The supersymmetric vacuum state coming in from infinity for infinitesimal small positive $m$ flows in the infrared to a supersymmetric vacuum with the $SU(N-1)_{k-{N-1\over 2}}$ TQFT. As the mass is increased this TQFT  merges with the $SU(N)_{k-{N+1\over 2}}$  TQFT at a second order transition, at the other side of which is the asymptotic positive mass $SU(N)_{k-{N-1\over 2}}$ TQFT.
  The new supersymmetric vacuum near $m=0$ accounts for the jump in the Witten index between the asymptotic large mass phases. 
 
 \item{2.}  $k= {N-1\over 2}$: At large positive mass there is a trivial (i.e no TQFT) supersymmetric vacuum. At large negative mass supersymmetry is spontaneously broken and the supersymmetry breaking vacuum has a Majorana Goldstino accompanied by the $U(1)_N$ TQFT (cf. \infraredintro). The vacuum state that appears from infinity  for infinitesimal small positive $m$ is a  trivial (i.e no TQFT) supersymmetric vacuum. As the mass is increased the vacuum that has appeared from infinity may be identified with the vacuum that we have found at very large positive mass. No transition is therefore necessary in this model. The new supersymmetric vacuum near $m=0$ accounts for the jump in the Witten index between the asymptotic large mass phases.

 \item{3.}  $k<{N-1\over 2}$: Supersymmetry is broken at large $|m|$. Thus, the asymptotic  phases have a Majorana Goldstino  with the TQFTs $U\left({N+1\over 2}-k  \right)_{{N-1\over 2}+k,N}$ and 
  $U\left({N-1\over 2}-k  \right)_{{N+1\over 2}+k,N}$ for large negative and large positive mass respectively.
 The vacuum state that appears from infinity  for infinitesimal small positive $m$  breaks 
 supersymmetry and contains a Goldstino and the $U\left({N-1\over 2}-k  \right)_{{N-1\over 2}+k,N-1}$ TQFT.

 \medskip
 
 One can similarly study the phases of the theory $U(N)_{k,k'}$ coupled to a single fundamental matter multiplet $\Phi$.
 This analysis leads us to propose the following duality, which is valid for $N\geq1$ and $k\geq 0$ \eqn\Dualityintro{ U\left(N\right)_{k+N/2+1/2,k+1/2}+\Phi \longleftrightarrow SU(k+1)_{-N-k/2}+\Psi~.}
 This duality\foot{The deformations map as $\Phi\Phi^\dagger \longleftrightarrow - \Psi\Psi^\dagger$. However, here we do not study the precise map 
 between the quartic deformations. This is left for the future. We note, however, that there exist~$\CN=2$ fixed points describing similar transitions (indeed, the SCFTs appear away from the walls, and hence the Witten index no longer jumps) and therefore in some situations there is possibly emergent $\CN=2$ supersymmetry in the infrared. We will say a few more words on the topic in the bulk of the paper. We thank D.~Gaiotto for discussions of this topic. } has interesting special cases,\foot{The case $N=0$ is trivial, as explained in item 2 above.} 
  for instance, $N=1,k=0$, which we will analyze in detail.\foot{The duality then reduces to an $\CN=1$ supersymmetric version of the duality between nonsupersymmetric $U(1)_{1/2}$  QED at level $1/2$ with a charge $1$ fermion and the XY model \refs{\ChenCD,\BarkeshliIDA,\SeibergGMD}.} It is crucial to follow the vacua 
 that appear from infinity at $m=0$ in order to see that this proposal has the right phases across the transition. In fact, the vacuum that comes from  infinity in one theory maps to a vacuum in an asymptotic phase of the dual theory. This is perfectly consistent with the duality, since the duality is valid in the infrared near the second order transition. See~\Dualityfig.
 \putfig\Dualityfig\
 
 The 't Hooft limit of this duality can be studied explicitly~\refs{\JainQI,\JainGZA,\Oferfuture} (and see references therein).
 In that approach, one discards the gauge kinetic term, i.e. one concentrates on the fixed point, which can be studied analytically in the 't Hooft limit. Our main new results concern with finding the supersymmetric vacua in the full theory (not necessarily in the 't Hooft limit or near the fixed point). 
 It is a nontrivial consistency check that in the regime where the techniques can be both applied, they lead to the same results about the 
 space of vacua of the theory.

Unlike in many of the non-supersymmetric dualities in the literature, where it is generally hard to establish that the transitions are second order, here it follows from the properties of the superpotential. Intuitively, since these are transitions involving (zero energy) supersymmetric vacua, a first-order transition is impossible. Hence the transitions have to be at least second order.

Similar methods can be employed to study many additional $\CN=1$ models. It is clearly of great interest to study models with more than one flavour multiplet, models with non-minimal charges and representations and so on. 
These studies are interesting in their own right, but they also naturally connect to questions about non-supersymmetric dynamics as well as to questions about theories with larger supersymmetry algebras.

\bigskip

The outline of the paper is as follows. In section~2 we review the pure $\CN=1$ vector multiplet dynamics. In section 3 we study the $\CN=1$ $SU(N)$ theory with a Chern-Simons term and an additional adjoint multiplet. We consider the large $k$ and small $k$ dynamics, explaining how the Witten index jumps and establishing the existence of a metastable supersymmetry-breaking state for small $k$. In section~4 we consider $SU(N)$ and $U(N)$ gauge theories with a Chern-Simons term  and fundamental matter representations, emphasizing duality and the important role played by the walls in establishing these dualities. Some details appear in three appendices.

\newsec{${\cal N}=1$  Vector Multiplet -- A Review~\GomisIXY}

  We consider Yang-Mills theory with $SU(N)$  gauge group and  a Chern-Simons term at level $k\geq 0$.\foot{Recall that $k<0$ is obtained by acting with time-reversal on the $k>0$ theory, which also reverses the sign of the fermion mass terms. We can therefore choose $k\geq 0$ without loss of generality.} The vector multiplet consists of the gauge field $A$ as well as a Majorana fermion $\lambda$ in the adjoint representation. The most general renormalizable Lagrangian is
\eqn\NoneVM{-{1\over 4g^2} \Tr F^2+i\Tr \lambda \slash D \lambda+{k\over 4\pi} \Tr\left( AdA-{2i\over 3}A^3\right)+m\Tr \lambda\lambda~. } 
The level $k$ is integral for $N$ even and half-integral for $N$ odd.

The Lagrangian \NoneVM\ has $\CN=1$ supersymmetry for  $m=-{kg^2 \over 2\pi}$ (cf. \IntroVM). We denote  this theory  as  the $\CN=1$ $SU(N)_k$   vector multiplet. Witten found that the Witten index of this theory is~\WittenDS\
\eqn\WI{I={\left(k+{N\over 2}-1\right)!\over (N-1)! \left(k-{N\over 2}\right)!}~.}
By writing the index as 
\eqn\WIi{I={1\over (N-1)!}  \left(k-{N\over 2}+1\right) \left(k-{N\over 2}+2\right)\cdots  \left(k+{N\over 2}-1\right)\,,}
one finds that the   index   vanishes for $0\leq k<N/2$ and it does not vanish for $k\geq N/2$. 
This holds for all the admissible values of $k$, namely integral values if $N$ is even and half-integral values if $N$ is odd.

Therefore supersymmetry is unbroken for $k\geq N/2$ and the number of states on $\bb T^2$ as counted by the Witten index is precisely consistent with the low energy theory in the supersymmetric vacuum being the  topological  $SU(N)_{k-N/2}$ Chern-Simons theory. For large $k$ the theory is weakly coupled and we can integrate out the fermion at one loop, obtaining at long distances the $SU(N)_{k-N/2}$ TQFT (cf. \levelmassshi).   It is worth  noting that for $k=N/2$ the  $SU(N)_{k-N/2}$ TQFT  trivializes, and indeed in that case the Witten index is 1, and there is a single trivial supersymmetric  ground state.\foot{The $\CN=1$ $SU(N)_k$   vector multiplet at $k=N/2$   has a   holographic dual~\MaldacenaPB (see also \GomisXW ).}

For $0\leq k<N/2$ the Witten index vanishes. The standard interpretation is that supersymmety is spontaneously broken and there is a massless Majorana Goldstino particle in the vacuum. However, this cannot be the whole story because a single massless Majorana particle cannot match various discrete anomalies in the system. In particular, the system has (for generic $k$) an anomaly in the one-form symmetry which precludes the supersymmetry breaking vacuum from being trivial. In addition, for $k=0$ there is a time-reversal anomaly which again would be sufficient to rule out a single Goldstino particle in the vacuum. 
In~\GomisIXY\ a scenario that matches all the anomalies and passes a number of additional nontrivial tests was proposed. The proposal is that the infrared consists of the Majorana Goldstino particle, $G_\alpha$, accompanied by the TQFT\foot{This TQFT is 
level/rank dual as spin TQFT to $U\left({N\over 2}+k\right)_{-{N\over 2}+k,-N}$.}
(cf. \infraredintro)
\eqn\VMNeqsone{U\left({N\over 2}-k\right)_{{N\over 2}+k,N}~.  }

In this paper we   build   on this recent understanding of the dynamics of the $\CN=1$ vector multiplet and study $\CN=1$ theories with matter multiplets.  

\newsec{${\cal N}=1$ Vector Multiplet with an Adjoint Matter Multiplet}

We consider the ${\cal N}=1$ supersymmetric theory with gauge group $SU(N)$ and an adjoint matter multiplet. Therefore, the theory consists of a gauge field $A$, two Majorana fermions $\lambda,\psi_X$   and a real scalar field $X$, 
all in the adjoint representation. 

The Lagrangian  includes the kinetic terms
\eqn\Kinetic{-{1\over 4g^2} \Tr F^2+i\Tr \lambda\slash D \lambda+i\Tr \psi_X \slash D \psi_X+\Tr (D X)^2~,}
the $\CN=1$ Chern-Simons term 
\eqn\CST{ {k\over 4\pi} \Tr\left( AdA-{2i\over 3}A^3\right)-{kg^2 \over 2\pi}\Tr \lambda\lambda~,} 
and the Yukawa coupling 
\eqn\Yukawa{\sqrt{2} ig \Tr[\lambda, X]\psi_X~.}

This theory enjoys $\CN=1$ supersymmetry and admits an $\CN=1$ preserving mass deformation, which can be written as a real superpotential term $W=m\Tr X^2$. This superpotential leads to the additional terms in the Lagrangian
\eqn\soft{\Tr (m^2X^2+m\psi_X\psi_X)~.} 
This $\CN=1$ theory  has a global $\Z_2$ symmetry which acts as $(X,\psi_X)\to (-X,-\psi_X)$. In addition, for $k=0$, $m=0$ the theory is  time-reversal invariant.
 For the value of the mass  
 \eqn\Nequalstwo{m=-{kg^2\over 2\pi}~ }
 the theory has  ${\cal N}=2$ supersymmetry and the Lagrangian is that of the 
  pure $\CN=2$ vector multiplet model with an $\CN=2$ Chern-Simons term. 
 In order to reflect the enhanced $\CN=2$ symmetry, we   sometimes use the notation $m_{soft}$, defined as \eqn\msoft{m=-{kg^2 \over 2\pi}+m_{soft}~.} We   use   $m_{soft}$ when it is particularly natural to think about the $\CN=1$ theory as a soft deformation of the $\CN=2$ theory.  The $\CN=2$ theory has an $SO(2)_R$ global symmetry rotating the two fermions but leaving the real adjoint boson invariant.   
Our goal is to determine   the infrared phases of this theory as a function of $k$ and $m$.\foot{Note that this adjoint theory appears naturally in the context of brane and domain wall dynamics~\AcharyaDZ. More generally, $\CN=1$ supersymmetric theories in 2+1 dimensions should arise naturally on BPS domain walls and branes of $\CN=1$ theories in 3+1 dimensions. For some such constructions 
see~\refs{\ArmoniVV\DieriglXTA\ArmoniJSA\TachikawaCHA-\DraperMPJ} and therein for additional references on these matters.} 

{\let\it=\bf  \subsec{Large Mass Asymptotic Phases}}

The infrared dynamics in the large mass region can be solved  for using semiclassical reasoning in conjunction with the results of the previous section. 

We can take the mass $m$ to be large and positive or large and negative. In this regime,   the scalar $X$  and the fermion $\psi_X$ are very heavy classically,  so that  $X$ is pinned at $X=0$  and quantum effects cannot change that. Integrating out the $(X,\psi_X)$ multiplet we obtain a theory of an $\CN=1$ pure vector multiplet. Using \levelmassshi\ we find that for positive $m$ this leads to  the $SU(N)_{k+N/2}$ $\CN=1$ vector multiplet while for  negative $m$ to the  $SU(N)_{k-N/2}$ ${\cal N}=1$   vector multiplet. 

Using the  results in the previous section,  the dynamics of these theories strongly depends on whether or not the effective $\CN=1$ level of the infrared pure vector multiplet is larger than $N/2$. 
Therefore, the discussion of the asymptotic phases splits into three cases depending on the value of $k$:

\medskip

\item{1.} $k\geq N$. Since in this range  $k\pm N/2\geq N/2$ the physics is that of  
the ``large $k$'' phase described in the previous section. Supersymmetry is unbroken for both large positive $m$ and large negative $m$.
In the first case the theory flows  to the $SU(N)_{k}$ Chern-Simons TQFT and in the second case to the $SU(N)_{k-N}$ TQFT. The Witten index in each phase is given (up to a sign) by the partition function of the corresponding TQFT on the torus. We therefore see that the non-vanishing Witten index jumps between the two asymptotic phases. 

\item{2.} $0<k<N$. Since in this range $|k-N/2|<N/2$,     supersymmetry is spontaneously broken for large negative $m$, but    remains unbroken for large positive $m$. For large negative $m$ the long distance theory consists of a Majorana Goldstino accompanied by the TQFT (cf. \VMNeqsone)
\eqn\TQFTleft{U\left({N}-k\right)_{{N}+k,N}~.}
 For large positive $m$ supersymmetry is unbroken and there is a supersymmetric vacuum supporting the $SU(N)_k$ TQFT. We observe once again that the Witten index jumps between the asymptotic phases, but  now it is zero for large negative $m$ and nonzero for large positive $m$. 

\item{3.} $k=0$. Since  $|k-N/2|=N/2$   supersymmetry is unbroken for large negative $m$ as well as for large positive $m$. Furthermore, the supersymmetric  vacuum in both phases is trivial and gapped. Therefore, the Witten index at large positive $m$ agrees with the Witten index at large negative $m$.
The dynamics in this case is the simplest and does not require new tools unlike the $k\neq 0$ cases.

\medskip

In summary, we have determined the asymptotic large positive and large negative mass phases for all values of $k$. There is a qualitative difference between $0<k<N$ and $k\geq N$. In the former case there is spontaneous supersymmetry breaking at large negative mass and in the latter case the large $|m|$ physics always has a supersymmetric ground state.  We have also noted  that for $k\neq 0$   the Witten index   jumps in absolute value between the two asymptotic phases. 

The next subsections will tackle the infrared dynamics  of the theory for small $m$ (i.e. not parametrically large), including the $\CN=2$ supersymmetric point~\Nequalstwo.  One central question will be to understand how the Witten index jumps between asymptotic phases. Our analysis in conjunction with general considerations~\WittenDF\ imply that the Witten index can only jump at $m=0$, as this is the only point where the classical asymptotics of the superpotential changes.  We therefore turn next to the  classical analysis of the point $m=0$ in preparation to analyzing the full quantum theory at $m=0$.

{\let\it=\bf  \subsec{Classical Moduli Space of Vacua at $m=0$}}

Since for   $m=0$ the adjoint scalar field $X$ is classically massless, the theory has a classical moduli space of $\CN=1$  supersymmetry preserving vacua parametrized by the eigenvalues of $X$  
\eqn\Xdiag{X=\left(\matrix{X_1&0&0&...&0\cr 0 & X_2 & 0 &...& 0\cr ... & ... & ... & ... &  ... \cr 0 & 0 & ... & X_{N-1} & 0 \cr 
0 & 0 & 0 &...& X_{N}}\right)~,}
where the $X_i$ are real and obey the     $SU(N)$ invariant constraint $\sum_{i=1}^N X_i=0$.\foot{This constraint does not really matter since the overall $U(1)$  part of $X$ and its fermionic partner $\psi_X$ are decoupled free fields.} The residual gauge symmetry leftover after diagonalizing $X$ implies that    the classical moduli space of vacua of the theory at $m=0$ is  
\eqn\modulivacua{
\R^{N-1}/S_N\,,} 
where $S_N$ is Weyl group of $SU(N)$, which acts on the eigenvalues $X_i$ by permuting them.  In a  generic vacuum the gauge symmetry is  Higgsed down to $U(1)^{N-1}$. When some eigenvalues coincide, corresponding to  singular loci  in the moduli space~\modulivacua, the unbroken gauge symmetry has at least one   non-Abelian factor. We now turn to a discussion of these various vacua. 

  \subsubsec{Classical Abelian Vacua}

We want to determine   the classical low energy theory around a generic vacuum, where the gauge symmetry is Higgsed down to $U(1)^{N-1}$.  The off-diagonal $W$-bosons and their fermionic massive partners in the $\CN=1$ massive vector multiplet acquire a mass from the Higgs mechanism, as well as from the Chern-Simons term. The remaining $U(1)^{N-1}$ gauge bosons  and fermions in the $\CN=1$ vector multiplet  only pick up a mass from the Chern-Simons term.\foot{The off-diagonal components of $X$ are eaten by the longitudinal components of the massive gauge fields.} The contribution to the mass of the off-diagonal gauge bosons $W_{ij}$ from the Higgs mechanism depends on $g^2 X_{ij}^2$, where   
\eqn\defdiff{X_{ij}\equiv X_i-X_j~.}
The very low energy theory therefore consists of the $N-1$ massless $\CN=1$ moduli multiplets $(X_i,\psi_{X_{i}})$, along with a topological theory that is associated with the unbroken $U(1)^{N-1}$ gauge symmetry.\foot{If $|X_{ij}|\gg gk$ we can write an effective field theory which includes the propagating $U(1)^{N-1}$ massive photons, since their mass scales like $g^2k$ which is much smaller than the mass of the off-diagonal $W$-bosons, which scales like $g|X_{ij}|$ in this limit. However, this would be unnecessary for our present purposes.}

The massless $\CN=1$ moduli multiplets $(X_i,\psi_{X_{i}})$ are trivial free fields. Let us now turn to determining the precise low energy Chern-Simons theory.
The infrared Chern-Simons theory  for the unbroken $U(1)^{N-1}$ gauge fields is induced from the original Chern-Simons term~\CST.  
Hence, there is a nontrivial matrix of Chern-Simons terms 
for the unbroken $U(1)^{N-1}$ gauge theory. This matrix can be found by simply plugging the matrix of unbroken gauge fields 
\eqn\Cartan{A=\left(\matrix{A_1&0&0&...&0\cr 0 & A_2 & 0 &...& 0\cr ... & ... & ... & ... &  ... \cr 0 & 0 & ... & A_{N-1} & 0 \cr 
0 & 0 & 0 &...& -(A_1+\dots+A_{N-1})    }\right) } into the Chern-Simons action~\CST. We can then immediately read off the $(N-1)\times (N-1)$ matrix $\bf k$
\eqn\kmat{{\bf k}=k\left(\matrix{2 & 1 & ... & 1 \cr 1 & 2 &...& 1 \cr . & . & . & . \cr 1 & 1 & ... & 2}\right)}
 of
Chern-Simons terms for the unbroken $U(1)^{N-1}$ gauge symmetry.
One can perform a change of basis with an $SL(N,\Z)$ transformation 
\eqn\freedom{{\bf k} \to A{\bf k}A^T~,\quad A\in SL(N,\Z)~}
in order to bring this theory to a  more familiar form:\foot{It is easy to prove   using~\freedom\ that one cannot bring ${\bf k}$ to  a diagonal form. Indeed, in a diagonal form all the entries on the diagonal have to be even (to avoid a dependence on the spin structure) and nonzero, and this is impossible for $N>2$ because $2^{N-1}>N$ for all $N>2$.}
\eqn\freedomi{{\bf k}=k\left(\matrix{2 & -1 & 0 &... & 0 \cr -1 & 2 & -1& 0 ...& 0 \cr . & . & . &. & . \cr 0 & 0 & ...&-1 & 2}\right)~.}
Upon this change of basis the  matrix ${\bf k}$ becomes the Cartan matrix of $SU(N)$ times $k$.
It is easy to see that $\det {\bf k} = k^{N-1}N$, which is the number of ground states of this $U(1)^{N-1}$ Chern-Simons theory on  the torus. 

This Abelian TQFT has a dual description for $k=1$:  it is dual to $U(1)_{-N}$ Chern-Simons theory~\FrohlichWE. This will be very useful for us below. We can think about this duality  as a level/rank duality 
\eqn\LR{\sum_{i,j=1...N-1}{k_{ij}\over 4\pi} A_i\wedge dA_j \longleftrightarrow -{N\over 4\pi} \tilde A\wedge d\tilde A~, }
with $k_{ij}$ being the entries of the matrix  ${\bf k}$ in~\freedomi\ for $k=1$.
 
In summary, classically, everywhere on the moduli space \modulivacua\ except at singular loci   where some $X_{ij}=0$, the theory flows to $N-1$ free, massless $\CN=1$  real multiplets  accompanied by the $U(1)^{N-1}$ Chern-Simons theory  with the  ${\bf k}$-matrix~\freedomi.

We   investigate later how quantum effects modify this classical analysis. We will find that quantum corrections generate a (super)-potential on the classical space of vacua  \modulivacua. This is a new phenomenon that occurs in $\CN=1$ theories which is not present in theories with more supersymmetry, where the superpotential does not receive any perturbative corrections.  We shall see, however, that the infrared Chern-Simons theory with the  ${\bf k}$ matrix   \freedomi\ will be an exact ground state for some range of parameters.

\subsubsec{Classical Non-Abelian Vacua}


We now turn to the classical analysis of  the non-generic vacua where some $X_{ij}=0$.
Such vacua will play a crucial role in unraveling the phase diagram of the theory.

We divide the matrix  $X$  into $L$ blocks such that the eigenvalue in each block is $X_I$ and the size of the corresponding block is $S_I\times S_I$, such that 
$$\sum_{I=1}^L S_I=N~,\qquad \sum_{I=1}^L S_IX_I=0~.$$
These are not to be confused with partitions of $N$. Indeed, because of the Weyl group the eigenvalues can be without loss of generality ordered. Thus, the order in which the summands $S_I$ appear is important. Those are called ``compositions'' of $N$. There are $2^{N-1}$ compositions of $N$. 

Assuming that all the $X_I$ $(I=1,...,L)$ are distinct, all the gauge fields (and   fermionic superpartners) away from these blocks acquire mass from the Higgs mechanism. The unbroken gauge symmetry in such a vacuum is thus
$$S[U(S_1)\times U(S_2)\cdots U(S_L)]~.$$

Let us now mention a useful way to think about the effective low energy $\CN=1$  gauge theory  in the infrared. We first extend the $SU(N)$ gauge  symmetry to a $U(N)$ gauge symmetry. That does not change the dynamics because the matter fields are in the adjoint representation.    In that case the low energy $\CN=1$ Chern-Simons couplings are simply those of the product theory
\eqn\produ{U(S_1)_{k,k}\times U(S_2)_{k,k}\cdots U(S_L)_{k,k}~.}
In order to go back to the $SU(N)$ theory,  we add another $U(1)$ gauge field $B$ which sets the overall trace to zero via the coupling
\eqn\LMCS{{1\over 2\pi }B\wedge  \sum_{I=1}^L S_I \Tr A_I\,.}

At   energies below the massive W-boson multiplets, we are left  with an  $\CN=1$ vector multiplet with gauge group and Chern-Simons levels~\produ\ with the additional constraint~\LMCS, accompanied by massless $\CN=1$ matter multiplets  
$(X_I,\psi_I)$ in the adjoint representation of the unbroken non-Abelian gauge group.

 In the vacua with Abelian gauge symmetry discussed above (i.e. $S_I=1$), there were no light charged particles and therefore the theory at long distances (if such vacua indeed exist in the full quantum theory) could be determined right away, as it is free. For non-Abelian vacua, the low energy theory just described  is still interacting, and the ultimate fate of these vacua necessitates further discussion. This will require us to understand the perturbative  (and ultimately also non-perturbative) corrections to the classical analysis described here.


{\let\it=\bf  \subsec{Semiclassical Moduli Space of Vacua}}

After the  discussion of the moduli space of vacua at $m=0$  and of the low energy theories that appear in each vacuum,   we now  turn to the important question of how quantum effects modify the classical analysis. 

In 2+1 dimensional  theories with $\CN=1$ supersymmetry there is no obstruction to the existence of perturbative corrections to the superpotential, in contrast  with more supersymmetric theories. Therefore, there may be a nontrivial superpotential which depends on the $N-1$ coordinates parametrizing  the moduli space \modulivacua
$$W(X_i)~, $$ 
with  $\sum_{i=1}^N X_i=0$. This can drastically change the classical picture (as we shall see, classical vacua get lifted).

Consider the classical vacua where the eigenvalues $X_i$  are distinct and very well separated so that the off-diagonal  degrees of freedom are very heavy. This allows to analyze what happens {\it far away} on the $N-1$ dimensional classical moduli space \modulivacua. Clearly, near the singular loci in \modulivacua, when some of the eigenvalues are close by, this expansion breaks down, and we  postpone  the discussion of the quantum behaviour of these vacua with non-Abelian gauge symmetry until later.

In order to understand how the radiatively induced  superpotential ought to behave, let us first appeal to dimensional analysis arguments. We consider for simplicity expanding around large $X_{ij}$ and let us assume that they  scale uniformly $X_{ij}\sim X$.
It is useful to canonically normalize the fields such that we have the following types of interactions: cubic interactions proportional to $g$, quartic interactions proportional to $g^2$ and we choose a gauge such that the Chern-Simons cubic vertex vanishes (e.g. $A_0=0$). The terms   containing $k$  therefore appear only in quadratic fermionic  and gauge field terms. These quadratic terms are proportional to $kg^2$. The mass of the heavy particles that run in the loops scales like $M\sim gX$.  This estimate is correct far away on the moduli space, i.e. as long as $X\gg gk$. Since we are interested in the structure of the effective superpotential far away on the moduli space, this estimate suffices.  

Consider now a vacuum $L$ loop diagram contributing to the quantum effective potential (i.e. the Coleman-Weinberg potential~\ColemanJX). Such a diagram is weighted by a factor of $g^{2L-2}$. Then, if we do not insert the quadratic vertices   depending on $k$, the sum over diagrams must vanish because   a superpotential cannot be generated  in  $\CN=2$ theories (recall that the theory with $m=k=0$ has enhanced $\CN=2$ supersymmetry). If we expand the effective scalar potential  in this manner, only even powers of $k$ can appear by virtue of parity. We therefore have at $L$ loops a perturbative series of the form 
\eqn\formalcw{V^{(L)}(X)= g^{2L-2}\sum_{n>0} d_{n;L}{(kg^2)^{2n}\over (gX)^{L+2n-4}}~,}
with some coefficients $d_{n;L}$. This is derived by imagining insertions of $k$ on the various edges of the diagrams. As we have explained, this representation of the effective potential is useful far away on the moduli space, and, more precisely, for 
$X\gg gk$.  The full scalar potential is of course given by summing over all loops
\eqn\fullpotential{V=\sum_L V^{(L)}(X)~.}
It turns out that the one-loop contribution vanishes
$$V^{(1)}=0~.$$
This follows from the fact
that the spectrum of heavy particles is supersymmetric at tree level and the one-loop potential is sensitive only to the supertrace of the classical spectrum\foot{The cubic and linear ultraviolet divergences vanish as they are proportional to $ \hbox{STr}(1)$ and $\hbox{STr}{\cal |M|}$ respectively.}
\eqn\oneloop{
\hbox{STr}{\cal |M|}^3\equiv 
\hbox{Tr}{|m_B|}^3- \hbox{Tr}{|m_F|}^3=0\,.
}
 
 As is well-known, integrating massive fermions at one loop  can induce a shift in the Chern-Simons levels (cf. \levelmassshi). A quick inspection of the mass matrix for the off-diagonal Majorana  fermions in $\CN=1$ vector and matter multiplets around the generic vacuum with non-degenerate eigenvalues shows that for each off-diagonal massive charged fermion there is a massive charged fermion with a mass of opposite sign.\foot{Consider  for simplicity the gauge group $SU(2)$, broken to $U(1)$ by the expectation value $X=\left(\matrix{x & 0 \cr 0 & -x}\right)$.
From the two adjoint Majorana fermions in the full theory   we obtain two complex fermions, $\lambda^{(+2)},\psi^{(+2)}$ of charge $2$ under the unbroken $U(1)$ gauge symmetry. Their mass matrix is proportional to  
\eqn\massM{{\cal M}\sim \left(\matrix{-kg^2 & igx\cr -igx & 0 }\right)~.}
Clearly, there is one positive and one negative eigenvalue.} Therefore, integrating
the  $\CN=1$ massive vector multiplets does not shift the levels of the ${\bf k}$ matrix \freedomi.\foot{In some places in the literature it is 
claimed that the Chern-Simons levels can also be shifted by integrating out heavy $W$-bosons. See for example~\DunneQY. This stems from a confusion between the Wilsonian and 1PI effective actions. } 
Furthermore, the Coleman-Hill theorem~\ColemanZI\  guarantees that the low energy Chern-Simons theory  is not modified by higher loop corrections.

In summary, at one loop the   $N-1$ $\CN=1$ matter multiplets remain massless and to all orders the infrared Chern-Simons theory is $U(1)^{N-1}$ with the ${\bf k}$ matrix \freedomi. Therefore, in order to unravel the leading quantum corrections to the classical moduli space we will need to go to two loops.

Equipped with this, the scalar potential~\fullpotential\ can be recast in terms of a superpotential
\eqn\fullsuppot{W(X)=kg^3\sum_{L>1}g^L\sum_{n>0}c_{n;L}{(kg^2)^{2n-2}\over g^{2n}X^{L+2n-5}  }\,, }
with some coefficients $c_{n;L}$ which can be, in principle, computed. From~\fullsuppot\ we can see the utility of the semiclassical 
large $X$ expansion: any given term in the $1/X$ expansion only receives contributions from {\it finitely} many loop orders in
perturbation theory. The leading term in the large $X$ expansion is at $L=2,n=1$, and it scales linearly $W(X)\sim X$. 
This term can only receive contributions from two-loop diagrams and determining whether this term is present or not, 
is absolutely crucial in order to understand the dynamics of these $\CN=1$ theories.

Fortunately, for rather different reasons, the full two-loop superpotential 
has been computed in~\refs{\ArmoniEE,\ArmoniSP}.\foot{In fact,~\refs{\ArmoniEE,\ArmoniSP}   computed the 
scalar potential and we are inferring the superpotential from that. But there is an ambiguity in doing this, aside from
the overall constant in $W(X)$ which cannot be determined and is insignificant. 
The sign of $W$ is very important as soon as we add back the mass perturbation~\soft\ $W=m\Tr X^2$ 
since the sign of the superpotential affects the interference between these two terms. 
The sign can be computed in principle by studying diagrams with external fermions. Here we simply assume that $W$ has a 
certain sign so that our overall picture for the dynamics is consistent.
} The result is\foot{To simplify notation, we have redefined the couplings in $W$ in order to absorb several unimportant factors.}
\eqn\twoloop{W=-\sum_{ij} g^3k\sqrt{g^2k^2+X_{ij}^2} ~, }
expressed in terms of the eigenvalue differences $X_{ij}\equiv X_i-X_j$   defined above. 

First, note that~\twoloop\ vanishes for $k=0$, as it should, since then the theory has  $\CN=2$ supersymmetry and the superpotential has no perturbative corrections.
 Second, note that if we rewrite~\twoloop\ in an expansion around large $X_{ij}$ then it agrees with the
general structure~\fullsuppot\ derived earlier. Finally, note that the superpotential appears to be regular even when 
the $X_{ij}$ go to zero, and where, classically, gauge symmetry would be enhanced. Usually, we expect the effective theory on the moduli space to indicate that it is breaking down on such loci due to 
the existence of new massless particles. But here the two-loop effective superpotential behaves perfectly regularly 
everywhere on the moduli space in spite of the fact that there are new massless particles when some of the eigenvalues coincide.

Understanding the   regime of validity of~\twoloop\ is of great importance in what follows. Very far on the moduli space, in the ``far zone" $X\gg gk$, 
as we have shown in~\fullsuppot, there may be contributions from three loops that scale like $X^0$ (i.e. it could be 
$\log X$) and contributions from four loops that scale like $1/X$. Therefore,  far out in the moduli space $X\gg gk$, the only reliable information captured by~\twoloop\ is the linear term 
\eqn\linear{X\gg gk:\qquad  W=-g^3k\sum_{ij}|X_{ij}|~.}
We will return to the ``near zone'' of the moduli space where $X$ is not large compared to $gk$ later. 

In summary, we see that the classical moduli space of supersymmetric vacua is lifted starting   at two loops~\twoloop\ and that there is a   radiatively induced asymptotically flat  (non-supersymmetric)  direction with non-zero energy density (cf. \linear). 
Next we analyze the  consequences of the two-loop superpotential for the phases of the theory around $m=0$.

{\let\it=\bf  \subsec{Semiclassical Abelian Vacuum near $m=0$}}

We  now proceed to study the fate of the semiclassical Abelian vacua  just described when the theory is deformed by a small mass term~\soft\
\eqn\supmass{\delta W=m \Tr X^2+\lambda \Tr X~,}
where we have also added a Lagrange multiplier $\lambda$ to enforce that $X$ is a traceless matrix.
Then, for infinitesimal $m$, by combining~\supmass\ and~\twoloop, we find that the superpotential on the
moduli space takes the form\foot{This formula is  valid to leading order in $m$ since the two-loop effective 
potential~\twoloop\ was computed in the massless theory.}  
\eqn\fullsupsmallm{W=-\sum_{ij} g^3k\sqrt{g^2k^2+X_{ij}^2}+m \sum_i X_i^2+\lambda \sum_i X_i~, }
with $i,j$ ranging over $1,...,N$.
In the perturbative ``far zone" regime  $X_{ij}\gg gk$  we can self-consistently approximate  the superpotential by 
\eqn\fullsupsmallmi{W=-g^3k\sum_{ij}|X_{ij}|+m\sum_i X_{i}^2+\lambda \sum_i X_i~.}

Supersymmetric vacua  (i.e zero energy states) of the deformed theory correspond to  solutions of the  equations ${\del W\over \del X_i}=0$.  The explicit equations are  (we define   $\sgn(0)=0$)  
\eqn\twoloopcp{-g^3k\sum_j \sgn(X_{ij})+mX_i+\half \lambda=0~,}
\eqn\twoloopcpi{\sum_i X_i=0~.}
Summing over all $i$ in the first equation we   find $\lambda=0$.
Therefore the equations can be simplified to
\eqn\twoloopcpii{-g^3k\sum_j \sgn(X_{ij})+mX_i=0~,}
\eqn\twoloopcpiii{\sum_i X_i=0~.}
The last equation implies that at least  one of the $X_i$ has to be positive. We can now use the residual $S_N$ group to order the eigenvalues  from the most positive one,
 $X_1$, to the most negative one, $X_N$. Then, the first equation with $i=1$ in~\twoloopcpii\ shows that  
  all the terms in the first term   are negative and   the second term is also negative for $\sgn(m)<0$, and hence there is no solution for $\sgn(m)<0$. Therefore, we conclude that there are no   supersymmetric vacua far on the moduli space for negative mass. We will discuss later the physics of the  vacuum at ``$X=0$,'' as it depends crucially on whether $k$ is large or small.

For small positive $m$   a supersymmetric solution  exists, and it is given  up to the action of the Weyl group by\foot{Note that this obeys~\twoloopcpiii.} 
 \eqn\solX{X_i={g^3k\over m}(N+1-2i)~.} 
The eigenvalue differences $X_{ij}$ are all parametrically large (compared to $gk$) 
for $m\ll g^2$,  and hence the existence of this supersymmetric vacuum is rigorously established in the full 
theory, that is, it is not an artefact of the two-loop approximation because as we proved around~\fullsuppot\ the 
higher-order corrections cannot compete far on the moduli space.  Therefore, we find that for $m$ small and positive there is an $\CN=1$ supersymmetric massive vacuum (all $N-1$ $\CN=1$ matter multiplets are massive) described by the $U(1)^{N-1}$ Chern-Simons theory with ${\bf k}$ matrix~\freedomi.

In summary, we have established that  for small positive $m$ there is a new supersymmetric gapped vacuum that comes in from infinity in field space that
supports the $U(1)^{N-1}$ Chern-Simons theory with ${\bf k}$ matrix~\freedomi. Instead, for small negative $m$,  there is no such Abelian vacuum. Intuitively, for small negative 
$m$, the potential  grows everywhere at large $X$, at $m=0$ it becomes asymptotically flat, and as we  make $m$ slightly positive, a new supersymmetric Abelian vacuum comes in from infinity.

We will use these facts to determine the different phases of the theory at finite $m$, connecting them to the 
asymptotic phases that we found in section 3.1. We will also establish by a  different semiclassical analysis   at large $k$ that the Abelian vacuum above is not the 
only one that appears from infinity in field space. We will find   additional supersymmetric vacua with degenerate eigenvalues that approach from infinity.

We now discuss in turn the complete phase diagram of the theory for $k\geq N$ and $0<k<N$, and $k=0$.

{\let\it=\bf  \subsec{Phases of the Theory with $k\geq N$}}

Above we studied the vacuum state of the theory with an Abelian gauge symmetry that can be reliably established 
for arbitrary $k$ by analyzing the theory far away in the space of vacua (i.e. large $X$). We now turn to the study of quantum vacua where some of the eigenvalues coincide. The low energy theory in such a vacuum has
non-Abelian gauge symmetry and the theory is interacting. 

Despite that the low energy theory is non-trivial around a non-Abelian vacuum, the infrared dynamics   can be determined for $k$ sufficiently large, that is for $k\geq N$. 

\subsubsec{Critical Points of the Superpotential}
At sufficiently large $k$ the physics is weakly  coupled, and 
unlike in section 3.4, we do not exclude vacua with degenerate eigenvalues. Indeed,  for  $k\geq N $ we can analyze them explicitly since the theory is ``semiclassical" and we can handle these interacting effective theories at these loci. 
The critical point equations are 
\eqn\crit{\half \lambda+mX_i=g^3k\sum_{j} {X_{ij}\over \sqrt{g^2k^2+X_{ij}^2}}~,}
\eqn\critii{\sum X_i=0~.}
Summing over $i$ in the first equation we find $\lambda=0$. Therefore the two equations can be simplified to
\eqn\critiv{mX_i=g^3k\sum_{j} {X_{ij}\over \sqrt{g^2k^2+X_{ij}^2}}~.}
For the sake of analyzing this equation it is useful to rescale $gk\tilde X= X$ such that the equation takes the form 
\eqn\critiii{{m\over g^2}\tilde X_i=\sum_{j} {\tilde X_{ij}\over \sqrt{1+\tilde X_{ij}^2}}~.}
There is clearly the solution $X_i=0$. 

Let us now look for solutions with at least one non-vanishing $X_i$. First let us make a general observation. There are solutions with some nonzero $X_i$ only for ${m\over g^2} \in (0,N)$. To prove that, assume that at least one $\tilde X_i$ does not vanish, so let $\tilde X_1$ be the largest positive eigenvalue. Then, $\tilde X_{1j}\geq 0$ for all $j$ and since 
$$\sum_{j} {\tilde X_{1j}\over \sqrt{1+\tilde X_{1j}^2}}\leq \sum_{j} \tilde X_{1j}=\sum_{j} \tilde X_1 - \sum_j \tilde X_{j}=N \tilde X_1$$
so that  $${m\over g^2}\tilde X_1\leq N\tilde X_1~,$$
and   therefore we conclude that nontrivial solutions exist only if ${m\over g^2}\in (0,N)$. The exact location where vacua with $\tilde X_i\neq 0$ disappear can get modified   by higher loop corrections. The estimate above is reliable for asymptotically large $k$, i.e. $k\gg 1$.

It is rather easy to write down all the solutions explicitly when the difference of eigenvalues in each distinct block are large.  These equations for small $m$ were
written in~\twoloopcpii, and the solutions are labeled by choosing blocks of sizes $S_I\times S_I$ and ordered eigenvalues $X_I$, $I=1,...,L$  
$$X_1>X_2>...>X_L~.$$
Using~\twoloopcpii\ we find that at very small positive $m$
\eqn\critgen{X_I={g^3k\over m}\left[\left(S_{I+1}+\cdots S_{L}\right)-\left(S_1+\cdots S_{I-1}\right)\right]~.}
We conclude that the vacua correspond to ordered partitions of $N$ (a.k.a compositions of $N$). There are 
$2^{N-1}$ vacua in total, including the one at the origin. The only supersymmetric vacuum with  small negative $m$ is the one at the 
origin.

%

In summary,  for $m<0$ we have one supersymmetric vacuum with TQFT $SU(N)_{k-N}$ and for (roughly) $m>g^2N$ we have one supersymmetric vacuum with TQFT $SU(N)_k$. The index jumps at $m=0$ and at small $m>0$ we have $2^{N-1}$ vacua with various TQFTs. As we increase $m$ the vacua gradually merge via second order transitions but the Witten index does not jump anymore. The transitions must be second order because these vacua correspond to zeroes of $W'$ with nontrivial Witten index. They cannot disappear without merging with other zeroes as we increase $m$.

How exactly 
these $2^{N-1}$ vacua merge into one vacuum is an interesting question. We find a complicated pattern where these $2^{N-1}$ vacua merge via a sequence of conformal field theories that appear away from the origin on the moduli space as we crank up $m$ from zero to $g^2N$. By the time we crank the mass up to $g^2N$ they will have all merged into a single vacuum and for $m>g^2N$ we have only the 
$SU(N)_k$ TQFT.\foot{We recall that the exact location where vacua away from origin disappear can be modified by higher loop corrections.} We will discuss in detail only one representative simple example of this phenomenon later.

Now we would like to analyze the vacua at small $m$ and show that in fact these $2^{N-1}$ vacua precisely account for the required
jump in the Witten index.
To simplify the computation we now take the original gauge group to be $U(N)$ rather than $SU(N)$. This would simplify the combinatorics while not making any difference for the physics since the $U(1)$ factor is anyway decoupled.
Let us consider the superpotential~\twoloop\ but now we omit the Lagrange multiplier term
\eqn\twoloop{W=-\sum_{ij} g^3k\sqrt{g^2k^2+X_{ij}^2} + m \sum X^2_i~.}
Let us expand around the critical point~\critgen. We take $X_i=X_i^0+\delta X_i$ and $X_i^0$ is given by~\critgen. 
Expanding the superpotential to second order in $\delta X_i$ (and dropping the constant piece) we find 
\eqn\twoloopexpanded{\eqalign{&W=-\half\sum_{ij} g^3k\sqrt{ g^2k^2+(X^0_{ij})^2} \left({2X^0_{ij}\delta X_{ij}+\delta X_{ij}^2\over g^2k^2+(X^0_{ij})^2}
-\left({X^0_{ij}\delta X_{ij}\over g^2k^2+(X^0_{ij})^2}\right)^2 \right) \cr&+2 m \sum X^0_i\delta X_i+m\sum \delta X_i^2~.}}
Omitting the linear piece in $\delta X_i$ which vanishes by virtue of the equations~\critiv\ we find after some simplifications
\eqn\twoloopexpandedi{W=-\half\sum_{ij} {g^5k^3\delta X_{ij}^2\over\left( g^2k^2+(X^0_{ij})^2\right)^{3/2}}
  +m\sum \delta X_i^2~.}
Now we restrict to infinitesimal $m$, where the solutions are given by~\critgen. Considering the first term in~\twoloopexpandedi\
we see that there are two cases -- if we consider $i,j$ to lie in the same block then $X^0_{ij}=0$ and the coefficient of $\delta X_{ij}^2$ in the first term scales like $g^2$, which is much larger than $m$. If we consider $i,j$ to lie in different blocks, the the coefficient of 
$\delta X_{ij}^2$ scales like $m^3/ g^4 $, which is much smaller than $m$. The conclusion is that for $i,j$ in the same block we should 
take into account the first term in~\twoloopexpandedi\ while for $i,j$ in different blocks we can neglect it. The mass matrix for the fermions 
$\psi_i$ (which are the partners of $X_i$) thus takes the form 
\eqn\massmat{{\cal M}=m\unit_{S_I\times S_I}+g^2\left(\matrix{-S_I+1 & 1 & 1 & ... & 1\cr 1 & -S_I+1 & 1 & ... & 1\cr 1 & 1 & -S_I+1 & ... & 1\cr
. & . & . & ... & . \cr 1 & 1 & 1 & ... & -S_I+1}\right) }
in each $S_I\times S_I$ block and is otherwise vanishing. Even though $m\ll g^2$, we have not neglected the first term on the right hand side of~\massmat\  since it lifts the zero mode $(1,1,..,1)$. The eigenvalues of this matrix are \eqn\eigens{(-g^2S_I+m,-g^2S_I+m,...,-g^2S_I+m,m)~.}

Therefore, we can now complete the exercise that we have embarked on in~\produ\ and compute the TQFT. For sufficiently small $m$, the eigenvalues are all negative other than the one that corresponds to the decoupled field $(1,1,..,1)$. Therefore, the charged fermions under the unbroken \eqn\producgauge{U(S_1)\times U(S_2)\cdots U(S_L)}
gauge symmetry all have a negative mass (we recall that the gauginos in the $\CN=1$ vector multiplet also have a negative mass). Since we are in the ``large $k$'' phase for each of the subgroups in~\producgauge, the long-distance theory can be read out by simply integrating out the fermions (matter fermions and gauginos) at one loop and supersymmetry is unbroken. Therefore  the TQFT is given by 
\eqn\prodgucaugei{U(S_1)_{k-S_1,k}\times U(S_2)_{k-S_2,k}\cdots U(S_L)_{k-S_L,k}~.}
The contribution of this vacuum to the Witten index is given by the number of states of this TQFT, which is simply 
$$\prod_I {k! \over  S_I! (k-S_I)!}~.$$
The contribution of each such vacuum (which corresponds to a composition of $N$) to the Witten index has to be weighted with the 
correct sign. The sign is simply given by counting how many fermions have a negative eigenvalue for any such given composition, and the 
answer is that, as we have seen above, in each block there are $S_I-1$ fermions with a negative eigenvalue. Therefore, the weight of 
this vacua in the Witten index is  
$$(-1)^{\sum_I(S_I-1)}=(-1)^{N-L}~,$$
where $L$ is the length of the partition. Therefore the Witten index at small positive $m$ is finally given by a sum over compositions $P$, $N=\sum_{I=1}^{L(P)} S_I$, where the length of each such composition is $L(P)$. We find that the Witten index is 
\eqn\Wittensmallm{I=\sum_{P} (-1)^{N-L}\prod_{I=1}^L {k! \over  S_I! (k-S_I)!}~.}
(Remember that this analysis is valid for $k\geq N$.)
Now we observe that there is an interesting combinatorial identity (proven in the Appendix C) for such compositions of~$N$
\eqn\identity{\sum_{P} (-1)^{N-L}\prod_{I=1}^L {k! \over  S_I! (k-S_I)!}={(N+k-1)! \over  N! (k-1)!}~.}
The right hand side is the  torus partition function of the TQFT $U(N)_{k,k}$, which is exactly the ground state of the theory at large positive $m$. 
This therefore nicely shows that the index jumps at $m=0$ by having $2^{N-1}-1$ vacua come from infinity, exactly reproducing the index of the large $m$ phase. Therefore, the total index no longer
changes as we continue to increase $m$. Instead, these $2^{N-1}$ vacua coalesce (not necessarily all at the same time, there could be
multiple second order phase transitions) until eventually they combine to the form the $U(N)_{k,k}$ ground state, which is visible semiclassically. 

We will now study the case of $SU(2)$ in more detail for concreteness and also because it is the simplest nontrivial 
case. 

\subsubsec{$SU(2)_k$}
The most general adjoint matrix $X$ can be brought to the form \eqn\SUtwo{X=gk\left(\matrix{x&0
\cr 0&-x}\right)} and we can plug this into~\critiii\ to obtain 
\eqn\finiteme{{m\over g^2}x={2x\over \sqrt{1+4x^2}}~.}

The solution $x=0$ always exists. For $m<0$ this is the only supersymmetric vacuum, namely the vacuum at the ``origin'' supporting the $SU(2)_{k-2}$ TQFT. For small positive $m$ we have two vacua. One is the one we just saw at $x=0$, supporting the $SU(2)_{k-2}$ TQFT. The other supersymmetric vacuum is the Abelian TQFT described in section~3.4, i.e. the $U(1)_{2k}$ pure Chern-Simons theory. 
Note that for small positive $m$ we have two more solutions, related by $x\to-x$. But $x\to-x$ is the generator of the Weyl group and hence these two solutions should be deemed equivalent.

As we increase $m$ these two vacua eventually meet at a second order phase transition (according to~\finiteme\ this happens at $m=2g^2$, an estimate that is reliable at asymptotically large $k$). Then, past this transition, there is again only one supersymmetric vacuum with an $SU(2)_k$ TQFT. 

Note that we can rigorously prove that the transition is second order. Indeed, for small positive $m$ we see two vacua ($SU(2)_{k-2}$ TQFT and $U(1)_{2k}$ TQFT). They correspond to two zeroes of $W'$. At large positive $m$ there is only one vacuum with a $SU(2)_{k}$ TQFT. The only way that this transition can occur is by the two vacua meeting. This is because the sign of $W''$ in the $SU(2)_{k-2}$ vacuum must change and this cannot happen without the zeroes meeting. 

Therefore, the conformal field theory at (approximately) $m=2g^2$ describes a phase transition (as we change the mass) between two isolated vacua carrying the $SU(2)_k$ and $U(1)_{2k}$ TQFTs and, on the other side of the transition, one isolated vacuum with $SU(2)_k$ TQFT. At large $k$ this conformal field theory can be studied systematically (and in that limit the Yang-Mills term and gaugino kinetic term can be dropped).

Strictly at $m=0$ there is an asymptotically flat direction and one supersymmetric ground state with $SU(2)_{k-2} $ TQFT near the origin.  At small positive $m$ the vacuum at the origin contributes to the Witten index $-(k-1)$ and the new Abelian vacuum contributes $2k$. Together they combine to $k+1$, which is precisely the Witten index of the vacuum at at asymptotically large positive mass. 

In the $SU(2)$ gauge theory there is therefore just one vacuum that appears from infinity for small positive mass $m$. And correspondingly, there is only one phase transition at positive $m$. The properties of this $\CN=1$ SCFT can be systematically computed in perturbation theory in $1/k$. Past this conformal field theory, the physics is smoothly connected to the large positive mass phase.

{\let\it=\bf  \subsec{Phases of the Theory with  $0<k<N$}}

  We now discuss the dynamics of the $\CN=1$ supersymmetric model $SU(N)_k$ with an adjoint multiplet and superpotential 
$$W=m \Tr(X^2)$$ 
in the regime of ``small'' Chern-Simons level $0<k<N$. In this regime non-perturbative effects dominate and the dynamics is quite rich.

First, we recall the basic facts about what happens at large $|m|$. At very large negative mass, integrating out the adjoint matter multiplet we get a pure $\CN=1$ vector multiplet with gauge group $SU(N)$  and Chern-Simons level $k-N/2$. Since $0<k<N$ then $|k-N/2|<N/2$ and hence this theory breaks supersymmetry spontaneously, leading to a massless Majorana Goldstino and a TQFT (according to~\VMNeqsone) \eqn\leftTQFT{U(N-k)_{k,N}\leftrightarrow U(k)_{-N+k,-N}~.}
We will see that it follows from our analysis that this continues to be true all the way to $m=0$, i.e. $m_{soft}={kg^2\over 2\pi}$. In particular, the $\CN=2$ supersymmetric point has the TQFT~\leftTQFT\ as well as a Dirac Goldstino particle (one Majorana fermion is lifted for nonzero $m_{soft}$ and thus we remain with one massless Majorana fermion away from $m_{soft}=0$).
At very large positive $m$ we get a pure $\CN=1$ vector multiplet with gauge group $SU(N)$  and Chern-Simons level $k+N/2$. Since for all non-negative $k$, $k+N/2\geq N/2$, the dynamics of the vector multiplet leads to a supersymmetric vacuum with the TQFT
\eqn\rightTQFT{SU(N)_{k}\simeq U(k)_{-N,-N}~.}
We therefore see that for  $0<k<N$ at very large negative mass we have a Majorana Goldstino and TQFT~\leftTQFT\ while for very large positive mass we have a supersymmetric vacuum with a (generally) different TQFT~\rightTQFT.
Clearly, the Witten index jumps and we have to understand how that comes about.

In~\critgen\ we have found that at small positive $m$ many new critical points of the superpotential appear. Those critical points are obtained by analyzing the two-loop+tree-level superpotential~\twoloop.  
These critical points correspond to compositions of the integer $N$, with unbroken gauge symmetry~\producgauge\ (after the center-of-mass $U(1)$ is removed). The effective field theory consists of the gauge group with bare Chern-Simons terms 
\eqn\prodgucaugei{S[U(S_1)_{k,k}\times U(S_2)_{k,k}\cdots U(S_L)_{k,k}]~,}
and we also have an adjoint matter multiplet for this gauge group. 
The mass terms for the adjoint multiplet around this critical point have negative eigenvalues~\eigens. 

For large enough $k$, i.e. as long as $k\geq S_I$ for all $I$, these supersymmetric critical points are not lifted and because the mass eigenvalues are negative we can simply integrate out these adjoint multiplets and arrive at~\prodgucaugei. 
It is guaranteed that $k\geq S_I$ for all $I$ for any composition $\{S_I\}$ as long as $k\geq N$. 

However, when $k<N$, in some of the vacua we will have at least one $S_I>k$ and therefore these vacua would be lifted. Namely, they are no longer critical points of the full quantum superpotential due to non-perturbative effects. The vacua that remain correspond to compositions of $N$ with all the $S_{I}\leq k$
\eqn\remainv{N=\sum_I S_I ~,\qquad S_I\leq k~.} 
In order to count the supersymmetric vacua that remain and their Witten index,  it is again 
useful to imagine that the gauge group is $U(N)$ instead of $SU(N)$, which of course makes no difference for the dynamics.
Therefore, those critical points that remain flow as before (since they are effectively in the ``large $k$'' phase and the mass term for the adjoint multiplet fermion and gaugino have a negative sign) to the TQFT 
\eqn\prodgucaugesmallk{U(S_1)_{k-S_1,k}\times U(S_2)_{k-S_2,k}\cdots U(S_L)_{k-S_L,k}~.}
The number of such compositions of $N$ is obtained from the coefficient of $x^N$ of the generating function, 
$${x(1-x^{k})\over 1-2x+x^{k+1}}~. $$

It is easy to see that summing over these restricted compositions the total Witten index still matches that of the single supersymmetric round state at large positive $m$. 
A simple way to see that is to consider the identity~\identity\ as an identity between two polynomials in $k$ of degree $N$. 
The terms on the left hand side that correspond to a composition $P$ for which at least one of the $S_I$ satisfies $S_{I}>k$ vanish. Therefore, the identity~\identity\ remains true for $k<N$ if we restrict the left hand side to compositions that satisfy~\remainv. 

We can now summarize that for $m\leq 0$ there are no supersymmetric ground states, but at $m=0$ an asymptotically flat direction with nonzero energy density opens and supersymmetric ground states appear. The supersymmetric ground states that appear at small positive $m$ correspond to compositions of $N$ restricted by~\remainv. As we increase $m$ these supersymmetric ground states coalesce and for sufficiently large $m$ there is only one supersymmetric ground state, which is visible semiclassically, with TQFT $SU(N)_k$. The Witten index at small positive $m$ therefore matches the Witten index at large positive $m$, as it should, since the asymptotic form of the potential does not change in this domain.

An interesting special case to consider is $SU(N)_1$. In that situation there is only one composition that remains of~\remainv, i.e there is only one supersymmetric ground state at small positive $m$, corresponding to the composition 
$$N=1+1+\cdots+1~.$$

The TQFT in that vacuum is Abelian and given by the ${\bf k}$-matrix~\freedomi\ with $k=1$. We have shown in~\LR\ that this theory has a  dual description in terms of the TQFT $U(1)_{-N}$. But now, using level/rank duality, we can also rewrite that model as
$SU(N)_1$. Therefore, the supersymmetric ground state at small positive $m$ supports the $SU(N)_1$ TQFT. Fortunately, this is precisely the TQFT in the supersymmetric ground state at large positive $m$ and hence in this particular case no phase transitions at 
positive $m$ are necessary at all. This is a nice consistency check since indeed there is only one gapped supersymmetric vacuum in this case and it has no other supersymmetric vacua to merge with. In other words, the deep infrared physics at small positive $m$ and large positive $m$ is essentially identical. 

In summary, for negative $m$ we have a supersymmetry breaking vacuum with a $U(1)_{-N}\simeq SU(N)_1$ TQFT accompanied by a Majorana Goldstino particle, and for positive $m$ we have the $SU(N)_1$ TQFT in a supersymmetric vacuum. This example will be important below in our discussion of metastable supersymmetry breaking. 
In Appendix A we discuss some further consistency checks of this particular dynamics of $\CN=1$ $SU(N)_1$ with an adjoint multiplet connecting our scenario to the ``duality appetizer.''

{\let\it=\bf  \subsec{Dynamical Metastable Supersymmetry Breaking at $0<k<N$}}

Here we consider the physics of the theory with $0<k<N$ at small $|m|$, i.e. near the wall (point) where the Witten
index jumps. We have seen that for small negative $m$ there is no supersymmetric ground state and we proposed a single supersymmetry-breaking ground state (which corresponds to the global minimum of the potential) carrying (as explained in \leftTQFT) the TQFT $U(N-k)_{k,N}$ and a Majorana Goldstino. Let us denote the energy density in this minimum by $f(m)$, defined for negative $m$. We can define the limit $$f(0)\equiv \lim_{m\to 0^-}f(m)~.$$

At $m=0$ we have an asymptotically flat direction with nonzero energy density~\fullsupsmallmi\ that is given by 
\eqn\eden{V_{asymp}=4g^6k^2\sum_i\left(\sum_j \sgn(X_{ij})\right)^2=4g^6k^2\sum_{i=1}^N\left(N+1-2i\right)^2={4\over 3}g^6k^2N(N^2-1)~.}
(We used the fact that the kinetic terms are  asymptotically approximately canonical  and we evaluated the energy for a generic direction far out on the moduli space. We also relaxed the traceless-ness condition on $X$ for simplicity.)
The scaling $g^6N^3k^2$ can be understood from general large $N$ considerations. Every insertion of $k$ corresponds to a factor of $k \lambda/N$, with $\lambda$ the usual 't Hooft coupling $\lambda=g^2N$. A generic planar two-loop  contribution to the vacuum energy scales like $\lambda N^2$ and hence with two insertions of $k$ we find 
\eqn\genest{V_{asymp}\sim k^2\lambda^3~,\qquad \lambda=g^2N~.}
This agrees with~\eden.

The question now is how does the energy density of the supersymmetry-breaking ground state at small negative $m$ compare to the energy density that opens up at infinity. There are a priori three options 

\eqn\options{\eqalign{ & f(0)>V_{asymp}\cr &f(0)< V_{asymp} \cr &f(0)= V_{asymp}\,.}}

We will now discuss these options in turn. First, let us consider the large $N$ scaling of $f(0)$. It is especially easy to estimate $f(0)$ if $k\sim N$, i.e. when $N/k$ is held fixed in the large $N$ limit. Then, the action is given by $N(\cdots)$ and the only dimensionful parameter in the action is $\lambda=g^2N$. Therefore, we should expect that 
$f(0)\sim \lambda^3N^2$. This is the same scaling as would be obtained from~\genest\ if $k$ scales like $N$. Hence, large $N$ considerations by themselves are not sufficient to decide among the options in~\options. 

First, let us argue that $f(0)>V_{asymp}$ is impossible. For small negative $m$, the superpotential has already a large region \eqn\farzoneIX{g\ll X\ll {g^3k\over m}~,}
where it is well approximated by $W=-g^3k\sum_{ij}|X_{ij}|$ and therefore it cannot be true that $f(0)>V_{asymp}$. 

Next we consider the possibility that $f(0)< V_{asymp}$. In this case, as we increase $m$ the supersymmetric vacua that 
come in from infinity must be separated by a potential barrier from the supersymmetry-breaking minimum. The distance between
these vacua must scale as $\sim 1/m$ and hence the supersymmetry-breaking vacuum is arbitrarily long lived. 

Finally, there is the most subtle case to consider $f(0)= V_{asymp}$. One way in which this can happen is that the supersymmetry-breaking vacuum in fact remains near the origin and the fact that the energy density coincides with what we computed
asymptotically is an accident. In this case there would have to be a potential barrier between the supersymmetry breaking state and the far region and hence the state will be metastable at small positive mass. This case is morally similar to the case of  $f(0)< V_{asymp}$. 

However, it could also be that as we tune $m$ to zero from the left, the supersymmetry-breaking vacuum just slides
to infinity and the equality $f(0)= V_{asymp}$ simply reflects the fact that the supersymmetry breaking vacuum now resides in the 
far zone. In particular, the supersymmetry breaking vacuum would be in the weakly coupled region~\farzoneIX. To rule this out we 
need to use a new element: that the TQFT in the supersymmetry breaking vacuum,  $U(N-k)_{k,N}$, cannot be obtained in the
semiclassical regime. Indeed, there is no weakly coupled description in the original degrees of freedom of a ground state with this TQFT. 

We therefore see that to establish the existence of a long-lived supersymmetry-breaking ground state in this theory, continuity arguments near the point where the index jumped were not quite sufficient, and we had to allude also to the topological degrees of freedom in the supersymmetry-breaking ground state. Because of that, our general argument may fail in some special cases where the TQFT in the supersymmetry-breaking vacuum does accidentally coincide with the theory far out on the Coulomb branch. For example, for $k=1$, 
$$k=1:\qquad U(N-k)_{k,N}\leftrightarrow U(1)_{-N}~,$$
and because of the duality~\LR\ this in fact coincides with the topological theory far out on the moduli space 
where no eigenvalues coincide. Another way to think about this special case is that if our gauge group was $U(N)$, then
for $k=1$ the vacuum far out on the moduli space with generic eigenvalues would be trivial but also the TQFT in the supersymmetry-breaking vacuum is trivial. Therefore, in the case of $k=1$ we essentially do not have the topology of the ground state that we used to protect the supersymmetry-breaking ground state from slipping to the weakly coupled region. 

The proof that we gave here for the existence of a metastable supersymmetry-breaking state depends on the assumption that
indeed our phase diagram is correct and for all negative $m$ there is a single supersymmetry-breaking state with the properties we discussed. This assumption is motivated by the fact that the phase diagram we proposed is the simplest that is consistent with  our detailed analysis of the ground states and with the 't Hooft anomalies. 

In $\CN=1$ theories, jumps in the Witten index are generic on co-dimension one hypersurfaces. 
If on one side of the hypersurface the ground state breaks supersymmetry spontaneously and on the other side new supersymmetric vacua come in from infinity, it is expected that supersymmetry breaking states would be metastable at least close enough to the hypersurface. The only way in which this conclusion can be avoided is if the supersymmetry breaking ground states slip to infinity as we get near the hypersurface. But as we saw, this can be in some cases ruled out by using the properties of the supersymmetry-breaking ground state.

{\let\it=\bf  \subsec{Phases of the Theory with $k=0$}}

The $\CN=1$ theory with an adjoint matter multiplet at $k=m=0$   has $\CN=2$ supersymmetry. This implies that the superpotential on the moduli space of vacua \modulivacua\ does not receive any perturbative corrections, and in particular,  the two-loop effective potential~\twoloop\
vanishes. As a consequence of this, the phase diagram of this theory is much simpler than for the theory with $k\neq 0$.

We recall that in section 3.1 we showed that the theory with large   positive and large   negative mass flows to  a trivial, gapped  supersymmetric vacuum (i.e. with no TQFT). Our goal here is to fill  in what happens between these asymptotic phases.

Let us start at the  $\CN=2$ supersymmetric point $m=m_{soft}=0$. While the superpotential is not 
renormalized in perturbation theory, it is well known~\AffleckAS\ that non-perturbative effects due to monopole-instantons lead to a   runaway  superpotential (i.e. there is no stable vacuum).

We now proceed to show that $m=0$ is the only singular point  in the phase diagram. We will demonstrate that 
immediately to the left and to the right of the $m=0$ point there is  a   trivial vacuum with unbroken supersymmetry. Therefore, the phases infinitesimally away from 
$m=0$ are identical to the asymptotic phases at large positive and negative $m$, and hence the phase diagram is particularly 
simple.

Let us first review some relevant parts of the derivation of the runaway behaviour for $SU(2)$ for simplicity. This theory has because of the enhanced $\CN=2$ supersymmetry at $m=0$   an $SO(2)_R$ $R$-symmetry, and
a classical  flat direction with unbroken $U(1)$   gauge symmetry parametrized by the eigenvalues of the scalar field 
$X= {\rm diag}(x,-x)$.
The Yukawa interaction~\Yukawa\ along this flat direction induces a coupling between $x$ and the charge two 
fermions $\psi^{(+2)}, \lambda^{(+2)}$ 
and their complex conjugates, as in~\massM\ (with $k=0$)
$$i gx \psi^{(+2)}\overline{ \lambda^{(+2)}}+c.c.~.$$
The $SO(2)_R$ symmetry acts naturally on the linear combinations $\Psi^{(+2)}=\psi^{(+2)}+i\lambda^{(+2)}$ and
 $\Theta^{(+2)}=\psi^{(+2)}-i\lambda^{(+2)}$. 
 
 Integrating out $\Psi^{(+2)},\Theta^{(+2)}$ does not generate 
a Chern-Simons term for the $U(1)$ gauge field, but  it does  generate a mixed Chern-Simons term    coupling  $SO(2)_R$ to the  unbroken $U(1)$ gauge symmetry. The mixed Chern-Simons term is (remembering that integrating out one fermion with charges (1,1) gives ${1\over 4\pi}\CB dA$) 
\eqn\CS{{2\over 2\pi} \CB dA~,}
where $\CB$ is the $SO(2)_R$ background gauge field and $A$ the unbroken $U(1)$ gauge field.  As a result, the minimal monopole operator picks up charge $-2$ under $SO(2)_R$. Denoting the corresponding chiral superfield by
$$Y=e^{x+i\tilde a}=e^u~,$$
we have that $\tilde a$, which  is the   scalar dual to the $U(1)$ gauge field,  is $2\pi$ periodic and transforms under $SO(2)_R$ rotations as $\tilde a \to \tilde a + 2\alpha$ with $\alpha$ a $2\pi$ periodic parameter. 

Therefore, the following   superpotential gets generated non-pertubatively\foot{For $SU(N)$ an $A_{N-1}$ Toda superpotential is generated, see Appendix B.}
\eqn\runsup{W={1\over  Y}=e^{-u}~.}
The kinetic terms are approximately linear in terms of $x$ and $\tilde a$ far out in the moduli space, where it is approximately true that $$K\sim (\log Y+\log\bar Y)^2~.$$
Therefore there is a runaway potential, scaling like $V\sim {1\over |Y|^2}\sim e^{-2x}$.

Let us now turn on a  small mass   deformation \soft\ for the  $\CN=1$ matter multiplet and determine where the theory flows to.  This can be done by writing the  deformation in the ultraviolet  using an $\CN=2$ spurion superfield $M$.  In terms of this, the mass deformation    \soft\  preserving  $\CN=1$ takes the form ($\Sigma$ is the $\CN=2$ chiral superfield constructed out of the $\CN=2$ vector multiplet, i.e. the field strength multiplet)
$$\delta\CL={1\over 2}\int d^4\theta\, M\Tr\left(\Sigma^2\right)~,$$
with $$M=m(\theta-\bar\theta)^2~.$$ 
This choice of $M$ preserves $\CN=1$ supersymmetry. This choice is of course non-unique;  We could have used the $R$-symmetry to relate any two such choices of $M$.

In the presence of $M$ the standard transformation from $\Sigma$ to the chiral superfield $u$  is modified
$${1\over 2} \int d^4\theta\,  (-1+M) \Sigma^2+  \Sigma(u+\bar u)~.$$
Integrating out  $\Sigma$  leads to the effective action in terms of   $u$ 
$$\half \int d^4\theta (1-M)^{-1}(u+\bar u)^2= \int d^4\theta\left(\ u\bar u+\half M(u+\bar u )^2+\cdots\,\right)~,$$
where on the right hand side we have only kept terms to   linear order in $M$. Expanding this action in components, and including the non-perturbative superpotential~\runsup,  we find (ignoring   terms with derivatives) 
$$e^{-u}F_u+c.c.-(e^{-u}\psi_u\psi_u+c.c.)+|F_u|^2-m(u+\bar u)(F_u+\bar F_u) - \half m(\psi_u+\bar\psi_u)^2~.$$ 
Solving for the auxiliary field we find the potential \eqn\potnok{|e^{-u}-(u+\bar u ) m|^2~.} 
This can be viewed as arising from the $\CN=1$ superpotential 
$$W_{\CN=1}=e^{-\Re u}\cos(\Im u)+m(\Re u)^2~.$$
For small positive $m$,  the minimum is  at $u\sim -\log 2m$ with $u$ real, while for small negative $m$ it is at $u\sim  -\log 2|m|+i\pi$. 
The scalar fields  are massive in these vacua, and since the vacua are  $\CN=1$ supersymmetric, so are the fermions. Therefore we have shown that the theory flows to a trivial phase for both positive and negative small $m$, leading to a very simple phase diagram, with a trivial massive vacuum everywhere except at $m=0$, where there is no stable vacuum. 

The analysis here straightforwardly generalizes to the case of $SU(N)$ gauge group, with identical  conclusions. Some  details are in 
Appendix B.

\bigskip
\newsec{${\cal N}=1$ $SU(N)_k$ and $U(N)_{k,k'}$ with Fundamental Matter}
\bigskip

We now discuss the phase diagram of $\CN=1$ gauge theories with matter multiplets in the fundamental representation. We study in turn $\CN=1$ gauge theories with $U(1),~SU(N)$ and $U(N)$ gauge groups. 
Based on this we put forward infrared dualities relating different $\CN=1$  theories.

{\let\it=\bf  \subsec{A $U(1)_k$ Warm Up}}

We take a U(1) ${\CN}=1$ gauge multiplet and couple it to matter multiplets with charges $q_i\in\Z$.  
The vector multiplet consists of $A_\mu,\lambda$ ($\lambda$ is neutral and Majorana) and the matter multiplets consist of 
$(\Phi_i,\Psi_i)$, carrying charge $q_i$ under the gauge symmetry, with $i=1,...,N_f$.
The Lagrangian with (classically) vanishing superpotential $W=0$ and with a Chern-Simons term with level $k$ is given by 
 \eqn\totalL{
  {\CL}=-{1\over 4g^2}F^2+i\lambda\slash\partial\lambda+{k\over 4\pi}AdA-{kg^2\over 2\pi}\lambda\lambda+\sum_i |D_\mu \Phi_i|^2+i\sum_i\bar \Psi_i\slash D\Psi_i +\sum_i \left(\sqrt{2}i\Phi_i\bar\Psi_i\lambda +c.c.\right)~.
}
Consistency requires  that 
$$k+\frac12 \sum_i q_i^2\in \Z~,$$
a condition which is  equivalent to  
$$k+\frac12 \sum_i q_i\in \Z~.$$

Let us now consider the symmetry group of the model. If we assume that the $q_i$ are arbitrary integers the symmetry group consists of $U(1)_i$ factors acting on the multiplets $(\Phi_i,\Psi_i)$ along with $U(1)_T$ which leads to a conserved magnetic charge, with conserved topological current ${1\over 2\pi}\star F$. One linear combination of the $U(1)_i$ and $U(1)_T$ is coupled to the dynamical gauge field $A_\mu$ and therefore the symmetry group is\foot{Let $(e^{is_1},...,e^{is_{N_f}},e^{it})$ be an element of $\prod_{j=1}^{N_f} U(1)_j\times U(1)_T$. The identification by 
the $U(1)$ in the denominator corresponds to 
$$s_j\to s_j+q_j\alpha~,\quad t\to t+\alpha k_{bare}$$
for all $\alpha$. Above we have defined 
$$k_{bare}=k+\frac12\sum_i  q_i^2~.$$}
 \eqn\symmg{ K = {\prod_{j=1}^{N_f} U(1)_j\times U(1)_T\over U(1)}~.}
If some of the charges coincide then the symmetry group~$K$ in \symmg\ is enhanced to a non-Abelian group.

There is a superpotential deformation that preserves all the symmetries
$$W=\sum_i m_i \Phi_i\Phi_i^\dagger~,$$
and again if some of the charges coincide then one can preserve the non-Abelian flavour symmetry by taking likewise some of the $m_i$ to coincide.

If we assume that the $m_i$ are large we can integrate out the superfields $\Phi_i$ and obtain a pure $U(1)$ vector multiplet   
with an integer Chern-Simons level 
\eqn\infr{k_{IR}=k+\frac12\sum_i q_i^2 {\rm sgn}(m_i)\,.}
The long distance theory is therefore a $U(1)_{k_{IR}}$ TQFT. The neutral gaugino is massive and at tree level its  mass is proportional to $g^2k_{IR}$. 

We can compute the Witten index as a function of these $m_i$, but we have to be careful about the sign of the index. We can take it to be positive (without loss of generality) when all the $m_i$ are positive and then if some $m_i$ crosses the origin we do not pick up a minus sign since the fermion $\Psi_i$ has two degrees of freedom. But if the sign of the gaugino $\lambda$ mass term changes its sign then we have to account for the change in fermion number. Hence, the Witten index of this theory is 
\eqn\indexxxx{I(m_i)={\rm sgn}(k_{IR})|k_{IR}|=k_{IR}~.}
Clearly, the Witten index jumps as we change the $m_i$.

In order to understand how this comes about it is useful to first study the theory with one charged multiplet $\Phi$ of charge $q$. Now the model has one mass parameter appearing in the superpotential $W=m\Phi\Phi^\dagger$. At very large positive $m$ we get the  $U(1)_{k+q^2/2}$ TQFT  and at very large negative $m$ we find the $U(1)_{k-q^2/2}$ TQFT. At $m=0$ there is a classical flat direction. There is once again a two-loop effective potential that would lift this flat direction~\toappear. 

At small positive $m$ we should have,  due to the 
two-loop potential,  another vacuum coming in from infinity with a condensed $\Phi$. In this vacuum there are no massless particles. 
It is crucial to understand that the gauge symmetry is Higgsed down to a $\Z_q$ subgroup. However, the effective theory is not just a $\Z_q$ gauge theory as we have a Chern-Simons term for the original gauge field, which becomes a Dijkgraaf-Witten-like term~\DijkgraafPZ\ for the $\Z_q$ gauge theory. The infrared effective theory can be written as
\eqn\qBF{{q\over 2\pi}A\wedge d B+{k_{bare}\over 4\pi}A\wedge dA}
with $A,B$ standard $U(1)$ gauge fields. The contribution of \qBF\
 to the Witten index is 
$$\Delta I = q^2~,$$
since this is the number of lines in this Dijkgraaf-Witten $\Z_q$ gauge theory. 

At small positive $m$ therefore we have two supersymmetric vacua. One near the origin with the  $U(1)_{k-q^2/2}$ TQFT and one very far out in the moduli space with the TQFT as in \qBF. The Witten index at small positive $m$ is thus 
$$I=k-{q^2\over 2}+q^2=k+{q^2\over 2}~,$$
which exactly coincides with the index at large positive $m$ \indexxxx\infr.

For generic $k$ the theory thus has a second order phase transition where the deformed $\Z_q$ gauge theory merges with a $U(1)_{k-q^2/2}$ TQFT and we get the $U(1)_{k+q^2/2}$ TQFT on the other side of the transition.
A special treatment is necessary for $k=q^2/2$, in which case for negative $m$ we have a vector multiplet with 
no Chern-Simons term. We dualize the vector multiplet  to a real multiplet 
$$(A,\lambda)\longrightarrow (\tilde\phi,\tilde\psi)\,,$$
with $\tilde \phi$  a pseudo-scalar and $\tilde\psi$ a Majorana fermion. The superpotential 
$W(\tilde\phi)$ vanishes because of the $U(1)_T$ symmetry. Therefore for negative $m$ we have an $S^1$ of vacua parameterized by the dual photon $\tilde \phi$. This vacuum contributes zero to the Witten index since 
$\tilde \psi$ is massless.\foot{Alternatively, since the Euler number of the circle is zero, the index vanishes.}

Let us now specialize to $q=1$, in which case the situation is simpler as the  $\Z_q$ gauge theories do not appear.
In the case of $q=1$, a trivial vacuum merges with the $U(1)_{k-1/2}$ TQFT to become the 
$U(1)_{k+1/2}$ TQFT. (The exceptional  case of $k={1\over 2}$ will be discussed in more detail soon.)

It is interesting to note that\foot{We thank D.~Gaiotto for discussions on this and some related topics.} the phase transition mentioned above 
$U(1)_{k-1/2}~\hbox{TQFT} +{\rm trivial}~\hbox{vacuum}$ to $U(1)_{k+1/2}$ TQFT for $k>1/2$ is similar to the one that appears in $\CN=2$ supersymmetry~\IntriligatorLCA.
In fact, for large enough $k$, the fixed point must have emergent $\CN=2$ supersymmetry in the infrared. This is not the case generically, but it is the case for $U(1)$ gauge theories~\refs{\AvdeevZA,\AvdeevJT}. For $k=1/2$ the transition involves a circle of supersymmetric vacua that have to disappear on the other side 
of the transition. By contrast, in $\CN=2$ supersymmetric theories the supersymmetric ground states are always complex manifolds. We will discuss this example more below.

{\let\it=\bf  \subsec{$SU(N)_k$ and $U(N)_{k,k'}$ with Fundamental Matter }}

We will now consider briefly a slightly different model: $\CN=1$ $SU(N)_k$ vector multiplet with a fundamental matter multiplet ($N_f=1$). The logic is quite similar. Imagine that we integrate out the matter multiplet with a large negative mass. Then we have a $SU(N)_{k-1/2}$ pure vector multiplet. For $k-1/2\geq N/2$ this flows to the $SU(N)_{k-1/2-N/2}$ TQFT and otherwise it breaks supersymmetry.
At large positive $m$ supersymmetry is unbroken if $k+1/2\geq N/2$ and we have a $SU(N)_{k+1/2-N/2}$ TQFT at long distances. For infinitesimal positive $m$ one finds~\toappear\ a new vacuum that is incoming from infinity, in which the fundamental matter field condenses. At low energies we remain with an $SU(N-1)_{k}$ pure vector multiplet. For $k\geq N/2-1/2$ again supersymmetry is unbroken and we have a $SU(N-1)_{k-N/2+1/2}$ TQFT at long distances. 

The Witten indices match as follows for $k-1/2\geq N/2$:
$$\left(\matrix{ N/2+k-3/2\cr N-1}\right)+\left(\matrix{ N/2+k-3/2\cr N-2}\right)=\left(\matrix{ N/2+k-1/2\cr N-1}\right)\,.$$
This is nothing but the standard Pascal identity for binomial coefficients.
The extension to $0<k-1/2< N/2$ is obvious: when the coefficients above vanish (written as polynomials in $k$) then the corresponding vacua break supersymmetry dynamically (but the formula continues to hold formally).

The $U(N)_{k+N/2,k}$ vector multiplet with a fundamental matter multiplet ($N_f=1$) behaves in a rather analogous fashion except for one additional small subtlety that needs to be taken into account. 
At large negative mass we get a pure vector multiplet $U(N)_{k+N/2-1/2,k-1/2}$, which (for $k\geq 1/2$) flows to the TQFT $U(N)_{k-1/2,k-1/2}$. At large positive mass we get the infrared TQFT $U(N)_{k+1/2,k+1/2}$. The vacuum that comes from infinity at infinitesimal positive $m$ is a little more subtle. The vacuum expectation value of the scalar field in the fundamental representation breaks the gauge symmetry to $U(N-1)$. The effective theory in this Higgsed vacuum is   that of an $U(N-1)$ vector multiplet with a Chern-Simons term at level $U(N-1)_{k+N/2,k+1/2}$. The shift of the $U(1)$ level by $+1/2$ requires an explanation. The point is that the Lagrangian of a $U(M)_{P,Q}$ Chern-Simons theory is 
$${P\over 4\pi }\Tr A\wedge dA+{Q-P\over 4\pi M} \Tr A\wedge \Tr dA~.$$
Therefore, if the $U(M)$ gauge symmetry is Higgsed to $U(M-1)$, we would find a $$U(M-1)_{P,Q+(P-Q)/M}$$ Chern-Simons theory. This is why we get the theory of an $\CN=1$ vector multiplet $U(N-1)_{k+N/2,k+1 /2}$ which in the infrared flows to the TQFT $U(N-1)_{k+1/2,k+1 /2}$. 
\smallskip

Using level/rank duality $$U(M)_{P,P}\leftrightarrow SU(P)_{-M}~,$$
we can relate all the phases that we found above in the dynamics of the $SU$ gauge theory to the phases of the $U$ theory. This leads us to the following duality 
\eqn\Dualityintroi{ U\left(N\right)_{k+N/2+1/2,k+1/2}+\Phi \longleftrightarrow SU(k+1)_{-N-k/2}+\Psi~.}
The duality map of deformations    is $\Phi\Phi^\dagger \longleftrightarrow - \Psi\Psi^\dagger$.

It is important to note that in this duality transformation the vacuum that comes from infinity in the SU theory is exchanged with a vacuum that is visible at large mass in the U theory and vice versa. This is allowed since the duality is valid near the conformal field theory (which occurs at finite distance away from the wall at $m=0$) where these vacua merge. We would like to mention that, for instance, if we take $k$ to be very large on the left-hand side of~\Dualityintroi, the conformal field theory is weakly coupled and can be analyzed explicitly~\refs{\AvdeevZA,\AvdeevJT} (both sides of~\Dualityintroi\ can be furthermore explicitly analyzed in the 't Hooft limit~\refs{\JainGZA,\Oferfuture}). There are typically several close-by fixed points, and one of them has emergent $\CN=2$ supersymmetry (and a $U(1)_R$ symmetry).\foot{If the gauge group is $U(1)$ and the Chern-Simons level is large enough, there is only one fixed point with emergent $\CN=2$ supersymmetry.} To understand the duality~\Dualityintroi\ in more detail, it is thus crucial to go beyond the analysis of the phases of the theory that we have undertaken here and study the precise mapping of the quartic operators on the two sides. This can be carried out along the lines of~\refs{\JainGZA,\Oferfuture}, where the same duality was studied in the 't Hooft limit. 
The duality~\Dualityintroi\ was summarized in~\Dualityfig\ in the introduction. 

The duality~\Dualityintroi\  is very similar to 
non-supersymmetric boson/fermion dualities and to $\CN=2$ Giveon-Kutasov dualities~\GiveonZN. It is not surprising that such a duality holds. 
We have arrived at this 
duality by studying in detail the walls in parameter space where the Witten index jumps.  That we find the correct space of ground states and phase transitions is a nontrivial consistency check of our methods. It would be very interesting to extend the analysis to 
a general collection of matter multiplets. The dynamics near the walls is then more complicated, as typically more than one new 
ground state appears from infinity in field space.

We would like to say a few words about the special case $N=1,k=0$. The duality reduces then to 
\eqn\Dualityintroii{ U\left(1\right)_{1/2}+\Phi \longleftrightarrow \Psi~.}
This appears to be a natural generalization of the non-supersymmetric duality between $U(1)_{1/2}+{\rm fermion}$ and the $O(2)$ model~\refs{\ChenCD,\BarkeshliIDA,\SeibergGMD}. 
  
 The phases of the model on the left hand side are a circle of vacua for negative $m$, a new trivial vacuum incoming from infinity at small positive $m$, and a trivial supersymmetric vacuum at large positive $m$. The Witten index for negative $m$ therefore vanishes and the Witten index for positive $m$ is one. 
On the other side of the duality we should interpret $\Psi$ as a complex superfield, namely two real $\CN=1$ multiplets.  

The transition of the theory on the left hand side occurs at some finite positive $m$, where we have a circle and a trivial vacuum on one side of the conformal field theory and a trivial vacuum on the other side of the conformal field theory. 
Since the Witten index of the circle vanishes, it can be that the circle disappears without merging with the trivial supersymmetric vacuum. The conformal field theory then would not involve the trivial vacuum in any essential way.\foot{We thank D.~Gaiotto for useful discussions of this point.}
Let us however consider the possibility that the circle merges with the trivial vacuum. This then leads to a very natural interpretation of the duality~\Dualityintroii. Indeed, let us add a quartic term in $\Psi$, namely $$W=- m\Psi\Psi^\dagger+{1\over 2}(\Psi\Psi^\dagger)^2~.$$ 
The critical point equation is $m\Psi=\Psi^2\Psi^\dagger$ and for   negative $m$ we see clearly that there is only one solution $\Psi=0$, while for positive $m$ we have two kinds of solutions $$m>0: \quad \Psi=0,\Psi=e^{i\tilde \phi}\sqrt m~,$$
with arbitrary $\tilde \phi$, parameterizing a circle. 
Therefore, the phase transition precisely agrees with what we have discussed for the $U(1)_{1/2}+\Phi$ theory, once we flip the sign of $m$ in the dictionary and include the term $|\Psi|^4$ in the superpotential.
From the point of view of the renormalization group, the theory $W=|\Psi|^4$ can be viewed as a marginally irrelevant deformation of the free $\CN=1$ theory of two real multiplets. Therefore, in the very deep infrared at the critical point we have emergent $\CN=2$ supersymmetry as in the other $U(1)$ gauge theories we discussed in the previous subsection. But here the quartic interaction is an important $\CN=1$ marginally irrelevant deformation that allows for a circle of supersymmetric vacua to exist on one side of the transition but not on the other.

\bigskip    
  \noindent {\bf Acknowledgments:}

We would like to thank  O.~Aharony, B.~Assel, C.~Choi, D.~Gaiotto, N.~Ishtiaque, H. Kim, B. Le Floch, N. Nekrasov, S. Pufu, M. Rocek and N. Seiberg for useful discussions.  
V.B.   acknowledges partial support from the MIUR-PRIN Project ÒNon-perturbative Aspects Of Gauge Theories And StringsÓ (Contract 2015MP2CX4). V.B.  and A.S. gratefully acknowledge  support from the Simons Center for Geometry and Physics, Stony Brook University at which some of the research for this paper was performed. This research was supported in part by
Perimeter Institute for Theoretical Physics. Research at Perimeter Institute is supported by the
Government of Canada through Industry Canada and by the Province of Ontario through the Ministry
of Research and Innovation. J.G. also acknowledges further support from an NSERC Discovery
Grant and from an ERA grant by the Province of Ontario.  Z.K. is supported by the Simons
Foundation grant 488657 (Simons Collaboration on the Non-Perturbative Bootstrap).
Any opinions, findings, and conclusions or recommendations expressed in this material are those of the authors and do not necessarily reflect the views of the funding agencies.

\appendix{A}{Relation to $\CN=2$ Duality}

We now relate our proposed  small $k$ behaviour of $\CN=2 $ $SU(N)_k $ vector multiplet in section 3.6 with  consequences that stem from  $\CN=2$ dualities that have appeared in the literature. This gives nontrivial evidence for our picture.  

Consider the duality in~\JafferisNS\foot{The $S^3$ partition function of both sides agree once a factor of $2$ that was missing in \JafferisNS\ is added, and that corresponds to the contribution of the $U(1)_2$ TQFT on the right hand side.}
\eqn\DualityAppetizer{
	\CN=2 \;SU(2)_1 + {\rm adjoint\;}X \longleftrightarrow {\rm free }  \ \CN=2 \ {\rm chiral\; multiplet\;}u + U(1)_2~\hbox{TQFT}~,}
	and its generalization\foot{We extend the duality from $U(N)$ to $SU(N)$ gauge group.} to arbitrary $N$ \KapustinVZ\ 
	\eqn\KapustinGeneralization{
\CN=2\; SU(N)_1 + \ {\rm adjoint\;} X \longleftrightarrow   \CN=2 \ {\rm free\ chirals } \ u_1,...,u_{N} \; + U(1)_{-N} ~\hbox{TQFT}~.}
We like to stress that these dualities   hold only after the TQFT on the right hand side is added.  This will be crucial for us in what follows.

Assuming~\DualityAppetizer\ then the following immediately follows: adding a superpotential mass for the adjoint chiral on the left hand side, we find that the left hand side   flows to the $\CN=2\;SU(2)_1$ vector multiplet. On the right hand side, this deformation amounts to adding a linear superpotential $ W\propto u $. This breaks $\CN=2 $ supersymmetry spontaneously, leading to a Dirac goldstino. Thus the right hand side flows to $U(1)_2~\hbox{TQFT} +\hbox{Dirac Goldstino}$.

Analogously, adding a superpotential mass for the adjoint chiral on the left hand side of \KapustinGeneralization\ we get the $\CN=2\;SU(N)_1$ vector multiplet. This is realized on the right hand side of \KapustinGeneralization\ by the linear superpotential   $ W\propto u_1 $. This   breaks supersymmetry,\foot{The rest of the chiral multiplets become massive by virtue of having a nontrivial K\"ahler potential.}
and the right hand side  flows to  $U(1)_{-N}~\hbox{TQFT} +\hbox{Dirac Goldstino}$

This should be contrasted with our general proposal for $0<k<N/2$ $$ \CN=2 \;SU(N)_k \longleftrightarrow U(k)_{k-N,-N} + {\rm Dirac\;goldstino}\,.$$
Setting $k=1$ we indeed obtain
\eqn\proposal{
\CN=2 \;SU(N)_1 \longleftrightarrow U(1)_{-N} + {\rm Dirac\;goldstino}\,,}
and our proposal for $k=1$  is precisely what  we  obtained from the dualities~\DualityAppetizer\  and \KapustinGeneralization\  by adding a superpotential deformation on both sides.

\appendix{B}{The General Case with $k=0$}

For  $SU(N)$  we have that the non-perturbative superpotential along the Coulomb branch is
$$W(U)\sim \sum_{i}   e^{-(U_i-U_{i+1})}\,.$$
This can be combined with the mass deformation of the $\CN=1$ adjoint multiplet to the following $\CN=1$ superpotential 
  $$W\sim \sum_i\ e^{-(\Phi_i-\Phi_{i+1})}\cos(\Gamma_i-\Gamma_{i+1})+ m\sum_i \Phi_i^2,$$
  where we have used that $U_i=\Phi_i+i \Gamma_i$.
From the supersymmetric vacua equations we find that $\Gamma_i=0$ and   
$$e^{\Phi_{i-1}-\phi_i}-e^{\Phi_{i}-\phi_{i+1}}+ m \Phi_i=0.$$
It is not difficult to prove that the system above can have only one solution and there is a unique massive supersymmetric vacuum for both signs of $m$.

\appendix{C}{Proof of identity~\identity.}

 We give an elementary proof of 
 \eqn\identity{\sum_{P} (-1)^{N-L}\prod_{I=1}^L {k! \over  S_I! (k-S_I)!}={(N+k-1)! \over  N! (k-1)!}~.}
 
 We start with the generating function $F(k)=-\sum_{S_I=1}^k \left(\matrix {k \cr S_I}  \right) x^{S_I}$. We next consider the function
 \eqn\sumofks{G_k=F_k+F_k^2+...}
 Next, we expand 
 $$G_k=\sum_m d_k^m x^m~.$$
 It is easy to observe that
 $$d_k^N=\sum_P (-1)^L\prod_{I=1}^L \left(\matrix {k \cr S_I}  \right)~.$$
 Indeed, considering~\sumofks, we see that the first term on the right hand side would correspond to a partition of $N$ into one term, 
 the term $F_k^2$ would correspond to a partition of $N$ into two terms, etc. The coefficient $(-1)^L$ comes from the minus sign in the 
 definition of $F_k$.
 
Now let us compute $G_k$ explicitly. We start from the binomial formula, which leads to $F_k(x)=-(1+x)^k+1$ and hence 
$$G_k={F_k\over 1-F_k}={-(1+x)^k+1\over (1+x)^k}=-1+(1+x)^{-k}$$
We then use the standard result 
$$(1+x)^{-k}=\sum{1\over j!}(-k)(-k-1)\cdots (-k-j+1) x^j=\sum_j (-1)^{j}\left(\matrix{k+j-1\cr j}\right) x^j$$
to infer that 
$$d_k^N=(-1)^{N}\left(\matrix{k+N-1\cr N}\right)~,$$
which finishes the proof.

\listrefs

\bye